\documentclass[twocolumn]{aastex62}

\usepackage{graphicx}   
\usepackage{grffile}       
\usepackage{amsmath}	
\usepackage{amssymb}	
\usepackage{indentfirst}
\usepackage{multirow}
\usepackage{caption}
\usepackage{subcaption}
\graphicspath{{./}{figures/}}

\received{2018 July 19}
\revised{2019 April 4}
\accepted{2019 April 27}
\submitjournal{ApJ}

\shorttitle{Comparison of Theoretical Starburst Photoionisation Models for Optical Diagnostics}
\shortauthors{D'Agostino et al.}

\begin{document}

\title{Comparison of Theoretical Starburst Photoionisation Models for Optical Diagnostics}

\correspondingauthor{Joshua J. D'Agostino}
\email{joshua.dagostino@anu.edu.au}

\author{Joshua J. D'Agostino}
\affil{Research School of Astronomy and Astrophysics, the Australian National University, Cotter Road, Weston, ACT 2611, Australia}
\affiliation{ARC Centre of Excellence for All Sky Astrophysics in 3 Dimensions (ASTRO 3D)}

\author{Lisa J. Kewley}
\affil{Research School of Astronomy and Astrophysics, the Australian National University, Cotter Road, Weston, ACT 2611, Australia}
\affiliation{ARC Centre of Excellence for All Sky Astrophysics in 3 Dimensions (ASTRO 3D)}

\author{Brent Groves}
\affil{Research School of Astronomy and Astrophysics, the Australian National University, Cotter Road, Weston, ACT 2611, Australia}
\affiliation{ARC Centre of Excellence for All Sky Astrophysics in 3 Dimensions (ASTRO 3D)}

\author{Nell Byler}
\affil{Research School of Astronomy and Astrophysics, the Australian National University, Cotter Road, Weston, ACT 2611, Australia}
\affiliation{ARC Centre of Excellence for All Sky Astrophysics in 3 Dimensions (ASTRO 3D)}

\author{Ralph S. Sutherland}
\affil{Research School of Astronomy and Astrophysics, the Australian National University, Cotter Road, Weston, ACT 2611, Australia}
\affiliation{ARC Centre of Excellence for All Sky Astrophysics in 3 Dimensions (ASTRO 3D)}

\author{David Nicholls}
\affil{Research School of Astronomy and Astrophysics, the Australian National University, Cotter Road, Weston, ACT 2611, Australia}
\affiliation{ARC Centre of Excellence for All Sky Astrophysics in 3 Dimensions (ASTRO 3D)}

\author{Claus Leitherer}
\affil{Space Telescope Science Institute, 3700 San Martin Drive, Baltimore, MD 21218, USA}

\author{Elizabeth R. Stanway}
\affil{Department of Physics, University of Warwick, Gibbet Hill Road, Coventry, CV4 7AL, UK}



\begin{abstract}

We study and compare different examples of stellar evolutionary synthesis input parameters used to produce photoionisation model grids using the \textsc{mappings v} modelling code. The aim of this study is to (a) explore the systematic effects of various stellar evolutionary synthesis model parameters on the interpretation of emission lines in optical strong-line diagnostic diagrams, (b) characterise the combination of parameters able to reproduce the spread of local galaxies located in the star-forming region in the Sloan Digital Sky Survey, and (c) investigate the emission from extremely metal-poor galaxies using photoionisation models. We explore and compare the stellar input ionising spectrum (stellar population synthesis code [Starburst99, SLUG, BPASS], stellar evolutionary tracks, stellar atmospheres, star-formation history, sampling of the initial mass function) as well as parameters intrinsic to the H \textsc{ii} region (metallicity, ionisation parameter, pressure, H \textsc{ii} region boundedness). We also perform a comparison of the photoionisation codes \textsc{mappings} and \textsc{cloudy}. On the variations in the ionising spectrum model parameters, we find that the differences in strong emission-line ratios between varying models for a given input model parameter are small, on average ${\sim} 0.1$ dex. An average difference of ${\sim} 0.1$ dex in emission-line ratio is also found between models produced with \textsc{mappings} and \textsc{cloudy}. Large differences between the emission-line ratios are found when comparing intrinsic H \textsc{ii} region parameters. We find that low-metallicity galaxies are better explained by a density-bounded H \textsc{ii} region and higher pressures better encompass the spread of galaxies at high redshift. 

\end{abstract}



\section{Introduction}

The ratios of emission lines have been used to separate extragalactic objects according to their dominant power source (referring to ionisation and excitation) since 1981, after \citet{BPT1981} pioneered the technique. \citet{BPT1981} used the combination of two ratios created from strong optical emission lines, including the classical [N \textsc{ii}]$\lambda$6584/H$\alpha$ versus [O \textsc{iii}]$\lambda$5007/H$\beta$ or `BPT' diagram, to demonstrate that the ionisation mechanism of emission-line regions could be determined. 

\citet{VO1987} revised and formulated this method, describing how diagnostic line ratios should be selected, including proximity in wavelength to minimise reddening corrections while maximising diagnostic power. They found that the BPT diagram, along with the [O \textsc{iii}]$\lambda$5007/H$\beta$ versus [S \textsc{ii}]$\lambda$(6716+6731)/H$\alpha$ and [O \textsc{iii}]/H$\beta$ versus [O \textsc{i}]$\lambda$6300/H$\alpha$ diagrams could clearly distinguish between galaxies dominated by star-formation from galaxies containing an active galactic nucleus (AGN). 

The theory underlining this clear separation was demonstrated by \citet{Kewley2001} using modelled stellar ionising spectra and photoionisation models. They found that the emission from star-forming galaxies (i.e.~H \textsc{ii} regions) could populate only a certain region of the diagnostic diagrams, even after accounting for variation in the gas-phase abundances and gas ionisation state. Only a hard ionising radiation field such as from the accretion disk surrounding the supermassive black hole could drive emission lines to the region in the diagnostic diagram populated by AGN. Using these models they parameterised an extreme starburst line, to be used in classifying galaxies into starburst or AGN type. They also suggested that the distance of a galaxy from this line could be used to determine the fractional contribution of an AGN to a galaxy spectrum. 

\begin{figure}
\centering
\includegraphics[width=1.1\columnwidth,height=6cm]{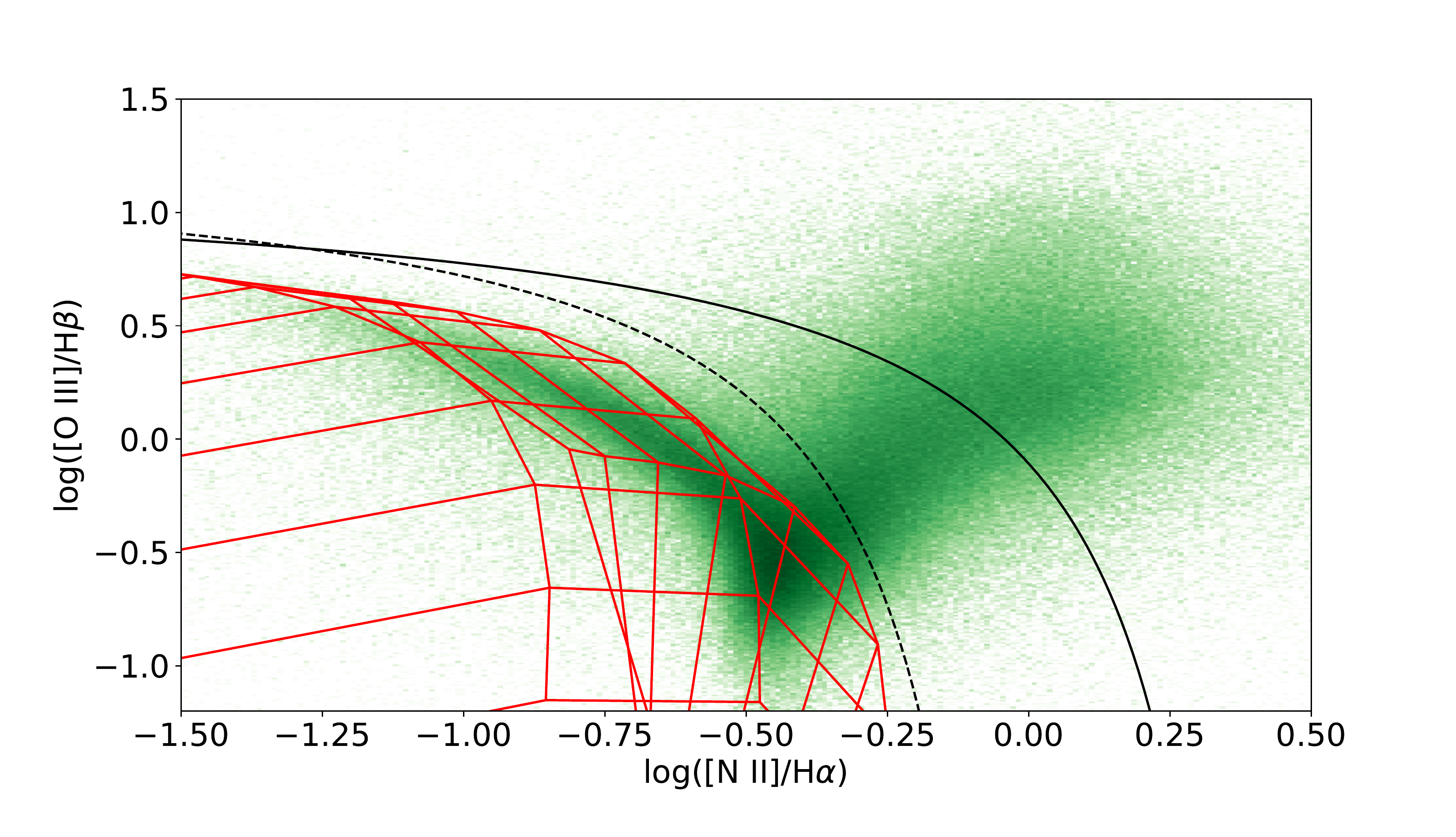}
\caption{Example of a photoionisation grid used by many in the astrophysics community in recent years, overlaid on the distribution of galaxies from SDSS DR7 \citep{SDSS7}.}
\label{fig:classicgrid}
\end{figure}

The large number of galaxy spectra from the Sloan Digital Sky Survey \citep[SDSS;][]{York2000,Stoughton2002} revealed fully where galaxies could lie in the [O \textsc{iii}]/H$\beta$ vs [N \textsc{ii}]/H$\alpha$ and [O \textsc{iii}]/H$\beta$ vs [S \textsc{ii}]/H$\alpha$ diagrams. Measuring the strong emission lines in a sample of roughly 122,000 galaxies in a redshift range of $0.02 < z < 0.3$, \citet{Kauffmann2003} found that galaxies showed a continuous distribution from being star-formation dominated to being AGN dominated (see Fig.~\ref{fig:classicgrid}). Based on this distribution, they created an empirical diagnostic line for separating AGN galaxies from star-forming galaxies on the BPT diagram, based on the theoretical shape of the \citet{Kewley2001} line, with this diagnostic developed further by \citet{Kewley2006}. 

While both the \citet{Kewley2001} and \citet{Kauffmann2003} lines signify the upper limit of star formation from a theoretical and empiral point of view, respectively, the line described by \citet{Kewley2001} predicts much larger BPT emission-line ratios for maximum star formation than the line by \citet{Kauffmann2003}. This difference is the result of using the PEGASE stellar population synthesis (SPS) code \citep{PEGASE}, which uses the \citet{CM1987} planetary nebula nuclei (PNN) atmosphere models to model Wolf-Rayet stars. These PNN atmospheres produce spectra that are harder in the 1-4 ryd range than spectra modelled using other W-R atmospheres explored by \citet{Kewley2001}, and hence these models were used to classify the extreme theoretical starburst line. Furthermore, the lines of \citet{Kewley2001} and \citet{Kauffmann2003} are slightly different in their interpretations; the \citet{Kewley2001} line signifies the extreme upper limit above which star formation can no longer be considered the dominant power source within the galaxy. The \citet{Kauffmann2003} line, however, was derived as a pure star formation line, above which the contribution to the emission-line ratios from star formation is no longer 100\%. The boundary of pure star-forming galaxies, which was reduced empirically by \citet{Kauffmann2003}, was further refined through photoionsation modelling by \citet{Stasinska2006} and was shown to be similar to the \citet{Kauffmann2003} demarcation line.

Contemporary work considers the region of the BPT diagram found below the \citet{Kauffmann2003} line as the pure star-forming classification region and the region found above the \citet{Kewley2001} line as the pure AGN classification region. Even with this clear separation of ionisation mechanism, galaxies show a large spread on diagnostic emission-line diagrams such as the BPT. Within each classification region, such as the star-formation-dominated region, this spread is driven by the variation of the physical conditions within the emission-line regions, such as gas-phase metallicity, ionisation state, and the age of the stellar population. Galaxies, or integral field unit (IFU) spaxels, found between the two lines of \citet{Kewley2001} and \citet{Kauffmann2003} are considered to have multiple ionisation mechanisms (such as AGNs, shocks and/or star-formation) occurring within the same galaxy.

Photoionisation modelling can be used to determine the physical conditions within the galaxies based on their emission-line ratios, including the gas-phase metallicity, density, and ionisation parameter. Yet the interpretation of emission lines relies greatly on the photoionisation models used to provide the calibrations, as demonstrated by \citet{Kewley2001}, \citet{Groves2004}, \citet{Stasinska2006}, and \citet{Levesque2010}.

These photoionisation models are heavily dependent on the input ionising spectrum. Without independent confirmation, the wide range of possible ionising spectra leads to large systematic uncertainties on the derived parameters from emission-line galaxies using these models. In the case of star-forming galaxies, large differences in resulting emission lines are observed between models owing to different stellar atmospheres, different stellar evolutionary tracks, and differences in treating stellar winds, star formation histories (SFHs) and even binary star evolution \citep[e.g.][]{Kewley2001,Morisset2016,Stanway2016,Wofford2016}. As such, differences between the resulting theoretical emission-line spectra are indications of the true systematic uncertainties in these models. 

Furthermore, despite the ongoing work and development of these stellar population models and photoionisation grids, as of yet no model can explain the entire sequence of local star-forming galaxies in the SDSS galaxy survey, particularly in the low-metallicity regime. This problem has been further highlighted by recent observations of high-redshift ($z \approx 2-3$) galaxies \citep[e.g.][]{Kewley2013a,Kewley2013b,Steidel2014}, which reveal that star-forming galaxies at these epochs appear offset to higher [O \textsc{iii}]/H$\beta$ values for given [N \textsc{ii}]/H$\alpha$ values when compared to SDSS galaxies.

Our work is motivated by three main aims: (i) to explore how the emission-line ratios on the BPT diagram are affected by the differences in the parameters used and physics within the SPS models; (ii) to explain the offset of high-redshift galaxies from the local star-forming spread on the BPT diagram; and (iii) to explain the position of extremely metal-poor star-forming galaxies on the BPT diagram. In the coming years, telescopes such as the \textit{James Webb Space Telescope} and \textit{Wide Field Infrared Survey Telescope} will allow strong optical lines to be obtained for unprecedented numbers of galaxies at intermediate and high redshift.  Understanding the effects of intrinsic interstellar medium (ISM) properties on emission lines is critical for the success of these future surveys. Furthermore, the growing wealth of data from recent IFU surveys such as SAMI \citep{Bryant2015}, MaNGA \citep{Bundy2015}, CALIFA \citep{CALIFA1,CALIFA2} and TYPHOON/PrISM (M. Seibert et al. in preparation) has made it increasingly necessary to perform an in-depth study into these photoionisation model grids and their systematics.

In this paper, we present a comparison of photoionisation model grids, extending the work of previous photoionisation model studies, such as those perfomed by \citet{Kewley2001} and \citet{Levesque2010}. We reference several galaxy samples throughout this paper, and describe the samples in Section~\ref{sec:sample}. We use the photoionisation modelling code \textsc{mappings v} and detail the updates made from the previous version, \textsc{mappings iv}, in Section~\ref{sec:mappings}. In Section~\ref{sec:radfield}, we introduce the various parameters associated with the input ionising radiation field that we explore, detailing differences amongst varying input models. Such parameters explored are the SPS code, stellar evolutionary tracks, stellar atmospheres, SFH, and stellar age, as well as the sampling of the initial mass function (IMF). In Section~\ref{sec:photmodels} we describe the photoionisation modelling input parameters used to produce the H \textsc{ii} region; several of the parameters shown in Section~\ref{sec:photmodels} are also explored throughout the scope of this paper, such as the H \textsc{ii} region pressure  and H \textsc{ii} region boundedness. Section~\ref{sec:modelcomp} showcases the systematic differences in each of the parameters explored, by firstly assessing the differences in the spectra in Section~\ref{sec:modelcomp1} before exploring differences in the resulting BPT diagrams in Section~\ref{sec:modelcomp2}. All BPT diagrams include the \citet{Kewley2001} and \citet{Kauffmann2003} lines for reference. We address our aims and raise further points for discussion in Section~\ref{sec:disc} before providing our concluding statements in Section~\ref{sec:conc}.

\section{Sample description}
\label{sec:sample}

In this paper, we discuss and compare photoionisation models to several datasets. The first is SDSS Data Release 7 \citep[DR7;][]{SDSS7}, featuring spectra from approximately 930,000 galaxies combined from the SDSS Legacy Survey \citep{SDSSDR1}, SEGUE \citep{segue}, and the SDSS-II SN survey \citep{sdsssn}.

Second, we also consider a combined sample of metal-poor galaxies, constituting the Small Isolated Gas Rich Irregular Dwarf galaxy (SIGRID) sample from \citet{Nicholls2014c}, and low-metallicity subsets of SDSS, described in \citet{Izotov2006,Izotov2012}. The sample taken from \citet{Izotov2006} is truncated to only include galaxies with $Z < Z_\odot$ to compare with the models used in \citet{Nicholls2014c}. The galaxies in the SIGRID sample were selected for observation based on the criteria detailed in \citet{Nicholls2011}. A further selection criterion was the detection of the [O \textsc{iii}]$\lambda$4363 line, making possible calculations of the electron temperature and the gas-phase oxygen abundance. A similar cut was made on the data in \citet{Izotov2006} and \citet{Izotov2012}, specifically stating that the [O \textsc{iii}]$\lambda$4363 line should have a detection greater than 1$\sigma$ and 2$\sigma$, respectively.

\section{MAPPINGS V}
\label{sec:mappings}

In this work, we make use of the photoionisation modelling code \textsc{mappings v}. This version is a significant improvement on the previous version \textsc{mappings iv}. A summary of the previous version of \textsc{mappings iv} can be found in \citet{Nicholls2014c} and references therein. Major changes from \textsc{mappings iv} to \textsc{mappings v} include updated atomic data, including version 8 of the CHIANTI database \citep{CHIANTI,CHIANTI8}, which gives atomic data for up to 80,000 spectral lines. Updated elemental depletion files are also included from \citet{Jenkins2014}. A much higher precision exponential integral function is included for calculation of the collisional ionisation rates, based on the functional form of \citet{CT1968} and \citet{CT1969}, giving enhanced stability and accuracy (to at least 18 significant figures) during the modelling. Optical, near-UV and near-IR H \textsc{i} and He \textsc{ii} wavelengths were reevaluated, although the wavelengths of hydrogen lines are largely similar. The new and former values of the optical, near UV, and near IR H \textsc{i} and He \textsc{ii} lines agree to four significant figures, and in some cases five significant figures. Updates were also made to input spectral libraries such as the AGN atmosphere library and the planetary nebula stellar library.

Photoionisation model codes such as \textsc{mappings} \citep[e.g.][]{SD1993} and \textsc{cloudy} \citep[e.g.][]{Ferland2013} require the input of nebula-specific parameters, as well as an input ionising spectrum. In the following sections we detail the parameters involved in the \textsc{mappings} simulations by first describing the parameters used to produce the input stellar ionising spectrum, followed by other necessary parameters to produce the models. Unless otherwise stated, the individual parameters mentioned in each section are to be assumed for all models in this work.

\section{Stellar radiation field}
\label{sec:radfield}

In this section, we compare various models that constitute the input ionising stellar radiation field. A summary of all the models used, along with references, can be found in Table~\ref{tab:stellar}.

\begin{table*}
\centering
\begin{tabular}{|c|c|p{0.3\linewidth}|}
\hline
\hline
\textbf{Parameter} & \textbf{Models} & \multicolumn{1}{c|}{\textbf{Reference}} \\
\hline
\multirow{3}{*}{Stellar population synthesis code} & Starburst99 (SB99) & \citet{Leitherer1999} \\
& Stochastically Lighting Up Galaxies (SLUG) & \citet{daSilva2012, Krumholz2015} \\
& Binary Population and Spectral Synthesis (BPASS) & \citet{Eldridge2008, ES2009, ES2012, Stanway2016} \\
\hline
\multirow{2}{*}{Stellar evolutionary tracks} & Geneva HIGH & \citet{Meynet1994} \\
& Padova TP-AGB & \citet{Girardi2000,VW1993} \\
\hline
\multirow{3}{*}{Stellar atmospheres} & Kurucz & \citet{Lejeune1997} \\
& Hillier + Miller & \citet{HM1998} \\
& Pauldrach & \citet{Pauldrach2001} \\
\hline
\hline
\end{tabular}
\caption{Various stellar radiation field models used throughout this paper, with references.}
\label{tab:stellar}
\end{table*}

\subsection{SPS Codes}
\label{sec:spscodes}

\subsubsection{Single Stellar Populations}

SPS codes compute the spectrum of a stellar population of given physical parameters such as SFH, age, and metallicity. These codes take theoretical stellar evolutionary tracks and populate these tracks with stars, distributed according to an IMF describing the number of stars per unit mass at the beginning of the formation of the stellar cluster, and assuming an age for the stars. The spectral output of the stellar population is determined through matching these stars with libraries of stellar atmosphere spectra. The final output is created by integrating simple stellar populations of given ages over the SFH of the stellar population -- the rate at which stars have formed over time. 

Starburst99 \citep[SB99;][]{Leitherer1999, VL2005, Leitherer2014} and Stochastically Lighting Up Galaxies \citep[SLUG;][]{daSilva2012, Krumholz2015} are two SPS codes that are quite different in nature. Whilst both codes produce ionising spectra and photometry for star clusters and galaxies given parameters such as the IMF, SFH, cluster mass function (CMF), cluster lifetime function (CLF), and possibly extinction, SLUG differs from SB99 in that it carries out explicit Monte Carlo sampling from the probability distribution functions (pdf's) describing the stellar population (such as IMF) and thus returns the pdf of output spectra rather than just the average \citep[see][for a detailed explanation on the methodology and technique of SLUG]{daSilva2012}. 

A major difference between the two codes is the interpolation method used to create isochrones (single age sequences) from the stellar evolutionary tracks and to interpolate from the stellar atmosphere libraries onto the stars populating these isochrones. These interpolation methods can have a large impact on the final spectrum, especially in the presence of poor sampling of the evolutionary tracks (described in Section \ref{sec:tracks1}) or spectral libraries (described in Section \ref{sec:atms1}). SB99 uses quadratic interpolation along the isochrones in the `isochrone synthesis' technique, described in detail in \citet{CB1991}. SLUG uses the same isochrone synthesis technique but uses the `Akima spline' \citep{Akima1970} in place of quadratic interpolation. The Akima spline is more robust to outliers in comparison to quadratic interpolation, thus ensuring that its isochrone interpolation is performed to a higher accuracy. Interpolation via the Akima spline requires five points, with the point of interest at the centre. In the absence of five points, SLUG reverts to linear interpolation. For comparison, SB99 only reverts to linear interpolation once only two points are available.

\subsubsection{Binary Populations}

We also explore the differences in output spectra between single stellar population models and the Binary Population and Spectral Synthesis code \citep[BPASS; see][]{Eldridge2008, ES2009, ES2012, Stanway2016}. The findings of \citet{Stanway2016} when considering binary populations, amongst other findings, include a boosted (50-60\%) ionising flux in stellar populations at low metallicities ($0.05Z_\odot \leq Z \leq 0.3Z_\odot$) and a more modest 10-20\% increase in the flux at higher (near-solar) metallicities, compared to single-star stellar populations. \citet{ES2009} show that binary populations tend to be bluer with fewer red supergiants than in single stellar populations, leading to a significantly smaller flux in the \emph{I} band and subsequent longer wavelengths. \citet{ES2009} also find W-R stars emerging over a wider range of ages, with populations of ages up to 10 Myr being found to host W-R stars.

BPASS models use a different set of model atmospheres for all stars other than OB stars than those used by SLUG and SB99, explained in Section~\ref{sec:stellaratmos}. For stars with surface temperatures $< 25,000$K, BPASS uses the BaSeL V3.1 model atmosphere library \citep{Westera2002}. Stars with surface temperatures greater than 25,000K are treated as OB stars and are modelled with the high-resolution atmosphere libraries of \citet{SNC2002} \citep[using the OB atmospheres of][]{Pauldrach2001}. The atmospheres of W-R stars (defined in BPASS as stars with a surface hydrogen mass fraction $X < 0.4$ and effective temperature $\mathrm{log}(T_\mathrm{eff}/\mathrm{K}) \geq 4.45$) are modelled using the Potsdam group's theoretical atmospheres \citep[see \citealt{ES2009} for a detailed summary on the use of the Potsdam libraries for differing W-R types]{HGL2006}.

The stellar evolution models used by BPASS are those from the Cambridge \textsc{stars} code \citep[and references therein]{Eggleton1971,Eldridge2008}. The \textsc{stars} models include not only a detailed set of single-star models, but also an extensive set of detailed binary star models. BPASS does not interpolate between the stellar tracks, as binary stellar evolution has more free parameters. Rather, each stellar model is weighted by a Salpeter IMF in order to calculate a stellar evolutionary track for stars of varying initial masses and binary parameters.

The various BPASS models available differ in the slope and endpoints of the IMF used to produce the spectral energy distribution. In order to accurately compare spectra of binaries produced using BPASS to single stellar models, we choose the BPASS model that resembles the IMF described in Section~\ref{sec:imfsec} as closely as possible. As a result, our BPASS model contains an IMF with power-law indices of $\alpha_1 = -1.30$ between masses of 0.1 and 0.5 $M_\odot$, and $\alpha_2 = -2.35$ between masses of 0.5 and 100 $M_\odot$. All instantaneous SFH BPASS models are for a $10^6 M_\odot$ cluster, and all continuous SFH BPASS models are for a cluster undergoing constant star formation at a rate of $1 M_\odot \mathrm{yr}^{-1}$.

\subsection{SFH and Age}

We consider two simplistic treatments for the SFHs of our stellar populations: an instantaneous burst of star formation (i.e.~ a single aged stellar population), and a constant SFH. In both cases we assume that the stellar population is of a single metallicity. Both types of SFH are applicable in different situations. A constant SFH is thought to be a reasonable approximation when modelling the integrated spectra of galaxies, as new stellar populations emerge repeatedly. An instantaneous burst, on the other hand, is assumed when studying individual H \textsc{ii} regions, where a single stellar population dominates the ionising spectrum. Cases where an instantaneous SFH may be applicable include galaxies whose star formation may be triggered by merger events, the sudden accretion of gas clouds within the IGM, or other situations where the star formation phase may be short-lived. 

For the constant SFH models we adopt an age of 5 Myr (that is, a constant star formation of 1 $M_{\odot}\, {\rm yr}^{-1}$ occurring over the past 5 Myr) for both
SB99 and SLUG (known as the `galaxy' model in SLUG). As shown by \citet{Kewley2001}, after 5 Myr, the spectrum of a constant SFH stellar population reaches equilibrium. However, when studying the effects of continuous SFH stellar cluster age on the emission-line ratios, we use cluster ages up to 10 Myr for reasons described in Section~\ref{sec:age2}. For the instanteous burst scenario we explore all ages up to 6 Myr, as ${\sim} 95\%$ of the total ionising radiation emitted by a stellar population over its lifetime has been emitted by this age \citep{Dopita2006}. By exploring the different ages, we demonstrate the impact of short-lived but active phases of a stellar population such as W-R stars on the stellar spectrum.

For our fiducial model we choose the constant SFH at 5 Myr following the work of \citet{Levesque2010}. They report that a continuous SFH produces a better agreement with the emission-line ratios found in their galaxy sample, made up of local ($z < 0.1$) star-forming galaxies from a sample of SDSS described in \citet{Kewley2006}, the Nearby Field Galaxy Survey \citep[NFGS;][]{Jansen2000}, a sample of blue compact galaxies from \citet{KC2002} and \citet{Kong2002}, and a sample of metal-poor galaxies described in \citet{BKG2008}.

\subsection{Initial Mass Function}
\label{sec:imfsec}

We use the IMF of \citet{Kroupa2002}, particularly Equation (5). The IMF describes the number of stars present in the cluster at a given stellar mass. The \citet{Kroupa2002} IMF is a broken power law, consisting of three different power-law segments. The segments start at 0.01$M_{\odot}$, with further breakpoints at 0.08 and 0.5$M_{\odot}$, and ending at 120$M_{\odot}$. The power-law indices within each segment are $\alpha_{1}$ = 0.3, $\alpha_{2}$ = 1.3, and $\alpha_{3}$ = 2.3, respectively. 

\subsection{Stellar Tracks and Metallicities}
\label{sec:tracks}

Stellar evolutionary tracks follow a star of a given initial mass as it evolves during its lifetime from the time on the main sequence until its end point as a supernova or white dwarf. SPS codes use these tracks to interpolate onto an isochrone, representing the distribution of a population of stars of a single age and metallicity.

During this work, we use the Geneva group's ``High" (HIGH) mass-loss tracks. First published in \citet{Meynet1994}, these tracks are intended to better reflect the mass-loss rates of low-luminosity W-R stars and blue-to-red supergiant ratios in both the Large and Small Magellanic Clouds \citep{Schaller1992,Meynet1993} by including higher mass-loss rates than those found in the Geneva group's ``Standard" (STD) mass-loss evolutionary tracks. These high mass-loss rates are derived by doubling the mass-loss rates in the ``standard" tracks for WNL and Population I stars (except W-R stars) found in \citet{deJager1988}, except for WNE W-R stars, as well as WC and WO W-R stars, which were left unchanged.

We compare the Geneva HIGH tracks with stellar evolutionary tracks from the Padova group. The Padova tracks we use during this paper contain thermally pulsing asymptotic giant branch (TP-AGB) stars and are described in \citet{Girardi2000} with an addition from \citet{VW1993}. We make a distinction between the TP-AGB Padova tracks and the non-TP-AGB option described in \citet{Bressan1993}, \citet{Fagotto1994a,Fagotto1994b,Fagotto1994c}, and \citet{Girardi1996}, yet we note that the AGB phase of stellar evolution begins at ${\sim} 100$ Myr \citep{VL2005}. Since the maximum age of our simulations in this work is 10 Myr, AGB stars are not present in the clusters generated. However, the TP-AGB Padova tracks are computed following major updates in the input physics from the non-AGB Padova tracks and so must be distinguished from one another. A summary of the differences in the input physics between the two sets of Padova tracks can be found in \citet{Girardi2000}. On occasion, the TP-AGB Padova tracks may be referred to as simply the ``Padova tracks," and the Geneva HIGH tracks as simply the ``Geneva tracks." The input physics between the Geneva and Padova tracks are different; some differences include the chemical composition and abundances, reaction rates and neutrino losses, equation of state, convection processes, and chemical opacities. \citet{VL2005} provide an excellent and detailed comparison between the two sets of stellar evolutionary tracks.

The Geneva and Padova sets of tracks immediately differ on the values of metallicity used. Both sets of tracks use five metallicities, with three ($Z = 0.004$, $Z = 0.008$, $Z = 0.020$) being equal. They differ at both the low-metallicity end ($Z = 0.001$ for Geneva, $Z = 0.0004$ for Padova) and at the high-metallicity end ($Z = 0.040$ for Geneva, $Z = 0.050$ for Padova), with the Padova tracks ultimately encompassing a larger range of metallicity.

The reaction rates also differ between the two sets, with the Geneva models having been calculated using the reaction rates from \citet{Caughlan1985}, whilst the Padova tracks utilised the rates calculated by \citet{CF1988}. The largest disagreement in the reaction rates between the two sets of tracks is in the $^{12}C(\alpha,\gamma)^{16}O$ rate as part of the CN cycle. The rate adopted by the Padova tracks provides the lower rate and hence provides a lower conversion rate of carbon to oxygen. Overall, this leads to an increase in the lifetime of the helium core burning of several percent.

Both Geneva and Padova use the \citet{deJager1988} parametrisation for mass-loss rates, although the Geneva HIGH mass-loss tracks differ slightly from this parametrisation by doubling the mass-loss rates for WNL W-R stars and Population I stars. Both use identical mass-loss rates for remaining W-R stars, using the mass-dependent mass-loss rates compiled by \citet{Langer1989}. The limit for initialisation of W-R evolution differs for both Geneva and Padova, who adopt a limit of hydrogen surface abundance below 40\% and 30\%, respectively.

The most significant difference between the Geneva and Padova tracks comes in the W-R phase. Due to W-R stars' extended atmospheres, the radius and effective temperature ($T_{\mathrm{eff}}$) of the stellar atmosphere become ambiguous. The calculation for such quantities depends on the definition of opacity in the atmosphere. The Geneva tracks aim to solve this issue by adopting a mean-weighted opacity $\kappa = \sigma_{\mathrm{e}} (1 + \mathrm{FM})$ from the radiation-driven wind theory. Here, $\sigma_{\mathrm{e}}$ is the electron scattering opacity, and FM is the ``force multiplier," which provides contribution from absorption lines. The radius and effective temperature are finally calculated by adding both the electron scattering and mean-weighted opacities. The Padova tracks, however, performed no such correction to the temperature of the W-R stars' atmospheres from outflows. As a result, the definition of effective temperature is the same for all stars in the Padova tracks, leading to an increase in the temperatures of W-R stars when compared to the Geneva models.

\subsection{Stellar Atmospheres}
\label{sec:stellaratmos}

The choice of stellar atmospheres used in the simulation is important in creating the ionising spectrum. Differences in atmospheric opacities can have a large impact in the final ionising spectrum, because emission lines of certain elements may be enhanced or hindered, depending on the favourability of the atmospheres.

Our chosen SPS codes, SB99 and SLUG, have several choices of stellar atmosphere libraries when synthesising the spectrum of a stellar population. The different stellar atmosphere choices we consider are formed from combinations of the three atmosphere models of the \citet{Lejeune1997} atmospheres (hereafter Kurucz), \citet{HM1998} W-R atmospheres (hereafter Hillier), and \citet{Pauldrach2001} OB atmospheres (hereafter Pauldrach). The Kurucz atmospheres can be selected as a stand-alone choice, with the option to include the Hillier W-R atmosphere models (Kurucz + Hillier) and the Pauldrach OB atmosphere models (Kurucz + Pauldrach). The final stellar atmosphere choice consists of the Kurucz atmospheres with both the Hillier W-R models and the Pauldrach OB models (Kurucz + Hillier + Pauldrach; hereafter the SB99 atmosphere). 

Both \citet{HM1998} and \citet{Pauldrach2001} expand on the Kurucz atmospheres modelled by \citet{Lejeune1997} by modelling stars with spherically expanding atmospheres. \citet{HM1998} further expand on the Kurucz atmospheres by including a technique to model line blanketing in W-R stars. The updated W-R atmospheres are shown by \citet{HM1998} to strengthen several optical CNO lines emitted by W-R stars by a factor of 2-5. \citet{Pauldrach2001} improve on the Kurucz atmospheres by including updated metal opacities in the OB stellar atmospheres. The inclusion of these updated opacities was aimed at solving the problems of line blanketing \citep[as with][]{HM1998} and line blocking. The \citet{Pauldrach2001} atmospheres also include an improved atomic data archive, as well as a revised EUV and X-ray radiation model as a result of shock cooling zones in the OB stellar winds.

\section{Photoionisation Models}
\label{sec:photmodels}
\subsection{Abundance Sets}

\begin{figure*}
\centering
\includegraphics[scale=0.6]{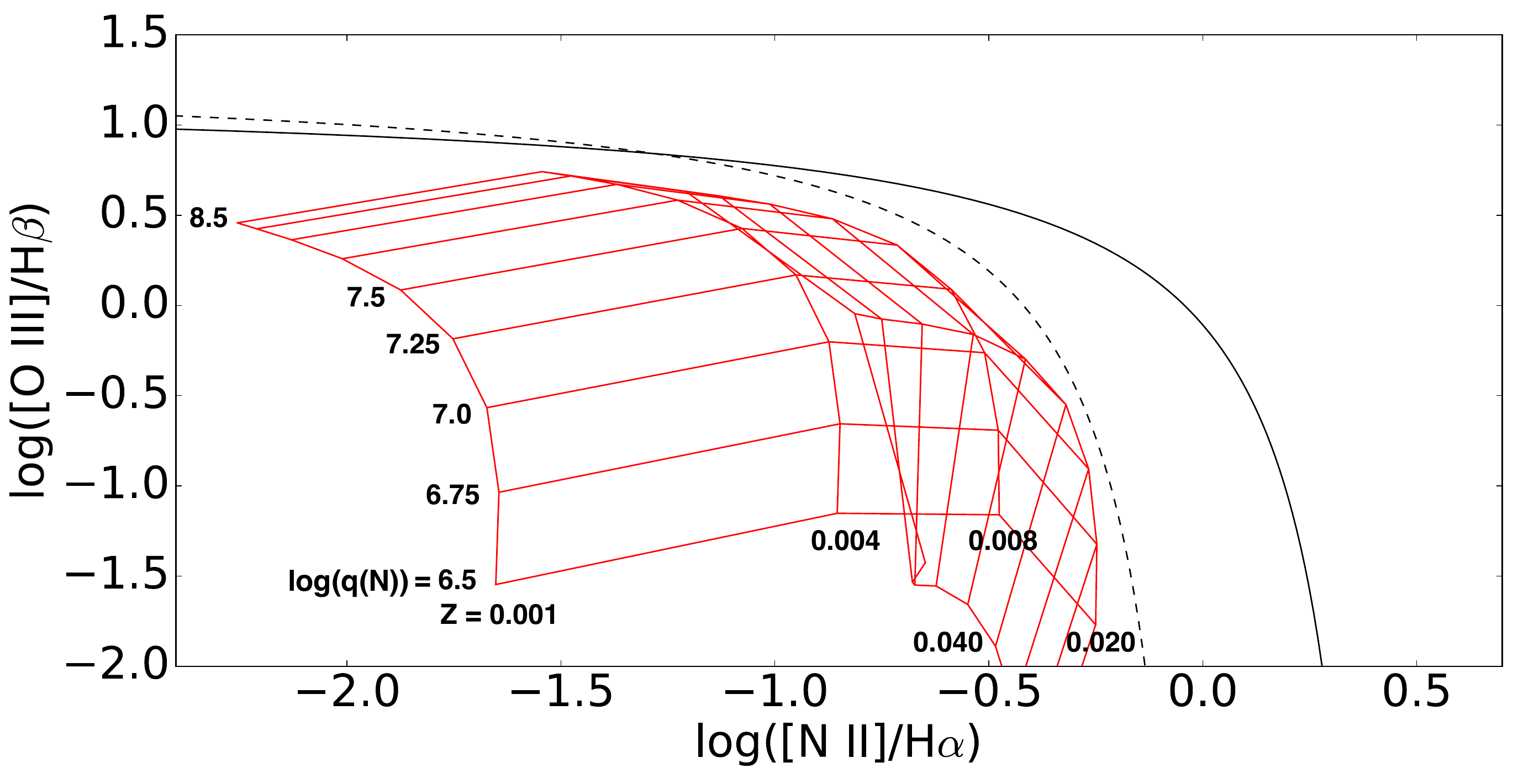}
\caption{Model grid computed with the SLUG SPS code, using the SB99 combination atmosphere and Geneva HIGH tracks, assuming a continuous SFH at 5 Myr, truncated at 99\% hydrogen recombination. The model is annotated showing directions of increasing metallicity $Z$ and ionisation parameter log($q(N)$). Five metallicities are used in the model grids: $Z$ = 0.001, 0.004, 0.008, 0.020, and 0.040. The values of the ionisation parameter used in the model grids range between log($q(N)/(\mathrm{cm}\;  \mathrm{s}^{-1})$) = 6.5 and 8.5 inclusive, in 0.25 dex increments.}
\label{fig:gridannotated}
\end{figure*}

For our nebular models, we adopt the solar reference abundances of \citet{AG1989}, corresponding to a solar value of $Z = 0.020 \equiv 12 + \mathrm{log(O/H)} = 8.93$. Abundances at other metallicities of $Z$ = 0.001, $Z$ = 0.004, $Z$ = 0.008, and $Z$ = 0.040 are obtained by applying the Local Galactic Concordance (LGC) abundance scaling prescription described by \citet{Nichollsabund}. Typically nebular models adopt a linear scaling for abundances with metallicity, with the exception of a few elements (e.g. He, C, and N). The LGC abundances model the individual scaling of elements with the overall nebular metallicity, based on empirical fits to stellar abundance data. These abundances account for the nonlinear scaling of alpha elements (including oxygen) with iron and account for primary and secondary production mechanisms for carbon and nitrogen. The variation with O/H of several elements, such as C, N, and Fe, is shown in \citet{Nichollsabund}. We note that this abundance scale is calibrated from MW stellar abundances and may not be appropriate for dwarf galaxies or galaxies with instantaneous SFHs \citep[see][for a detailed explanation of abundance scaling with metallicity]{Nichollsabund}. The abundance values are found in Table \ref{tab:lgc}. Through our simulations, we match the stellar abundance to the overall metallicity of the stellar evolutionary tracks (with the exception of $Z = 0.050$ Padova tracks, for which we use the $Z = 0.040$ abundance set). Whilst the Padova tracks also differ at the low-metallicity end with a metallicity of $Z = 0.0004$ rather than $Z = 0.001$, SLUG sets the low-metallicity end of the Padova tracks to $Z = 0.001$ for reasons described in Section~\ref{sec:tracksbpt}). The direction of increasing metallicity in the model grids can be seen from Figure~\ref{fig:gridannotated}.

\begin{table}
\centering
\begin{tabular}{|c|c|c|}
\hline
\hline
\textbf{Element} & $\mathbf{12 + \mathrm{\textbf{log}}(X/H)}$ & \textbf{Depletion value} \\
\hline
H & 12.00 & 0.00 \\
He & 10.99 & 0.00 \\
 Li & 1.16 & -0.22 \\
 Be & 1.15 & -0.40 \\
 B & 2.60 & -0.58 \\
 C & 8.56 & -0.16 \\
 N & 8.05 & -0.04 \\
 O & 8.93 & -0.11 \\
 F & 4.56 & -0.09 \\
Ne &  8.09 & 0.00 \\
Na & 6.33 & -0.42 \\
Mg & 7.58 & -0.70 \\
Al & 6.47 & -0.70 \\
Si & 7.55 & -0.71 \\
P & 5.45 & -0.11 \\
S & 7.21 & 0.00 \\
Cl & 5.50 & -0.09 \\
Ar & 6.56 & 0.00 \\
K & 5.12 & -0.62 \\
Ca & 6.36 & -1.95 \\
Sc & 3.10 & -0.69 \\
Ti & 4.99 & -1.95 \\
V & 4.00 & -2.17 \\
Cr & 5.67 & -1.45 \\
Mn & 5.39 & -1.27 \\
Fe & 7.67 & -1.50 \\
Co & 4.92 & -1.64 \\
Ni & 6.25 & -1.57 \\
Cu & 4.21 & -0.90 \\
Zn & 4.60 & -0.20 \\
\hline
\hline
\end{tabular}
\caption{Solar abundance reference set from \citet{AG1989}, and depletion values using the method from \citet{Jenkins2014} for each element with a logarithmic base depletion of iron of -1.50.}
\label{tab:lgc}
\end{table}

\subsection{Depletion Factors}
\label{sec:depfactors}

The depletion factors used are those from \citet{Jenkins2014}, using a logarithmic base depletion of -1.50 for iron. Depletion values for each element are detailed in Table \ref{tab:lgc}. Throughout our simulations, we keep the depletion factors constant as a function of metallicity and set to the diffuse ISM values within the Milky Way. Hence, we neglect the possible variation in the depletion of each element amongst H \textsc{ii} regions.

\subsection{Model Geometry}

The geometry of the H \textsc{ii} region is determined by the radiation source's position relative to the molecular cloud that it ionises. In the simplest case, the geometry of the region may be considered to be spherical if the ionising source is found within the cloud. If the ionising source is external to the molecular cloud that it ionises, the geometry may be considered to be plane-parallel. Further, the geometry of the model can be either `open' or ``closed." Closed-geometry models are applicable when all ionising radiation is assumed to be incident on a molecular cloud (i.e. for a stellar cluster within an H \textsc{ii} region), whereas open-geometry models imply that a fraction of ionising radiation is lost to the ISM (e.g. when modelling a narrow-line region photoionised by the accretion disk of a supermassive black hole). In all simulations, we assume a closed spherical shell assumed to be empty or optically thin, such that the inner surface of the H \textsc{ii} region receives the same unobscured flux from the ionising central cluster. 

\subsection{Pressure and Density}
\label{sec:pandd}

The pressure of the H \textsc{ii} region is important for the resulting emission from the nebula. Differences in pressure alter the density structure of the H \textsc{ii} region. For a constant-pressure (isobaric) model, density increases towards the edge of the H \textsc{ii} region owing to the decrease in temperature farther away from the central stellar cluster, resulting in different mean densities amongst differing ions throughout the nebula. We adopt an isobaric model for our simulations, with a total pressure of $P/k = 8 \times 10^5\;\mathrm{cm}^{-3}\;\mathrm{K}$, derived from an initial total density of $n = 100$ cm$^{-3}$ and initial temperature of $8000$ K.

Setting the initial total density $n = 100$ cm$^{-3}$ is widely supported, within the order of magnitude. \citet{Dopita2006} put a constraint of $n \leq 100 \;\mathrm{cm}^{-3}$ when computing H \textsc{ii} region spectra for pressure values of $\mathrm{log}(P/k)= 6$. Meanwhile, \citet{Kewley2001} found an average electron density of $n = 350 \;\mathrm{cm}^{-3}$ for the \citet{Kewley2000} sample of warm infrared starburst galaxies (maximum redshift $< 0.1$). An average value of $n = 222^{+172}_{-128} \;\mathrm{cm}^{-3}$ was found by \citet{Kashino2017}, using their sample from the FMOS-COSMOS survey ($1.4 \leq  z \leq  1.7$), and also by work done by \citet{Shimakawa2015} and \citet{Sanders2016} ($n = 291$ and $n = 225^{+119}_{-4}|_{\mathrm{[O \textsc{ii}] \;doublet}}, n = 290^{+88}_{-169}|_{\mathrm{[S \textsc{ii}] \;doublet}}$ cm$^{-3}$, at redshifts of $z = 2.5$ and 2.3, respectively). 

However, there have also been reported findings of galaxies with electron densities on the order of $n = 1000 \;\mathrm{cm}^{-3}$ from the likes of the FMOS-COSMOS sample presented in \citet{Kashino2017}, and from high-redshift ($z \approx 3.3$) galaxies shown in \citet{Onodera2016}. This leads \citet{Kashino2017} to conclude that H \textsc{ii} regions in high-redshift galaxies do have an electron density a few to several times larger than that of local galaxies on average. This increase in electron density in high-redshift galaxies is thought to be linked to the increase in star-formation rate (SFR) of galaxies at high redshift \citep[e.g.][]{MD2014}. \citet{Mel2017} show that any offset in the electron density between local and high-redshift star-forming galaxies disappears if comparisons are performed between local and high-redshift galaxies matched in SFR or mass. They show an electron density in local galaxies ($z < 0.1$) in their COSMOS-[O \textsc{ii}] sample of $26.8_{-0.2}^{+0.2}$ cm$^{-3}$, which increases to $98_{-4}^{+4}$ and $98_{-5}^{+5}$ cm$^{-3}$ when choosing galaxies matched to high-redshift galaxies in SFR and both SFR and mass, respectively. For the entire sample at $z {\sim} 1.5$, they measure an average electron density of $114_{-27}^{+28}$ cm$^{-3}$, leading to an agreement within errors for the electron densities at both low and high redshift. Throughout their sample, \citet{Mel2017} also show instances of high-redshift galaxies with electron densities calculated to be ${\sim} 1000$cm$^{-3}$.

\subsection{Ionisation Parameter}

For the ionisation parameter $q(N)$ (ionising photon number density relative to the number density of all ions in the nebula), we use a range of values for $\mathrm{log} \;(q(N)/(\mathrm{cm}\;  \mathrm{s}^{-1})) = 6.5 - 8.5$ with 0.25 dex increments. These values of $\mathrm{log} \;q(N)$ correspond to values of the dimensionless ionisation parameter $\mathrm{log} \;U(N) = q/c \approx $ -4 to -2, with 0.25 dex increments (nine values in total). The direction of increasing ionisation parameter in the model grids can be seen from Figure~\ref{fig:gridannotated}.

\citet{RR2004} find a range of $ -3.2 \lesssim \mathrm{log}\;U(N) \lesssim -0.9$ within local starburst galaxies. Whilst our values of the ionisation parameter are slightly lower than those found by \citet{RR2004}, they compare well with those used by \citet{Kewley2001} within the errors, who use a range of $ -4.3 \lesssim \mathrm{log}\;U(N) \lesssim -2.5$. 

\subsection{Boundedness}

Boundary conditions can have a profound effect on the final emission-line fluxes. Radiation-bounded models represent an H \textsc{ii} region where ionisation and excitation within are limited by the amount of ionising radiation from its ionising source, and it is assumed that all ionising photons are absorbed by the nebula. Radiation-bounded simulations are terminated once a certain fraction of H \textsc{ii} has recombined to H \textsc{i} (usually considered to be a high fraction, such as 95-99\%). 

It is possible that some fraction of the ionising radiation escapes the nebula, meaning that the gas column is not sufficient to absorb all photons. This is known as a `density-bounded' nebula. These density-bounded situations are associated with `leaky H \textsc{ii} regions' \citep[e.g.][]{Pellegrini2012,Zastrow2013}, which may give rise to diffuse ionised gas (DIG; described in Section~\ref{sec:discother}) and play a key role in the reionisation of the universe. Our density-bounded models are truncated at a given optical depth at 13.6 eV, corresponding to the ionisation potential of hydrogen. 

For individual H \textsc{ii} regions, the escape of ionising photons is an important consideration. On typical galaxy scales, estimates place the escape fraction at $< 3$\% \citep[e.g.][]{Inoue2006}. However, for dwarf galaxies, where a star forming region can have a large impact, density-boundedness may be important. \citet{Nicholls2014c} explored the ISM conditions within a sample of dwarf galaxies. They found that these dwarf galaxies were unable to be explained by radiation-bounded photoionisation models. The emission lines of the dwarf galaxies indicated a low metallicity, but with line ratios outside of the standard model predictions. By moving to a density-bounded regime, they found that lower optical depths may in fact offer explanations for the observed emission lines. Through the mass-metallicity relation \citep{Tremonti2004}, the low-metallicity galaxies seen in \citet{Nicholls2014c} are also at a low mass, leading to a weaker gravitational pull on the material within the galaxy. Thus, supernova explosions in these low-mass galaxies have a greater effect on the surrounding material, by expelling and clearing gas far more easily. The result is a higher ionising photon escape fraction due to a larger mean free path, which ultimately results in a decrease in the optical depth of the galaxy \citep{Trebitsch2017}. The same principle can be applied to H \textsc{ii} regions, because supernovae in low-mass galaxies cause more porous H \textsc{ii} regions, thus increasing the ionising photon escape fraction.

\subsection{Fiducial Model}
\label{sec:fiducial}

\begin{figure}
\centering
\includegraphics[width=1.1\columnwidth]{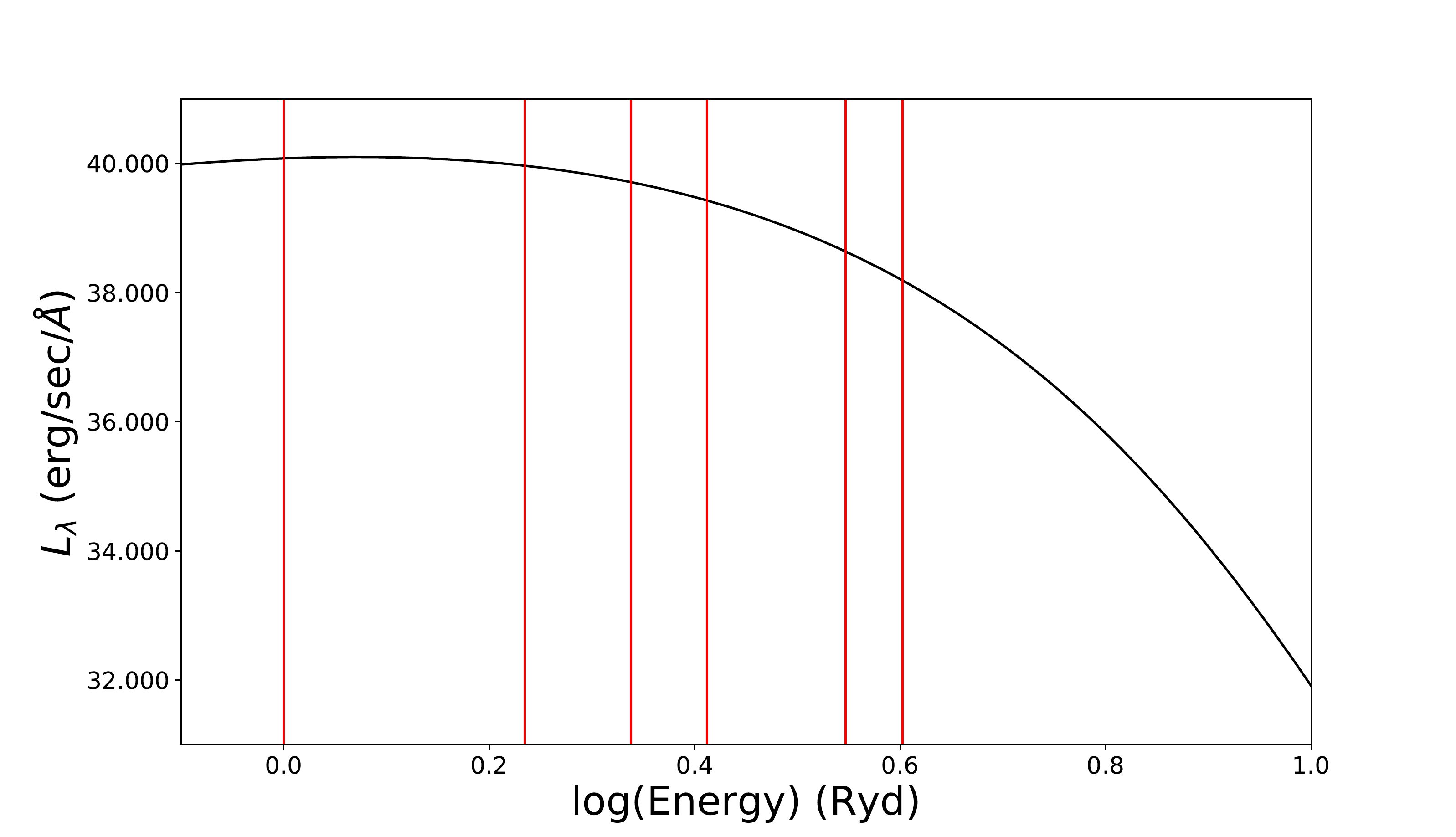}
\caption{SB99 spectrum of a 5 Myr cluster undergoing constant star formation at $1\;M_\odot \;\mathrm{yr}^{-1}$, treating all stars as pure blackbodies. This is the spectrum to which all spectra in Section~\ref{sec:modelcomp1} are first normalised.}
\label{fig:refspec}
\end{figure}

Throughout this work, we adopt and compare various models for the parameters listed in Sections~\ref{sec:radfield} and~\ref{sec:photmodels}. When exploring the systematic differences amongst variations in a single parameter associated with the photoionisation grids, we keep all other parameters constant. Unless otherwise stated, we assume a solar metallicity ($Z = 0.020$), constant star formation rate ($1 M_\odot \;\mathrm{yr}^{-1}$) with an age of 5 Myr, created using SLUG with a Kroupa IMF, the Geneva HIGH tracks, and combination (Kurucz + Hillier + Pauldrach)
atmospheres. All spectra shown in Section~\ref{sec:modelcomp1} are first normalised to an SB99 spectrum of a 5 Myr cluster undergoing constant star formation at $1\;M_\odot \;\mathrm{yr}^{-1}$, treating all stars as pure blackbodies. This reference spectrum is shown in Figure~\ref{fig:refspec}. The SFR and the age of the cluster provide a cluster mass of $5 \times 10^6 \;M_\odot$. The H \textsc{ii} region model assumes a spherical, radiation-bounded, isobaric structure with $P/k = 8 \times 10^5\;\mathrm{cm}^{-3}\;\mathrm{K}$, terminated once 99\% of hydrogen recombination has occurred. Our abundances are solar, and the models include dust, following the depletion factors detailed in Section~\ref{sec:depfactors}. Our models do not allow grain destruction, and the grain size distribution follows the MRN distribution \citep{MRN1977}. Finally, we include polycyclic aromatic hydrocarbons (PAHs), with a carbon dust depletion fraction of 0.3.

\section{Model Comparisons}
\label{sec:modelcomp}
\subsection{Ionising Radiation Field}
\label{sec:modelcomp1}
\subsubsection{Cluster Age}
\label{sec:age1}

\begin{figure}
\centering
\includegraphics[width=\columnwidth]{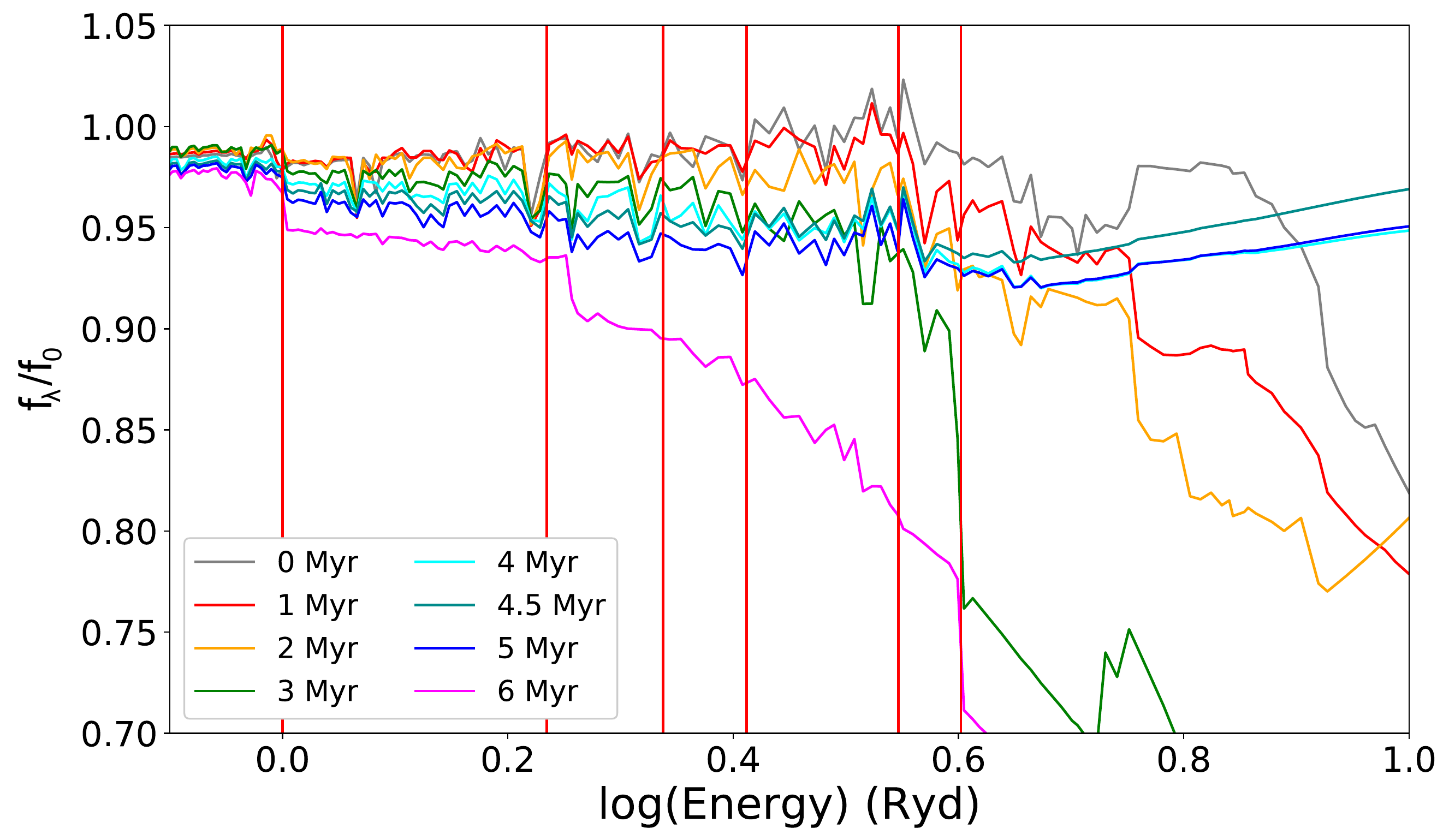}
\caption{SLUG spectra for an instantaneous SFH cluster of mass $10^6 M_\odot$ for ages $0 - 6$ Myr in steps of 0.5 Myr. $f_{\lambda}/f_0$ is the input spectrum normalised to the blackbody spectrum shown in Figure~\ref{fig:refspec}. All spectra are at a metallicity of $Z = 0.020$. The red vertical lines represent the ionisation potentials of important ISM species; from left to right: H$^0$, S$^+$, N$^+$, O$^+$, C$^{2+}$, He$^+$.}
\label{fig:instspec}
\end{figure}

The spectra of instantaneous SFH stellar clusters at various ages are shown in Figure~\ref{fig:instspec}. When assuming an instantaneous SFH, we end the simulation at 6 Myr. Beyond the ages of 6 Myr, the flux of H$^0$-, He$^0$- and He$^+$-ionising photons from a single stellar cluster decreases considerably \citep[e.g.][]{Wofford2016}. Hence, we neglect ages beyond 6 Myr. 
A gradual decrease in the flux of the ionising spectrum continues up until 3.5 Myr. At this point, the beginning of a distinct and large increase in the ionising spectrum at high energies can be seen, corresponding to the evolution of the W-R stars, as a subset of evolved OB stars \citep{Conti1976,Lamers1991,Groh2013}. 

Using spectroscopic techniques, \citet{Vacca1996} calculated the masses of OB stars of many spectral types (O3-9.5, B0-0.5) and luminosity classes (Ia, III, and V). Averaged over the spectral type and luminosity class, \citet{Vacca1996} find an average OB star mass of $35.5 \pm 24.8 M_\odot$. This value is in very close agreement with the value calculated by \citet{Martins2005}, who calculate an average OB star mass of $34.99 \;M_\odot$ with an average error of approximately $15 \; M_\odot$, using a very similar technique. At masses of ${\sim} 40 M_\odot$, \citet{Lamers1991} calculate the main-sequence lifetime of OB stars to be $2.5 - 4$ Myr, coinciding with the W-R emergence time frame shown in Figure~\ref{fig:instspec}. The spectra of W-R stars contain very strong broad emission lines of specifically helium, carbon, nitrogen, oxygen, and silicon, with hydrogen lines either weak or absent \citep{Crowther2007}. \citet{Lamers1991} showed W-R stars to be late stages in the evolution of massive stars, which have removed their hydrogen-rich outer layers in a stellar wind. The loss of the outer layers exposes a bare core, leading to an increased detection of heavier elements.

The increase in the hardness of the spectrum continues for the duration of the lifetime of W-R stars \citep[${\sim} 0.5$ Myr at a mass of $40 M_\odot$;][]{MM2005}; Figure~\ref{fig:instspec} shows an the beginning of the distinct increase in hardness of the spectrum at 4 Myr, peaking at 4.5 Myr before decreasing at 5 Myr. 

\subsubsection{Stellar Evolutionary Tracks}
\label{sec:tracks1}

\begin{figure*}
\centering
\includegraphics[width=\textwidth]{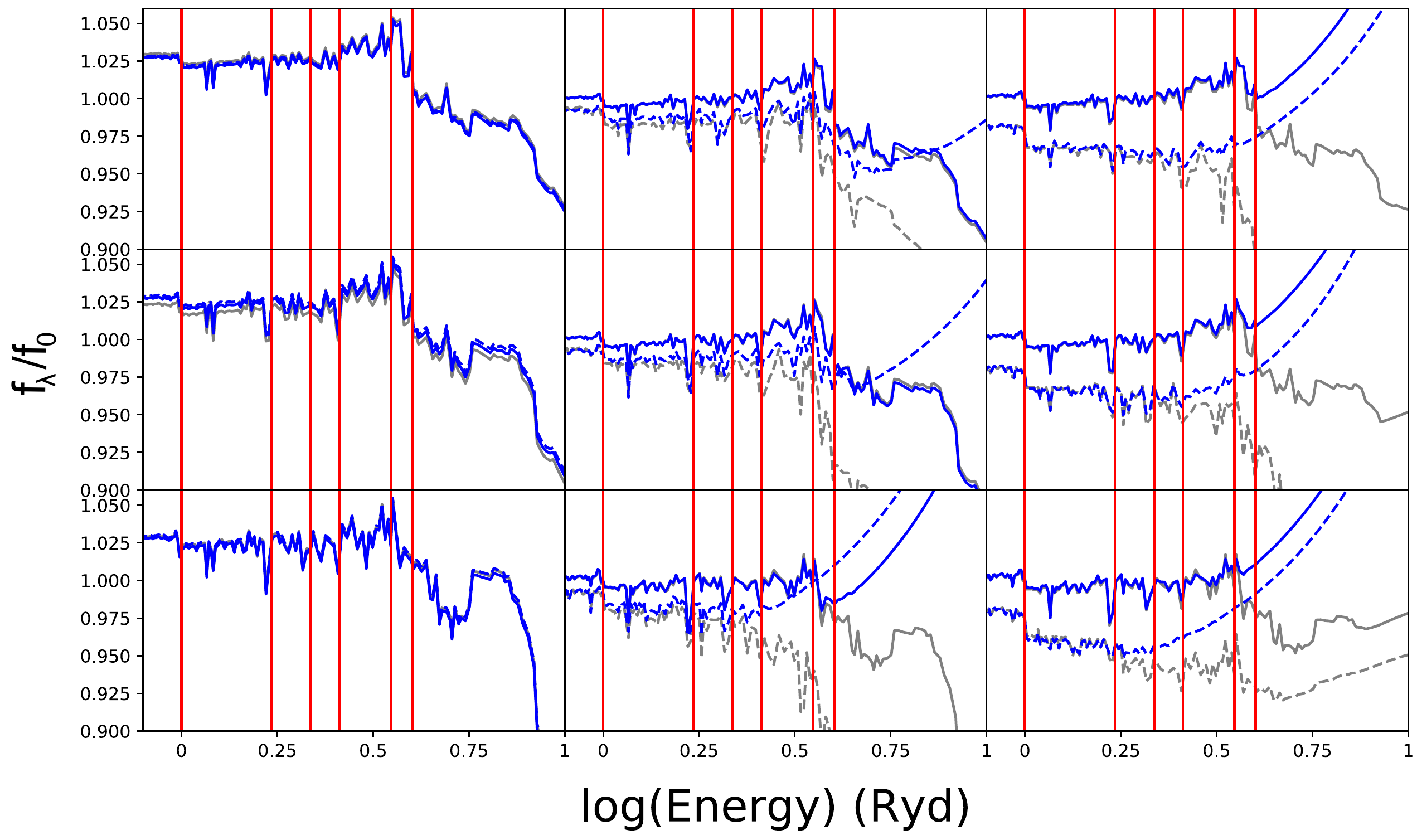}
\caption{Input spectra (flux in units erg sec$^{-1}$ \AA$^{-1}$ vs. energy) of the Geneva HIGH tracks (grey) and the Padova tracks with thermally pulsing AGB stars (blue). $f_{\lambda}$ represents the individual input specta, and $f_0$ is the blackbody spectrum shown in Figure~\ref{fig:refspec}. Solid lines are for continuous SFH clusters undergoing star formation at a rate of $1 M_\odot \;\mathrm{yr}^{-1}$. Dashed lines are for instantaneous SFH clusters of mass $10^6 M_\odot$. The continuous SFH spectra shown at 0 Myr (10 kyr) are for clusters undergoing star formation at a rate of $100 M_\odot \;\mathrm{yr}^{-1}$, ensuring that the continuous SFH cluster has a mass of $10^6 M_\odot$ for direct comparison to the instantaneous SFH clusters. Each column shows a different cluster age; from left to right: 0 Myr (10 kyr), 3 Myr, 5 Myr. Each row shows a different metallicity; from top to bottom: $Z = 0.004$, $Z = 0.008$, $Z = 0.020$. The red vertical lines represent the ionisation potentials of important ISM species; from left to right: H$^0$, S$^+$, N$^+$, O$^+$, C$^{2+}$, He$^+$.}
\label{fig:speccomp}
\end{figure*}

The spectra for the Geneva and Padova stellar evolutionary tracks are shown in Figure~\ref{fig:speccomp}, at continuous and instantaneous SFH cluster ages of 0 Myr (10 kyr), 3 Myr, and 5 Myr. Each spectrum in Figure~\ref{fig:speccomp} is shown normalised to a continuous SFH spectrum at 5 Myr treating all stars as blackbodies, produced using SB99. Only the spectra at metallicities where the tracks are directly comparable are shown ($Z$ = 0.004, 0.008, and 0.020; the tracks differ at the lowest and highest metallicities). Spectra for continuous and instantaneous SFH are shown at identical ages. Seen in Figure~\ref{fig:speccomp}, the spectra at varying ages between the continuous SFH Geneva and Padova models are extremely similar for energies below log($E/\mathrm{ryd}) \lesssim 0.6$, agreeing in flux to within ${\sim} 0.5$\%. In this energy regime, any noticeable trend in the difference in the spectra between the two sets of tracks with age or metallicity is difficult to determine. The difference between the spectra produced from both sets of tracks is more easily seen when using the spectra from an instantaneous SFH cluster. Without continually emerging young stellar populations contributing to the overall spectrum, the dependence of stellar evolution on the choice of stellar evolutionary tracks becomes clearer. The Padova tracks are seen to produce a consistently higher flux than the Geneva tracks when considering the evolution of an instantaneous SFH cluster over time. The systematically higher flux from the Padova tracks is a result of the larger zero-age main-sequence (ZAMS) effective temperature and luminosity of the Padova tracks by ${\sim} 0.03$ and ${\sim} 0.02$ dex, respectively. The evolution of $T_\mathrm{eff}$ and luminosity is also similar over time in the two sets of tracks, with the exception of the W-R phase \citep{VL2005}. 

At high energies, more noticeably as age increases, the difference between the spectra of the two sets of tracks becomes extremely large. This large difference in the flux between the Padova and Geneva tracks typically begins at log($E/\mathrm{ryd}) \gtrsim 0.6$ ($E \gtrsim 4\;\mathrm{ryd} \Rightarrow \lambda \lesssim 366$\AA), increasing in size with energy. \citet{VL2005} attribute this large discrepancy in ionising flux between the two tracks to the presence of W-R stars. The different definition of $T_\mathrm{eff}$ between the two models, as well as the subsequent higher effective temperature in the Padova tracks, has large ramifications once W-R stars begin to emerge in the cluster. This is reflected in the relative ionising fluxes of photons at shorter wavelengths, capable of ionising He$^0$ and He$^+$ ($\lambda \lesssim 504$ and 228\AA\;, respectively). At a cluster age of 4 Myr, \citet{VL2005} show the relative difference in the number of ionising photons in the He \textsc{ii} continuum to be roughly $4 - 8$ orders of magnitude in favour of the Padova tracks, depending on the choice of stellar atmosphere. At ${\sim} 6$ Myr, this difference is increased to roughly 10 orders of magnitude. Beyond ${\sim} 6.3$ Myr, the number of W-R stars left within the cluster becomes negligible \citep[see also][]{Leitherer1999}, and hence the relative difference in the number of He$^+$-ionising photons decreases rapidly towards zero. Figure~\ref{fig:speccomp} shows a sustained large difference in the ionising flux at high energies in the continuous SFH spectra, as W-R stars continue to emerge for the duration of the simulation. SLUG calculates the flux for a minimum wavelength of 91\AA. At a wavelength of 91\AA, the difference between the total integrated ionising fluxes from the Geneva and Padova tracks calculated by SLUG is roughly 8 orders of magnitude.

\subsubsection{Stellar Atmospheres}
\label{sec:atms1}

Spectra produced using each of the atmosphere models we consider are shown in Figure~\ref{fig:specplot} for 0 Myr clusters on the left and clusters of ages 5 and 4.5 Myr assuming a continuous and instantaneous SFH, respectively, on the right. An age of 4.5 Myr for the instantaneous SFH clusters in Figure~\ref{fig:specplot} was chosen to coincide with maximum W-R activity shown in Figure~\ref{fig:instspec}. 

The W-R atmospheres added to the Kurucz atmospheres by \citet{HM1998} make little difference to the overall shape of the ionising spectra of the stellar clusters, including once W-R activity is at a maximum at 4.5 Myr. The atmospheres used to model OB stars produced by \citet{Pauldrach2001} improved on the OB stellar models used in the Kurucz atmospheres by introducing metal opacities into the atmospheres of the OB stars. The metals introduced into the OB stellar atmospheres undergo excitation and ionisation through the absorption of internal stellar photons, and subsequent emissions from these metals result in differences in the strength of particular emission lines from the surrounding nebula. This can be seen in the left  panels of Figure~\ref{fig:specplot}. The spectra that include the \citet{Pauldrach2001} OB atmospheres show increased absorption or emission at various photon energies throughout the spectrum. The spectra at 0 Myr in Figure~\ref{fig:specplot} show an increased luminosity across all metallicities at the ionisation potential for O$^{+}$ (log($E$/ryd) $\approx 0.6$), whereas the luminosity at the ionisation potential for N$^0$ (log($E$/ryd) $\approx 0.3$) remains relatively constant with metallicity.

\begin{figure*}
\begin{subfigure}{0.49\textwidth}
\includegraphics[width=1.1\columnwidth]{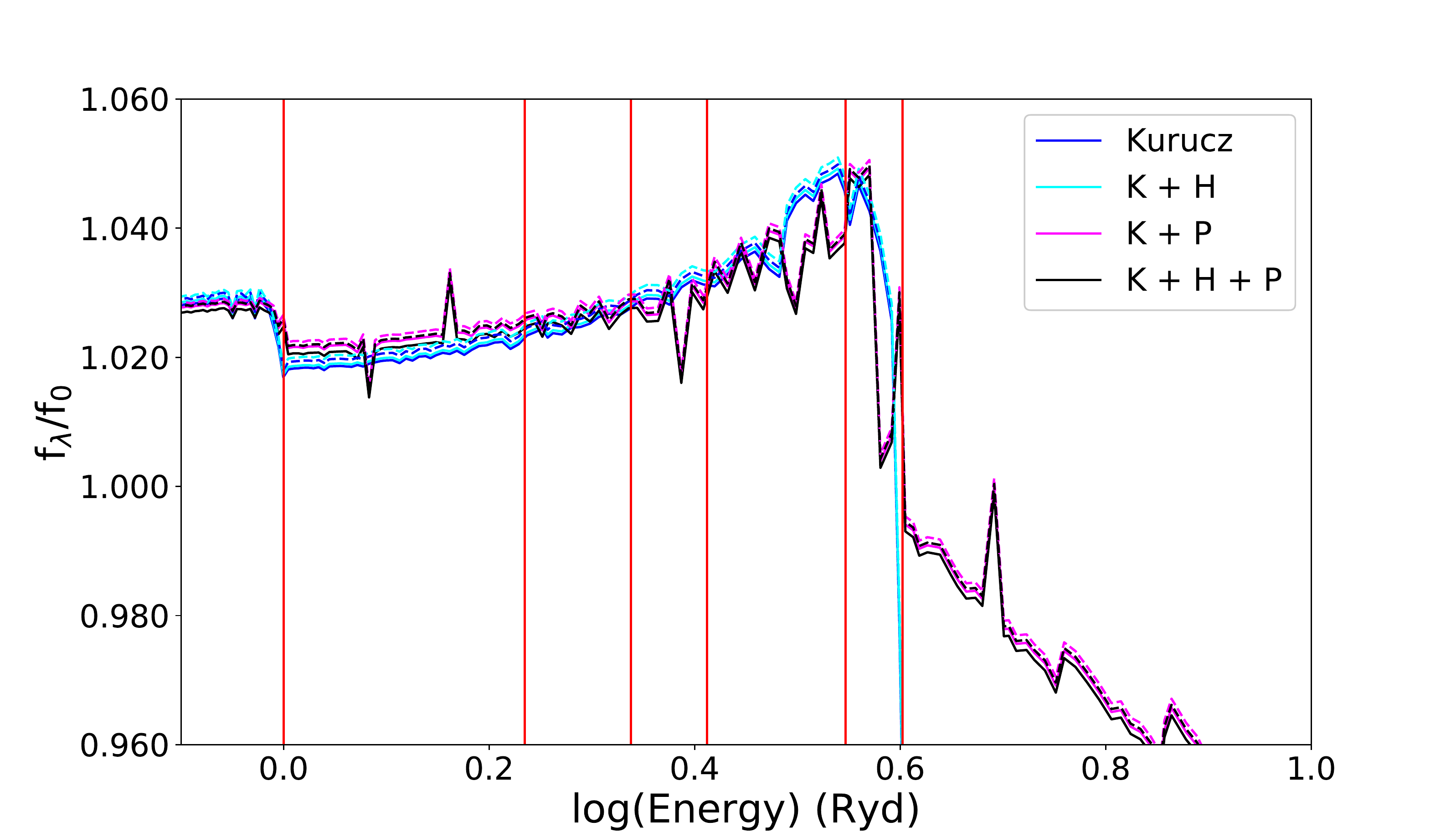}
\caption{Metallicity $Z = 0.001$}
\label{fig:specplota}
\end{subfigure}
\begin{subfigure}{0.49\textwidth}
\includegraphics[width=1.1\columnwidth]{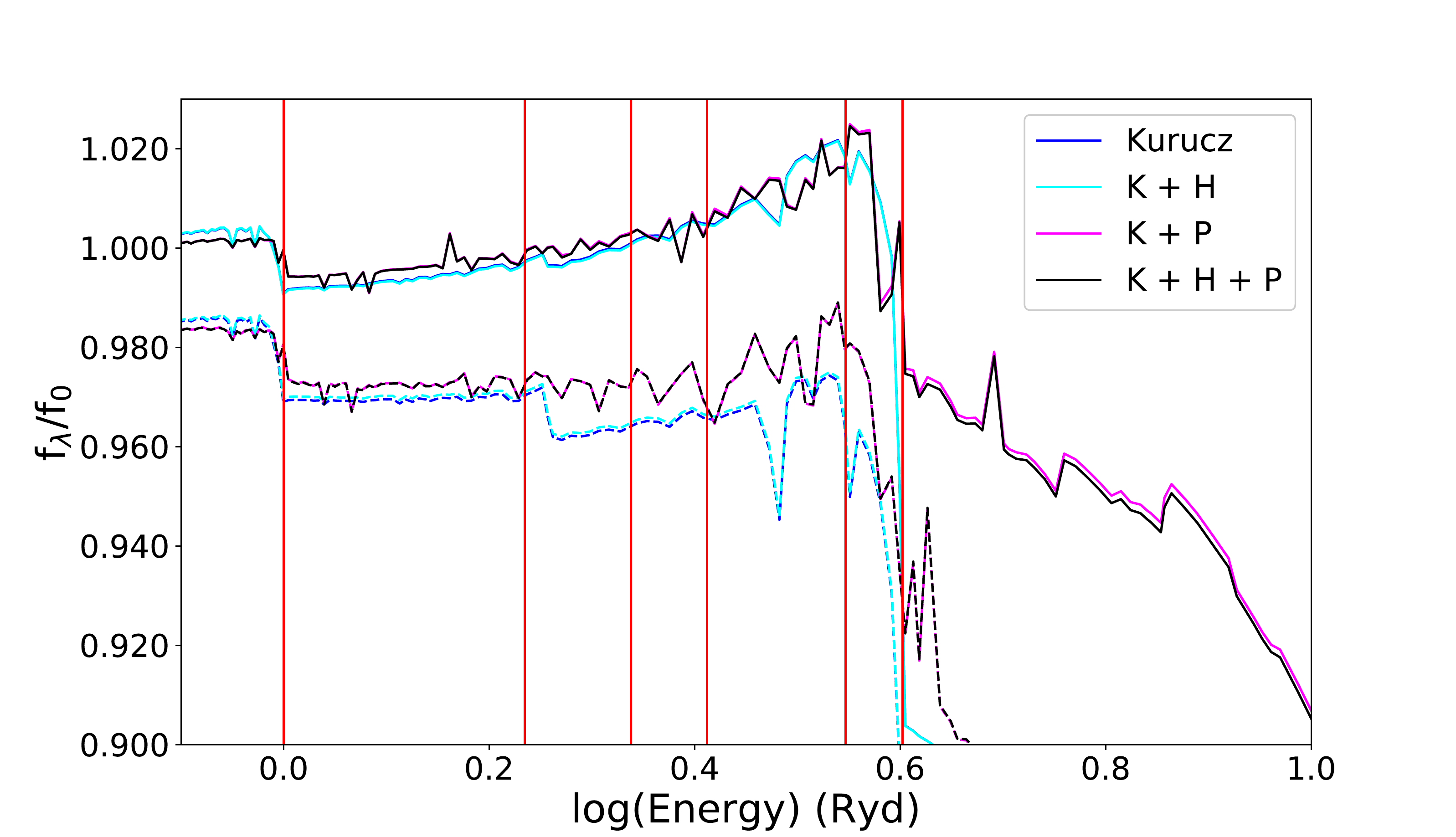}
\caption{Metallicity $Z = 0.001$}
\label{fig:specplotb}
\end{subfigure}
\begin{subfigure}{0.49\textwidth}
\includegraphics[width=1.1\columnwidth]{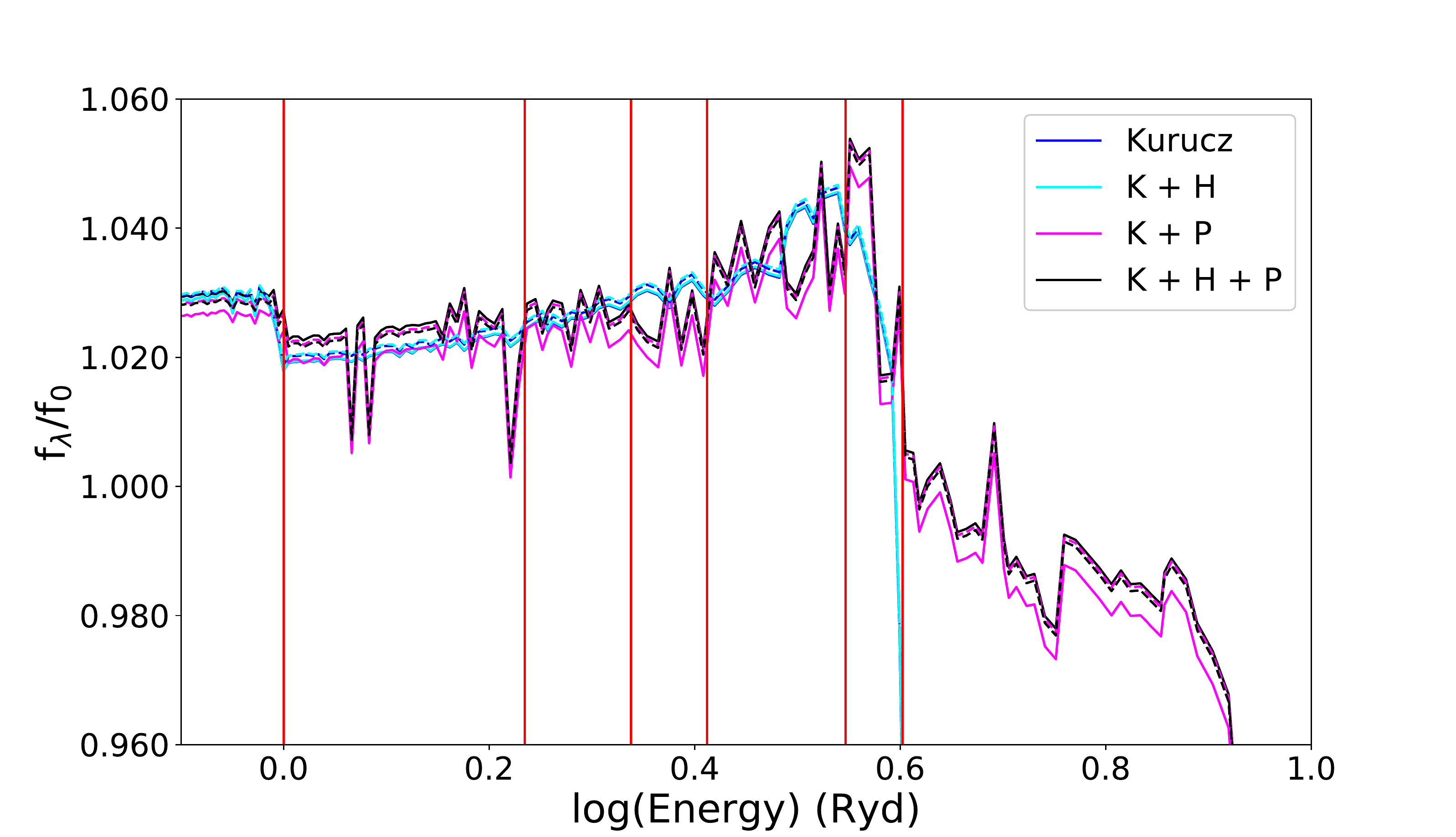}
\caption{Metallicity $Z = 0.004$}
\label{fig:specplotc}
\end{subfigure}
\begin{subfigure}{0.49\textwidth}
\includegraphics[width=1.1\columnwidth]{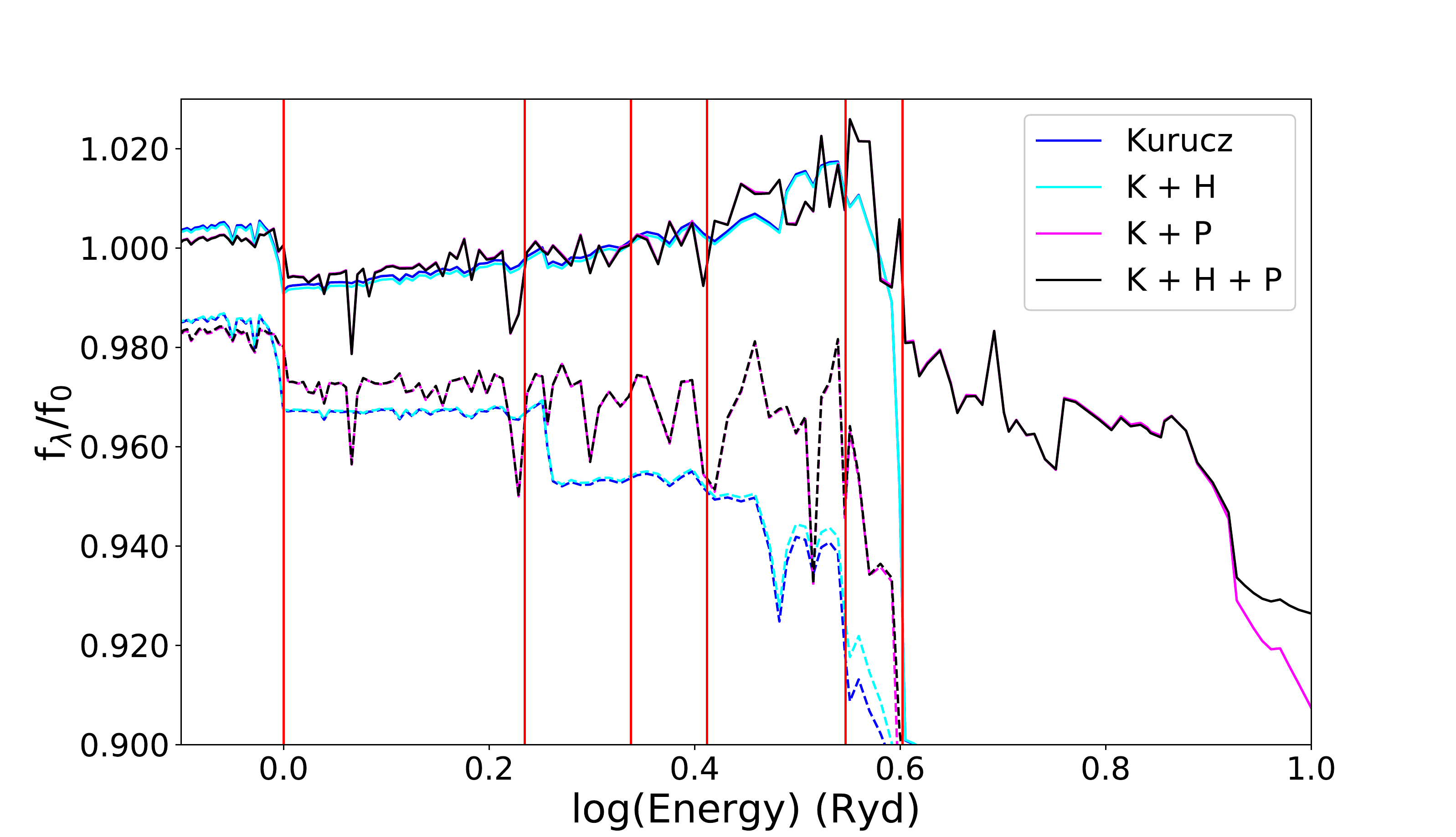}
\caption{Metallicity $Z = 0.004$}
\label{fig:specplotd}
\end{subfigure}
\begin{subfigure}{0.49\textwidth}
\includegraphics[width=1.1\columnwidth]{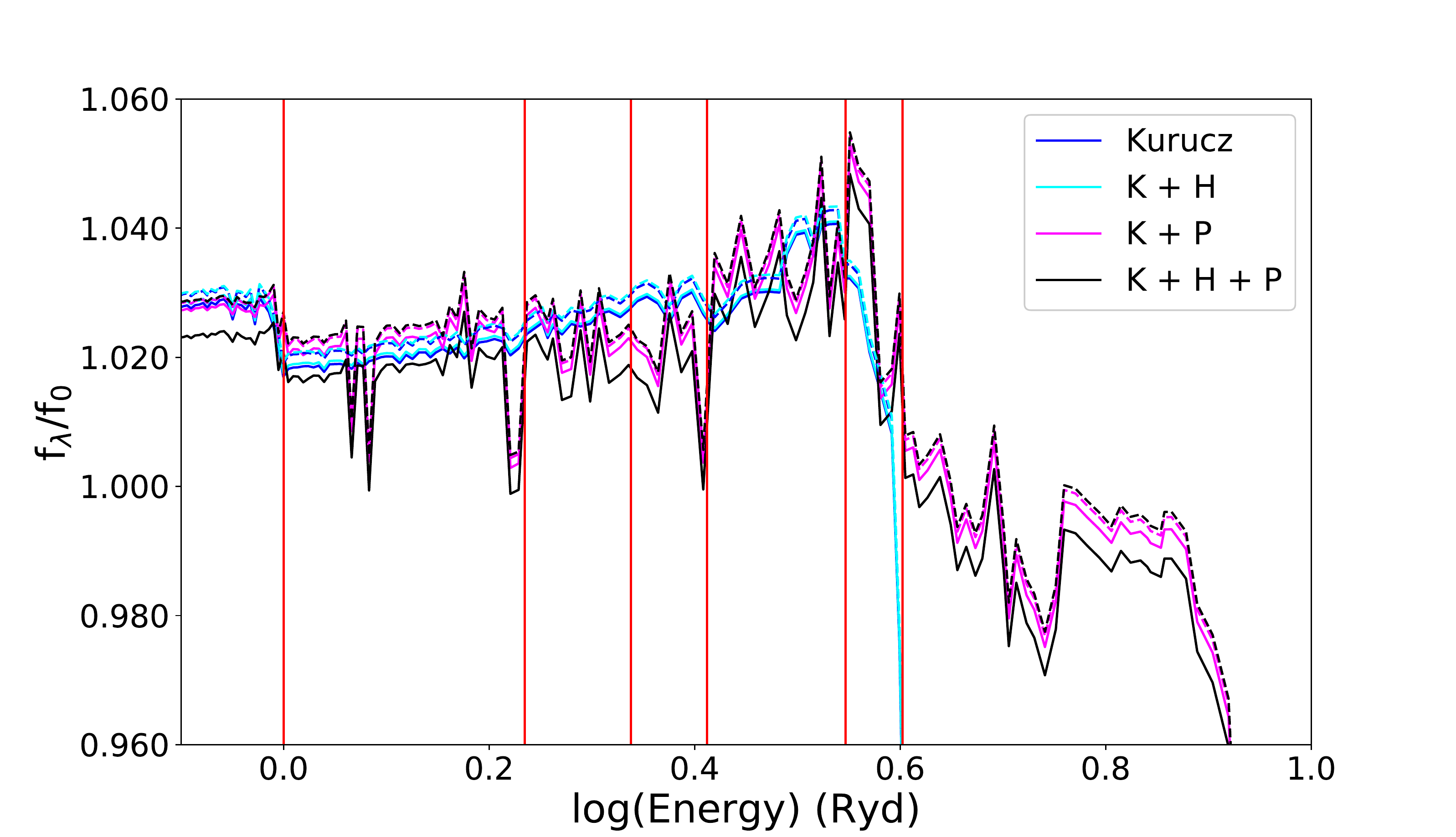}
\caption{Metallicity $Z = 0.008$}
\label{fig:specplote}
\end{subfigure}
\begin{subfigure}{0.49\textwidth}
\includegraphics[width=1.1\columnwidth]{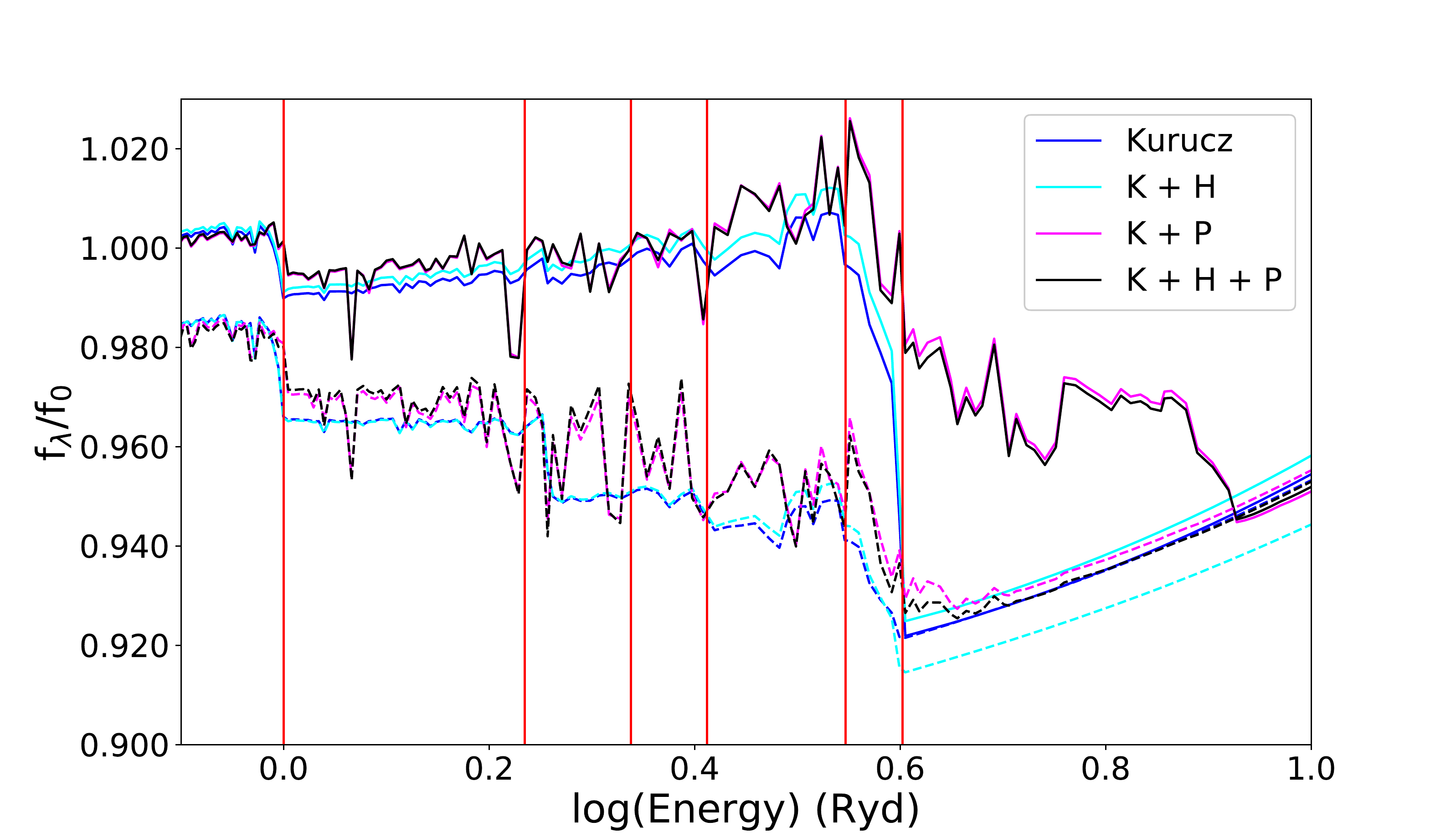}
\caption{Metallicity $Z = 0.008$}
\label{fig:specplotf}
\end{subfigure}
\begin{subfigure}{0.49\textwidth}
\includegraphics[width=1.1\columnwidth]{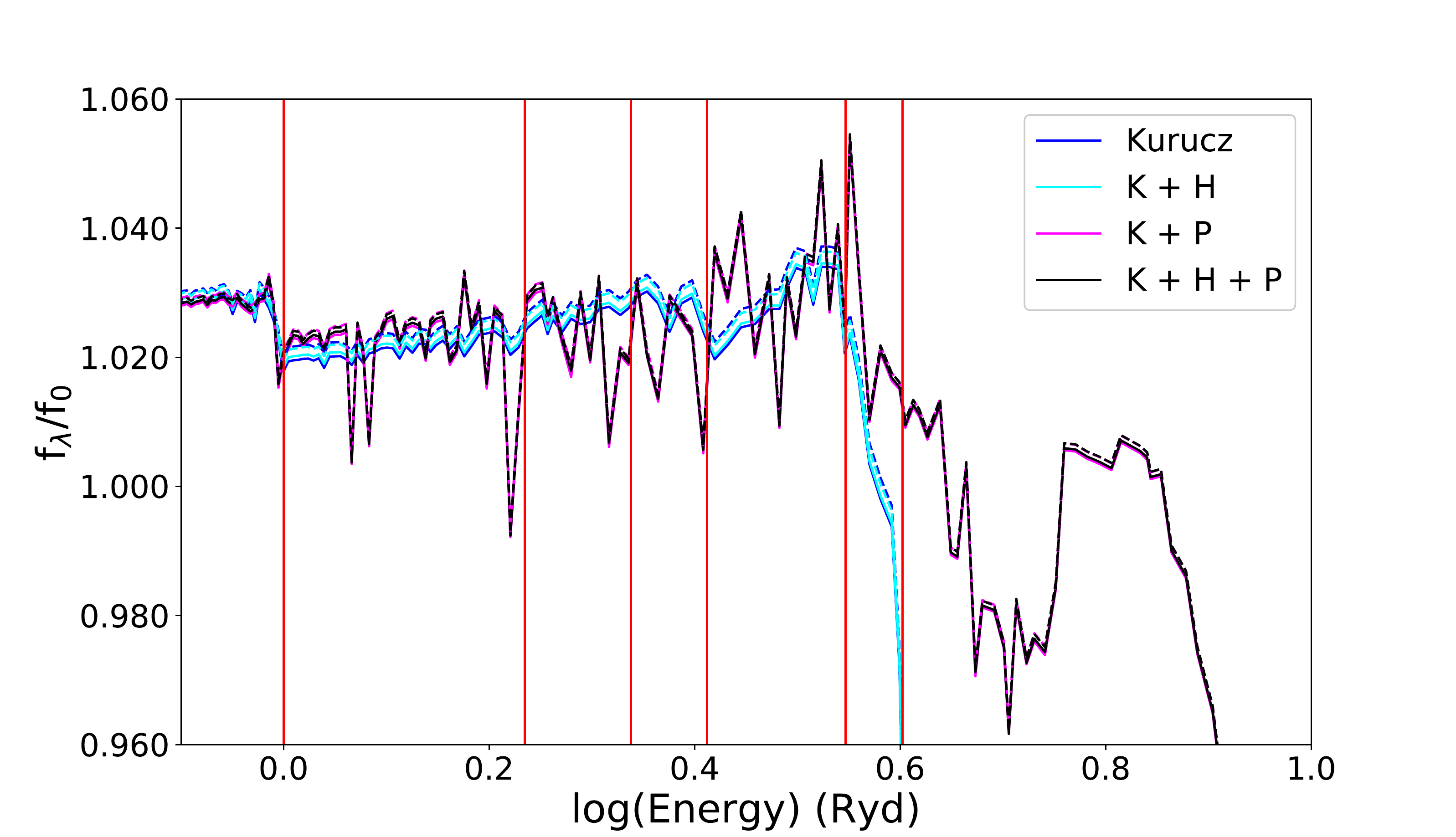}
\caption{Metallicity $Z = 0.020$}
\label{fig:specplotg}
\end{subfigure}
\begin{subfigure}{0.49\textwidth}
\includegraphics[width=1.1\columnwidth]{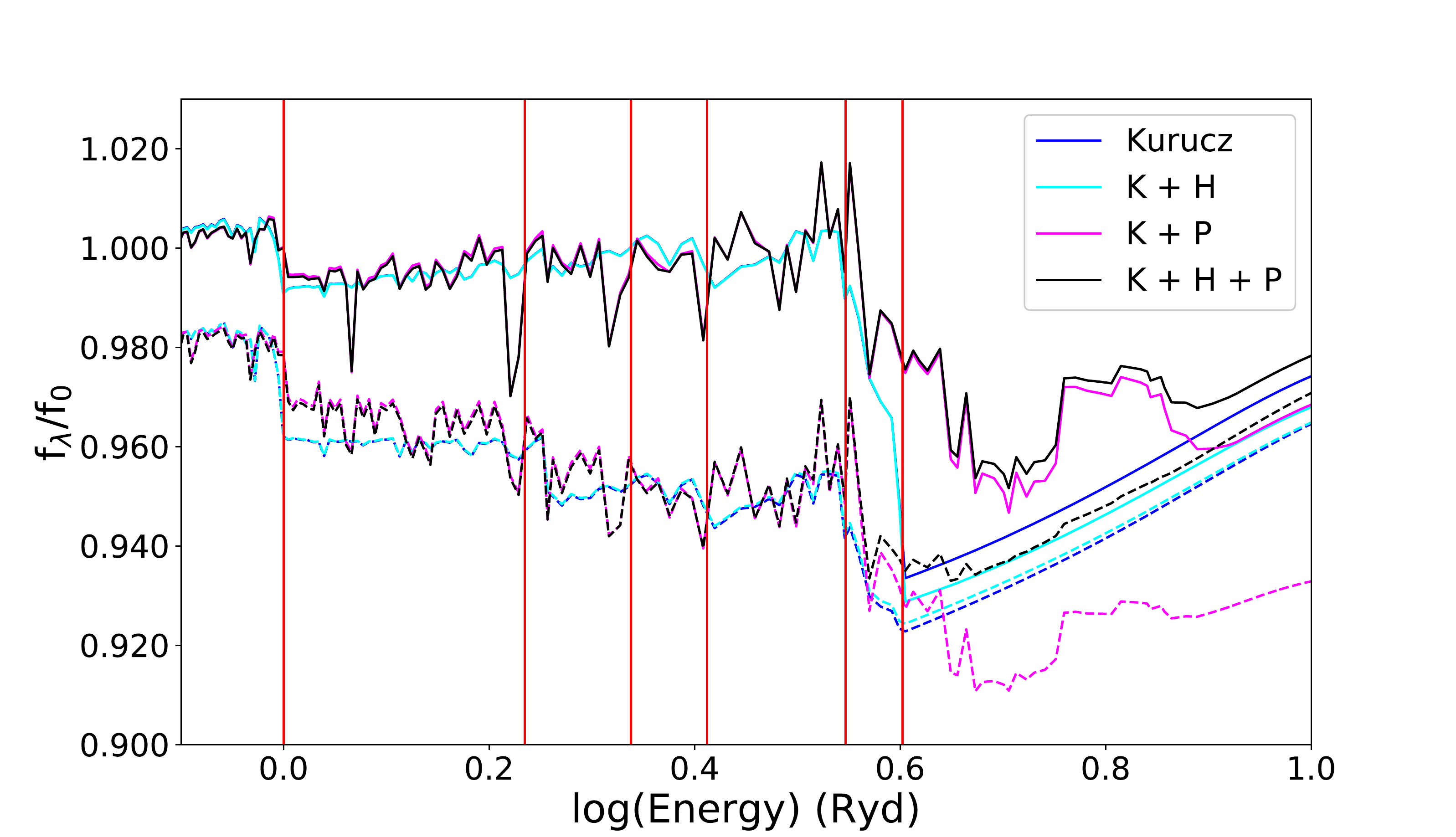}
\caption{Metallicity $Z = 0.020$}
\label{fig:specploth}
\end{subfigure}
\end{figure*}

\begin{figure*}\ContinuedFloat
\begin{subfigure}{0.49\textwidth}
\includegraphics[width=1.1\columnwidth]{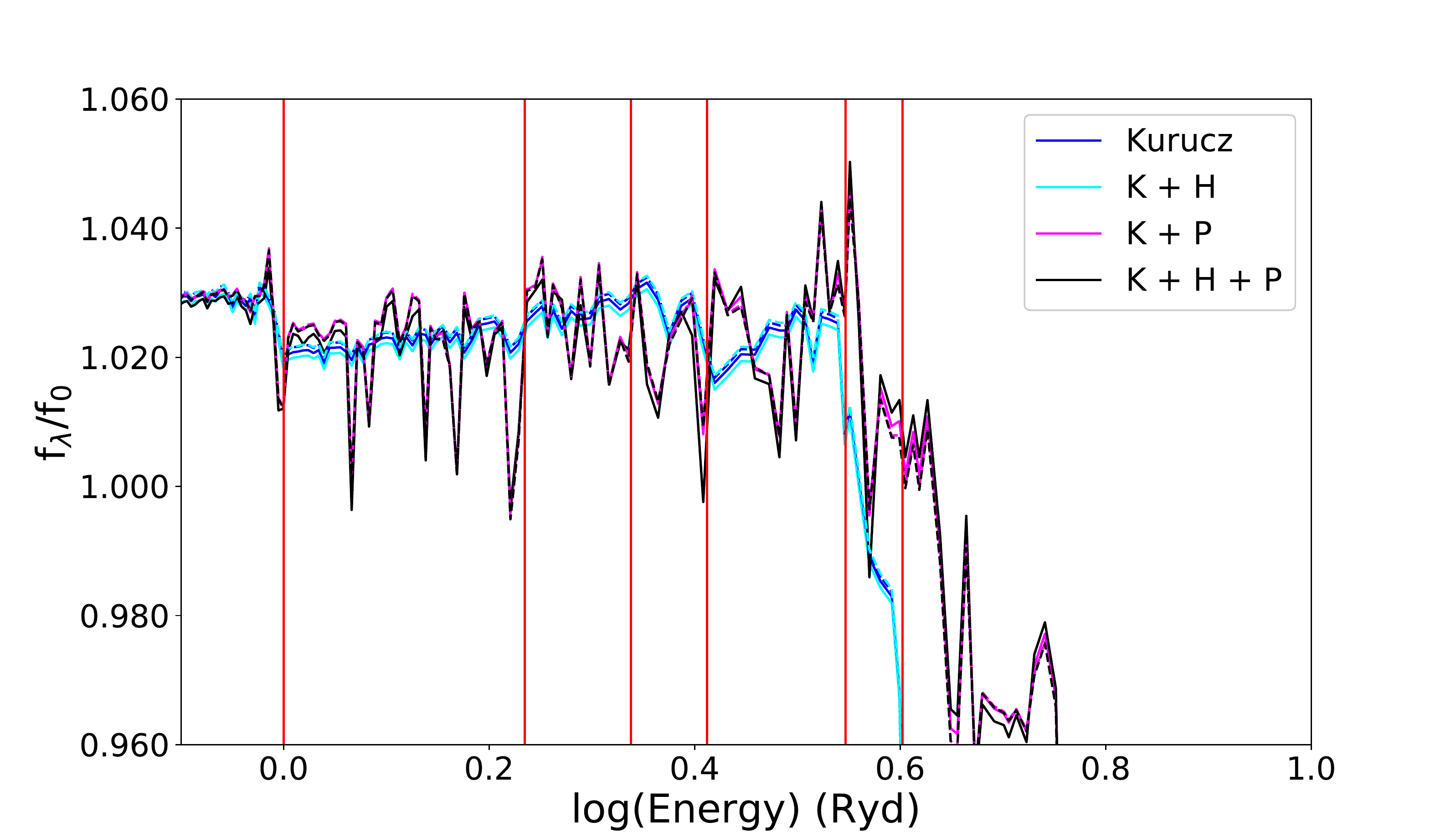}
\caption{Metallicity $Z = 0.040$}
\label{fig:specploti}
\end{subfigure}
\begin{subfigure}{0.49\textwidth}
\centering
\includegraphics[width=1.1\columnwidth]{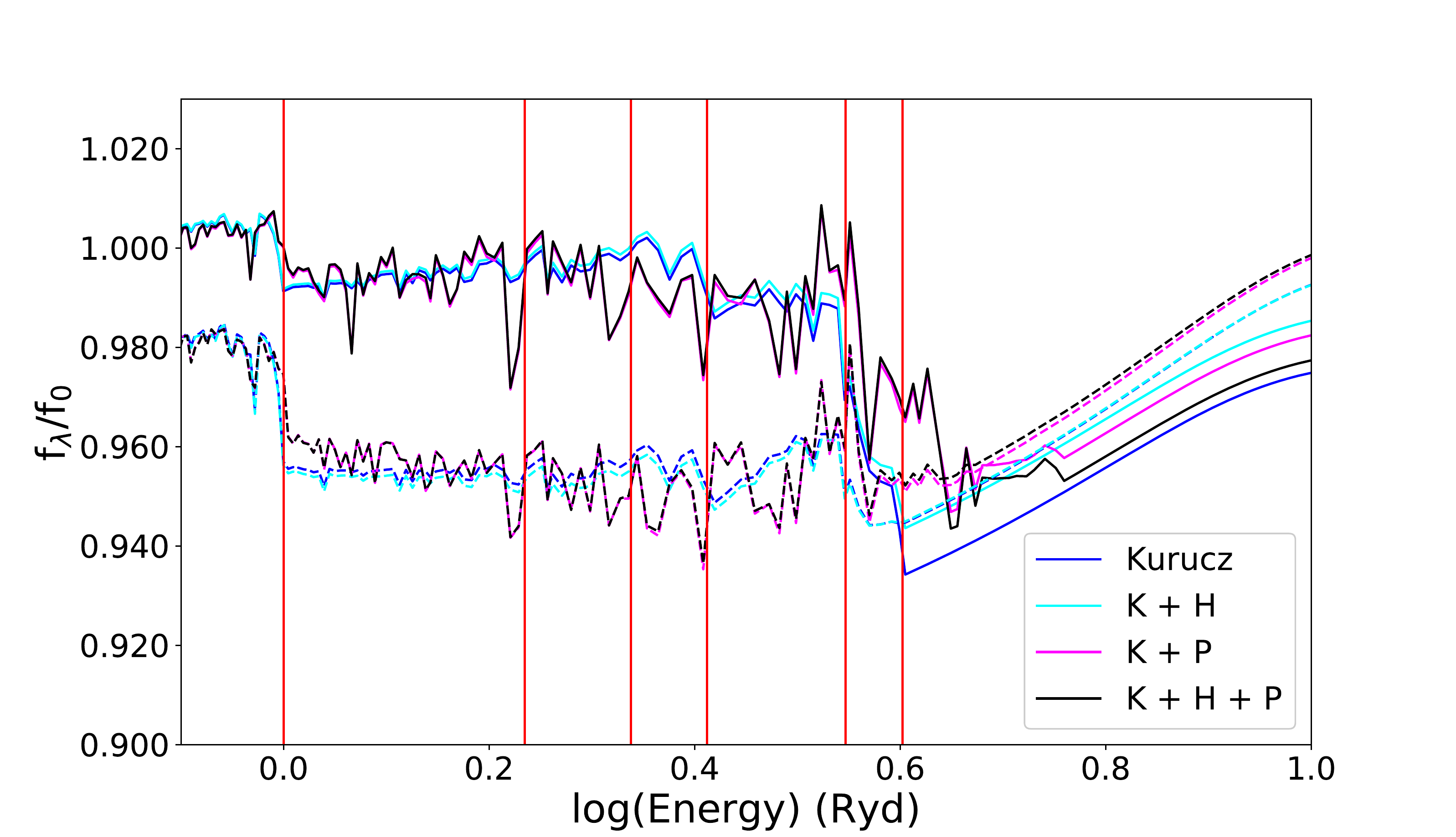}
\caption{Metallicity $Z = 0.040$}
\label{fig:speplotj}
\end{subfigure}
\caption{Input spectra (flux in units erg sec$^{-1}$ \AA$^{-1}$ vs. energy) of each of the four SLUG atmospheres for each metallicity value used in the grids. $f_{\lambda}$ represents the individual input specta, and $f_0$ is the blackbody spectrum shown in Figure~\ref{fig:refspec}. The atmospheres used in each plot are Kurucz, Kurucz + Hillier (K + H), Kurucz + Pauldrach (K + P), and Kurucz + Hillier + Pauldrach (K + H + P). The left panels show both continuous SFH (solid lines) and instantaneous SFH (dashed lines) clusters of mass $10^6 M_\odot$ at ages of 0 Myr (10 kyr). The right panels show continuous SFH clusters at 5 Myr and instantaneous SFH clusters of mass $10^6 M_\odot$ at 4.5 Myr. The age of 4.5 Myr for the instantaneous SFH cluster was chosen to coincide with maximum W-R activity shown in Figure~\ref{fig:instspec}. The red vertical lines represent the ionisation potentials of important ISM species; from left to right: H$^0$, S$^+$, N$^+$, O$^+$, C$^{2+}$, He$^+$.}
\label{fig:specplot}
\end{figure*}

\subsubsection{Stellar Population Synthesis Codes}
\label{sec:spscodes}

\paragraph{Single stellar populations}

\begin{figure*}[p]
\begin{subfigure}{0.49\textwidth}
\includegraphics[width=1.1\columnwidth]{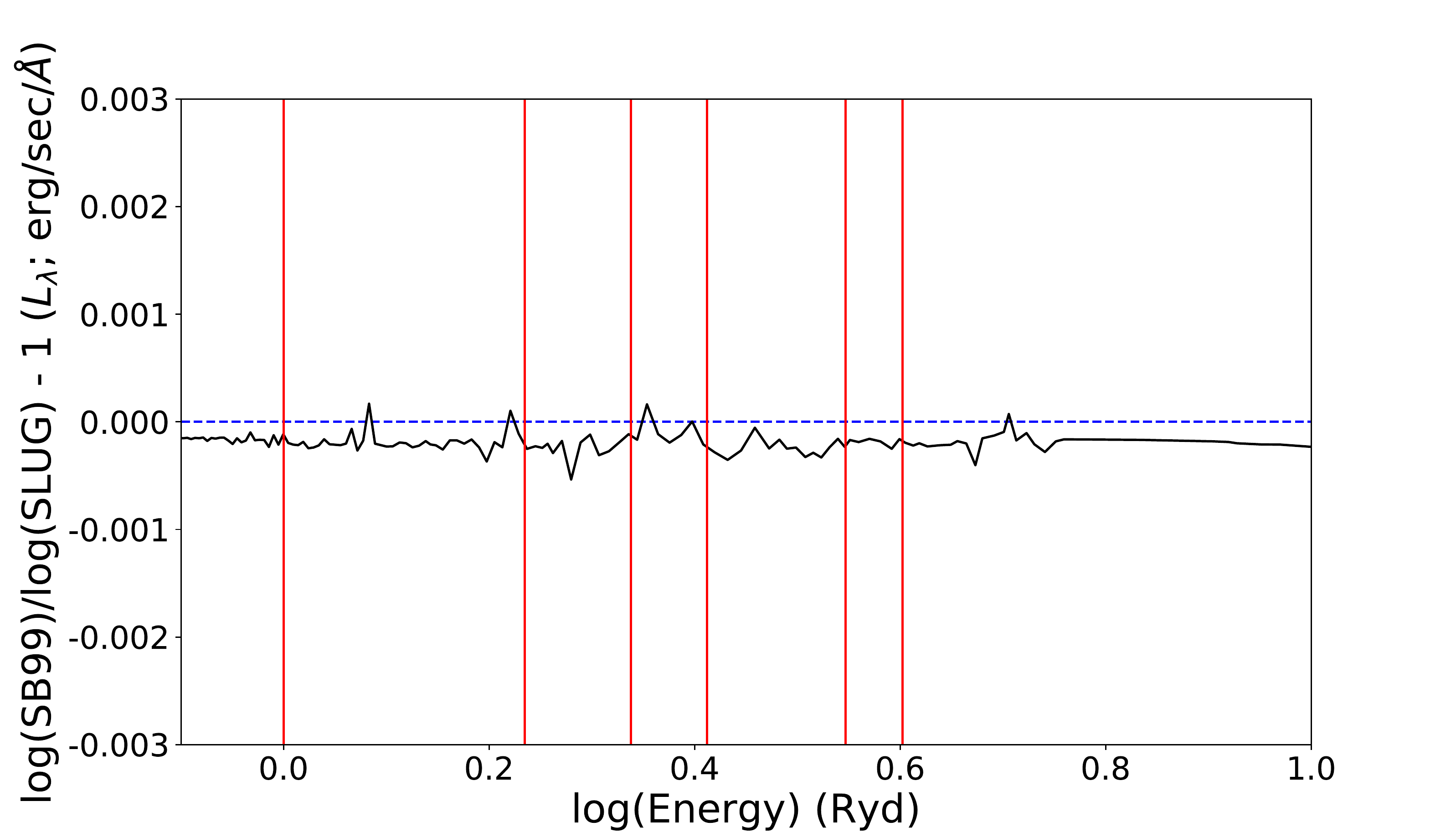}
\caption{0 Myr (10 Kyr)}
\label{fig:reldiff_inst0}
\end{subfigure}
\begin{subfigure}{0.49\textwidth}
\includegraphics[width=1.1\columnwidth]{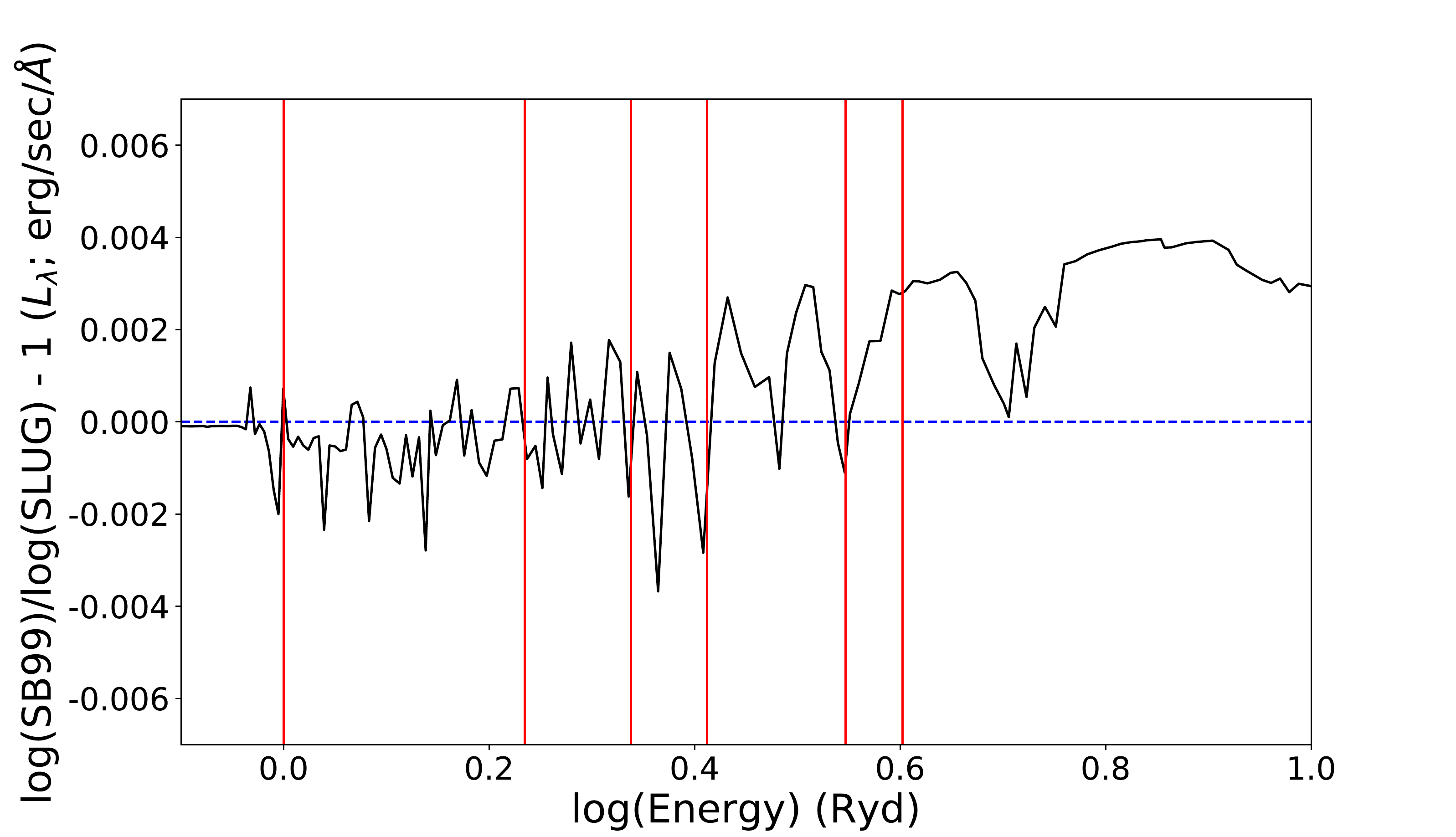}
\caption{1 Myr}
\label{fig:reldiff_inst1}
\end{subfigure}
\begin{subfigure}{0.49\textwidth}
\includegraphics[width=1.1\columnwidth]{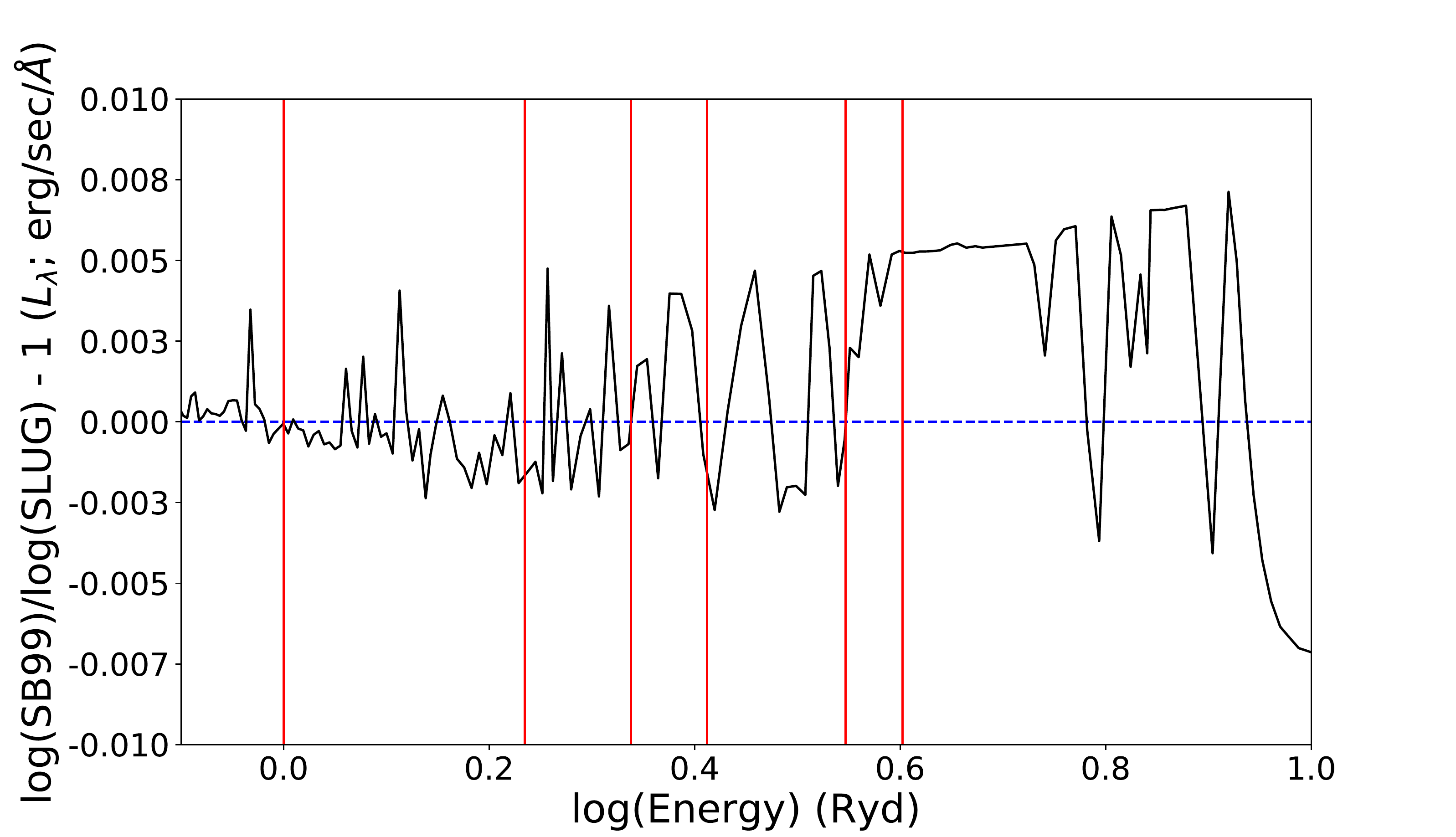}
\caption{2 Myr}
\label{fig:reldiff_inst2}
\end{subfigure}
\begin{subfigure}{0.49\textwidth}
\includegraphics[width=1.1\columnwidth]{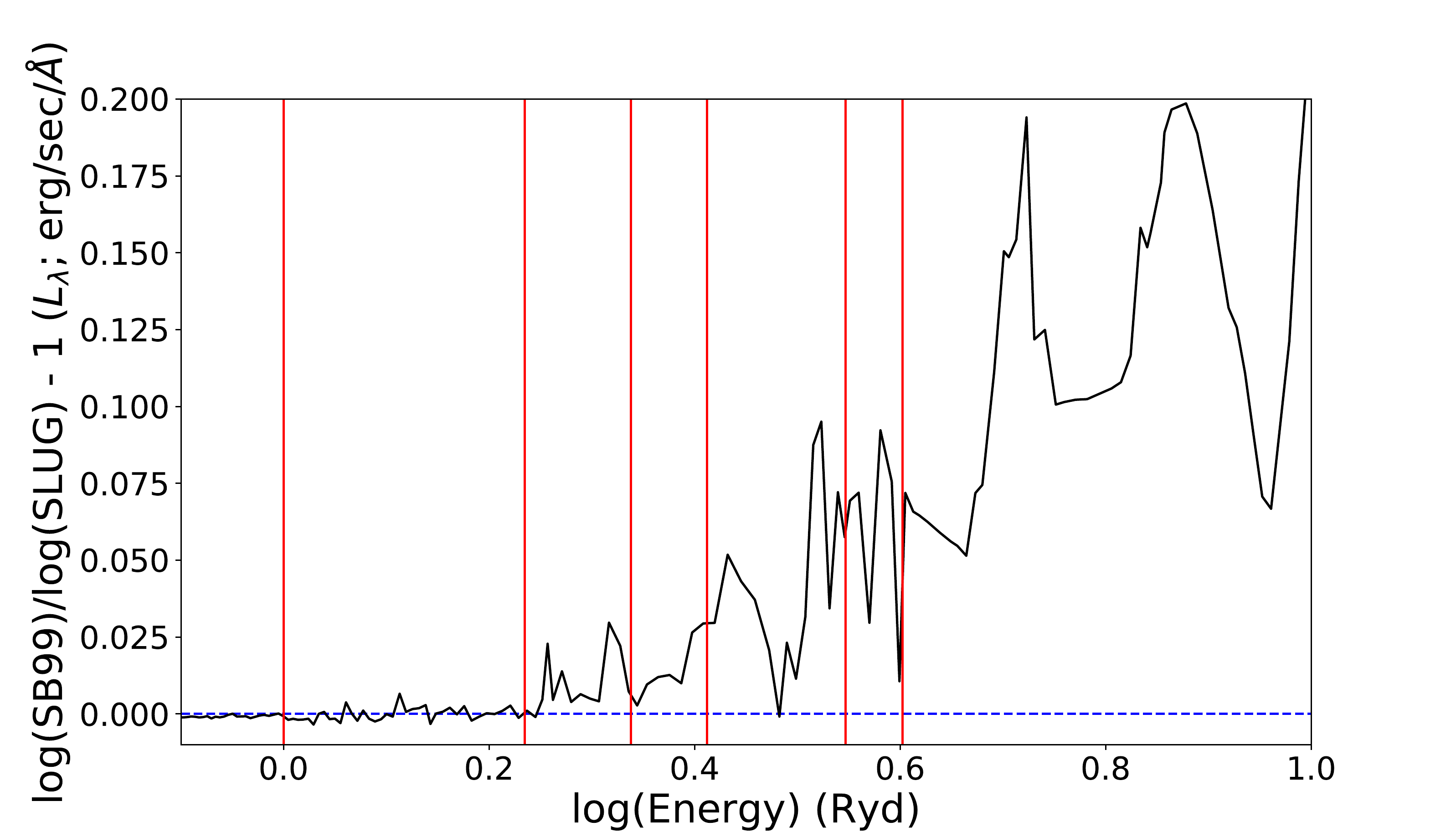}
\caption{3 Myr}
\label{fig:reldiff_inst3}
\end{subfigure}
\begin{subfigure}{0.49\textwidth}
\includegraphics[width=1.1\columnwidth]{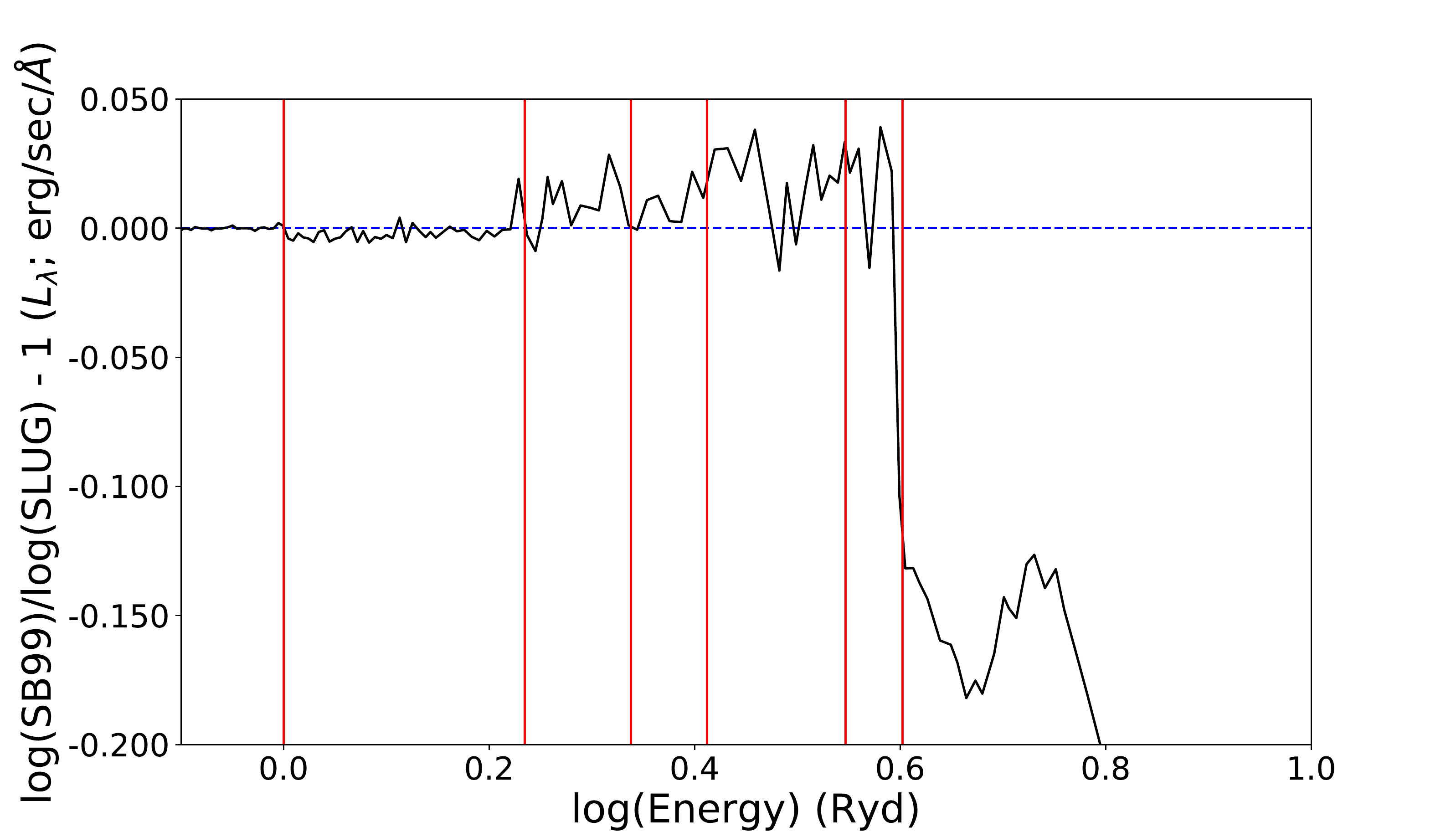}
\caption{4 Myr}
\label{fig:reldiff_inst4}
\end{subfigure}\hspace{0.012\textwidth}
\begin{subfigure}{0.49\textwidth}
\includegraphics[width=1.1\columnwidth]{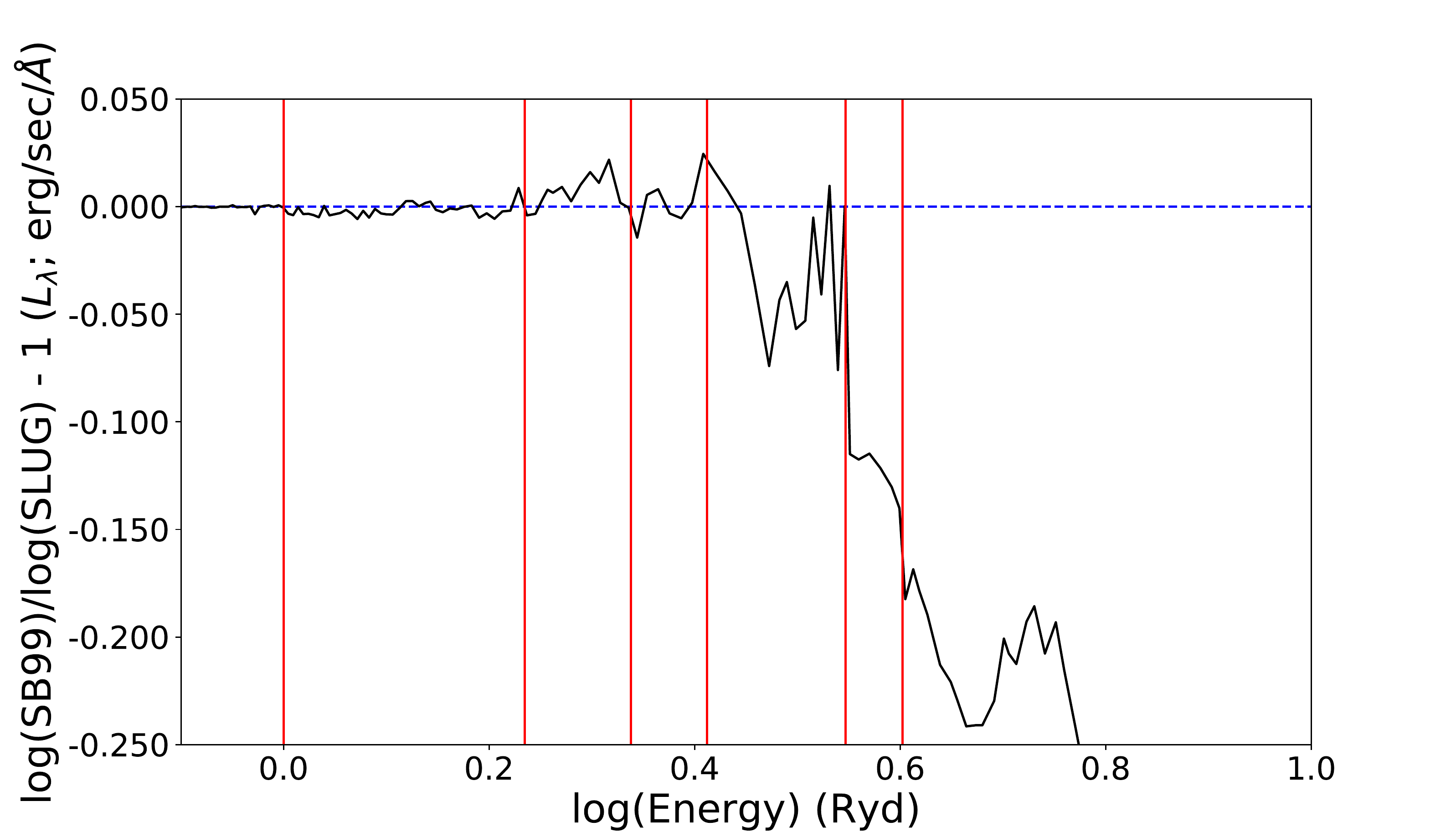}
\caption{5 Myr}
\label{fig:reldiff_inst5}
\end{subfigure}
\end{figure*}
\begin{figure*}\ContinuedFloat
\begin{subfigure}{0.49\textwidth}
\includegraphics[width=1.1\columnwidth]{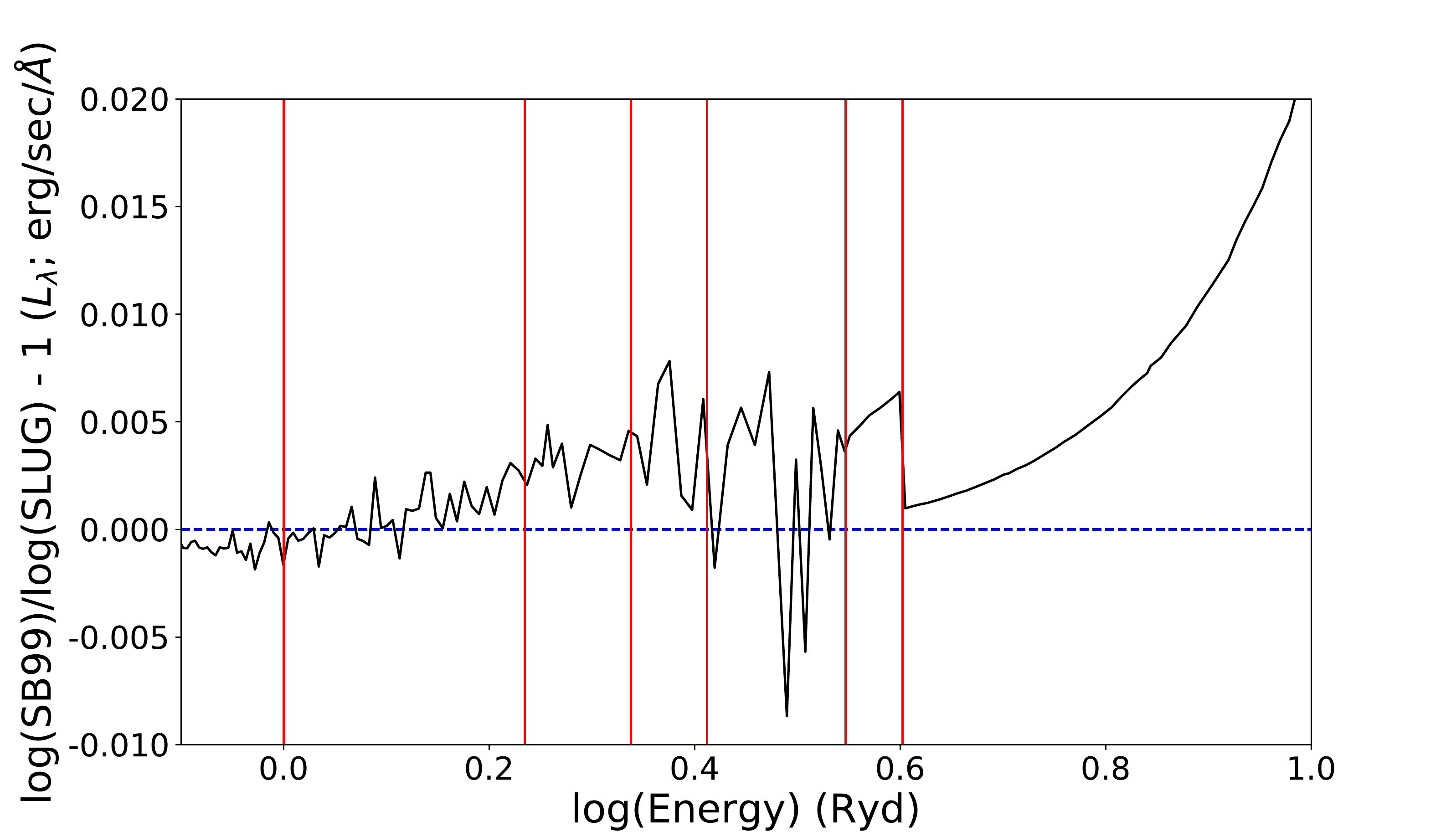}
\caption{6 Myr}
\label{fig:reldiff_inst6}
\end{subfigure}
\begin{subfigure}{0.49\textwidth}\vspace{0.7cm}
\centering
\includegraphics[width=\columnwidth]{spec_1e6_ikr_Z100_new_newlabel.pdf}
\caption{Cluster spectra, $10^6 M_\odot$}
\label{fig:spec_sluginst}
\end{subfigure}
\caption{Relative difference between spectra of an instantaneous SFH $10^6 M_\odot$ cluster computed with SB99 and SLUG. The relative difference is shown at ages of 0 Myr (10 Kyr) to 6 Myr inclusive, in increments of 1 Myr. Instantaneous SFH cluster spectra produced with SLUG for each age are shown in panel (h). $f_{\lambda}$ in (h) represents the individual input specta, and $f_0$ is the blackbody spectrum shown in Figure~\ref{fig:refspec}. The blue dashed line represents a relative difference of 0. The red vertical lines represent the ionisation potentials of important ISM species; from left to right: H$^0$, S$^+$, N$^+$, O$^+$, C$^{2+}$, He$^+$.}
\label{fig:reldiff_inst}
\end{figure*}

\begin{figure}
\centering
\includegraphics[width=1.1\columnwidth]{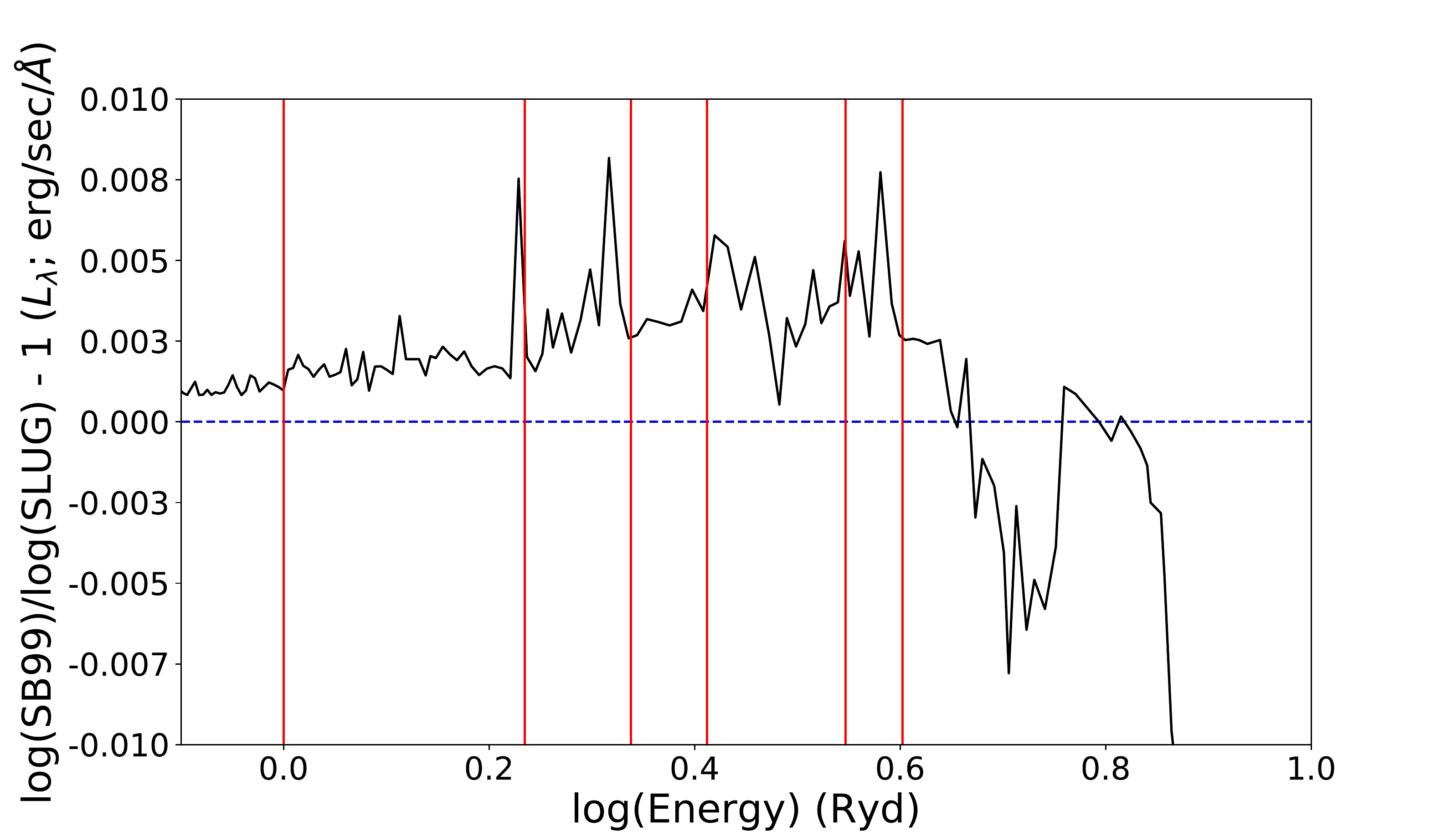}
\caption{Relative difference between the spectra of SB99 and SLUG for a continuous SFH cluster at $Z = 0.020$ and at an age of 5 Myr. The red vertical lines represent the ionisation potentials of important ISM species; from left to right: H$^0$, S$^+$, N$^+$, O$^+$, C$^{2+}$, He$^+$.}
\label{fig:reldiff}
\end{figure} 

\begin{figure}
\centering
\includegraphics[width=1.1\columnwidth]{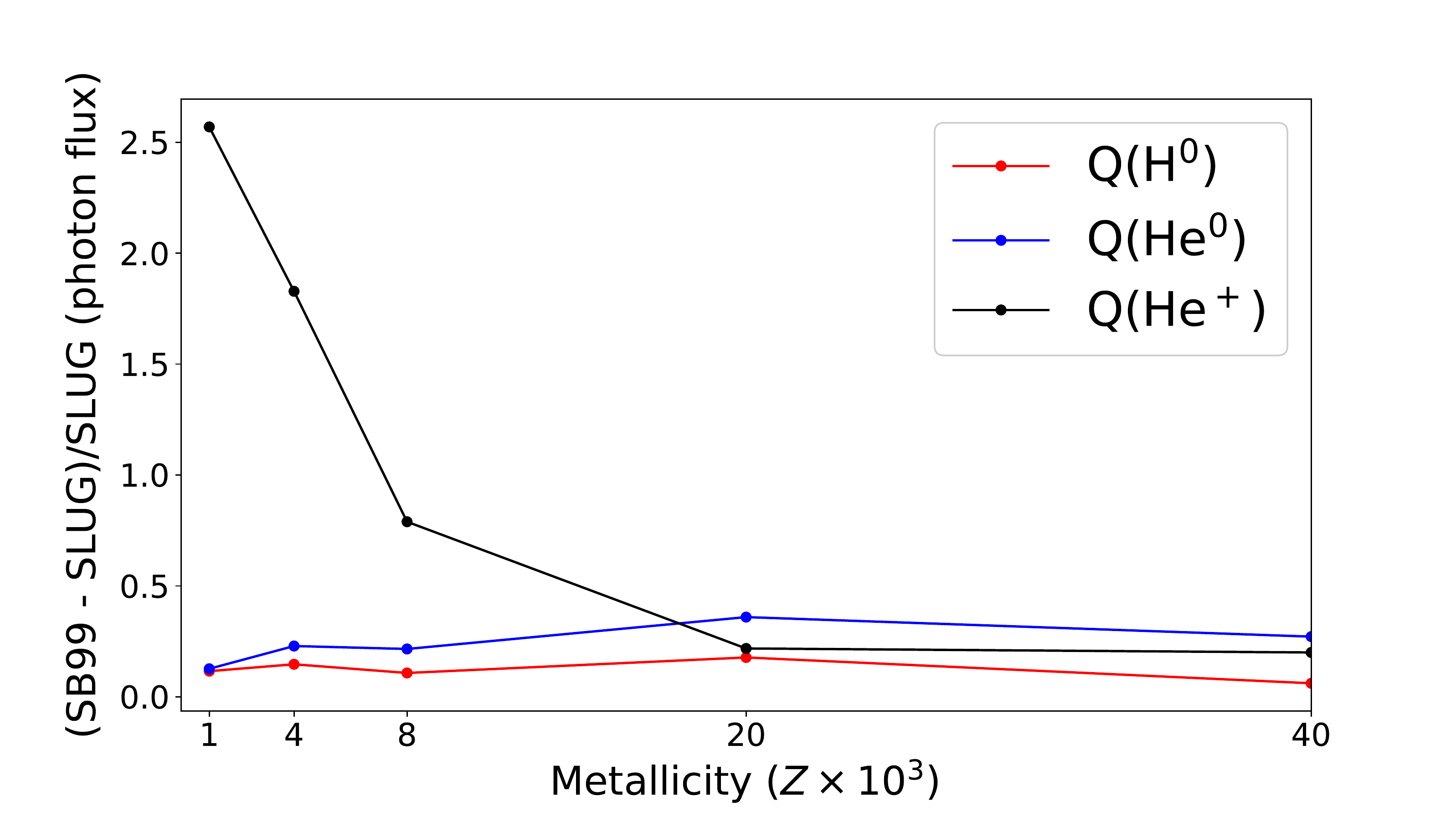}
\caption{Relative difference between SB99 and SLUG in the number of H$^0$-, He$^0$- and He$^+$-ionising photons vs. $Z$. All model parameters and input are equal to those in Figure~\ref{fig:reldiff}.}
\label{fig:spsflux}
\end{figure}  

We first explore the differences in the ionising spectra produced by the two SPS codes SB99 and SLUG. Even with the same input parameters and libraries, differences in the spectra between two SPS codes can still arise owing to the different assumptions and computational methods within the codes. In Figure~\ref{fig:reldiff_inst} we show the relative difference between the two ionising spectra generated using SB99 and SLUG for an individual cluster of mass $10^6 M_\odot$ at ages $0 - 6$ Myr, in 1 Myr increments. It should be noted that for all our simulations using SLUG, 10 kyr has been used as an approximation to 0 Myr; SLUG requires a simulation starting time greater than 0. Both codes produce spectra that agree at the ZAMS (0 Myr; Figure~\ref{fig:reldiff_inst0}), because the main sequence is well sampled and understood. 

However, at later ages, differences in the ionising spectra arise owing to the difficulty in modelling high-mass stars (Figures~\ref{fig:reldiff_inst}b-g). High-mass stars ($M {\sim} 120 M_\odot$) evolve rapidly off the main sequence and go through several phases of evolution (e.g. red supergiant, blue supergiant post helium flash, with also possibilities for W-R, luminous blue variables, yellow hypergiants). As a result, their position on the Hertzsprung-Russell (H-R) diagram can alter drastically and quickly \citep[see, e.g., Figure 5.2 of][]{BM1998}. Furthermore, mass sampling of massive-star evolutionary tracks used for isochrone interpolation is sparse, due to the difficulty in modelling, and obtaining constraining observations of high-mass stars (more on this in Section~\ref{sec:disclife}). Hence, the resulting stellar spectrum is heavily dependent on the specific stellar masses of the evolutionary tracks, as well as the interpolation method itself.

Shown in Figure~\ref{fig:reldiff} is the relative difference between SB99 and SLUG for a 5 Myr old cluster undergoing continuous star formation at a rate of $1 M_\odot \;\mathrm{yr}^{-1}$. The difference in luminosity between the two spectra is less than 1\%, with SB99 producing the more luminous ionising spectrum for most energies. At energies of log($E$/ryd) $>$ 0.6, SLUG is shown to produce the harder ionising spectrum. Figure~\ref{fig:spsflux} also shows the more luminous ionising spectrum overall produced by SB99, by showing a relative increase in the flux of H$^0$-, He$^0$- and He$^+$-ionising photons for all $Z$.

\subparagraph{Stochastic sampling of IMF}

A main difference between SB99 and SLUG is the method by which each SPS code infers values for the parameters of the stellar cluster. SLUG regards each parameter as a probability distribution and will draw values for each parameter through Monte Carlo simulations. One such parameter for which this applies is the IMF. Here we show the effect of stochastic sampling of the IMF on the resulting stellar spectrum by comparing the spectra of SLUG and SB99 -- the latter of which does not include stochasticity -- at varying cluster masses. The size of the cluster is known to make a difference to the cluster's final ionising spectrum, because at low cluster masses undersampling from the IMF may lead to a skew in the stellar masses. 
 
\begin{figure}
\centering
\includegraphics[width=\columnwidth]{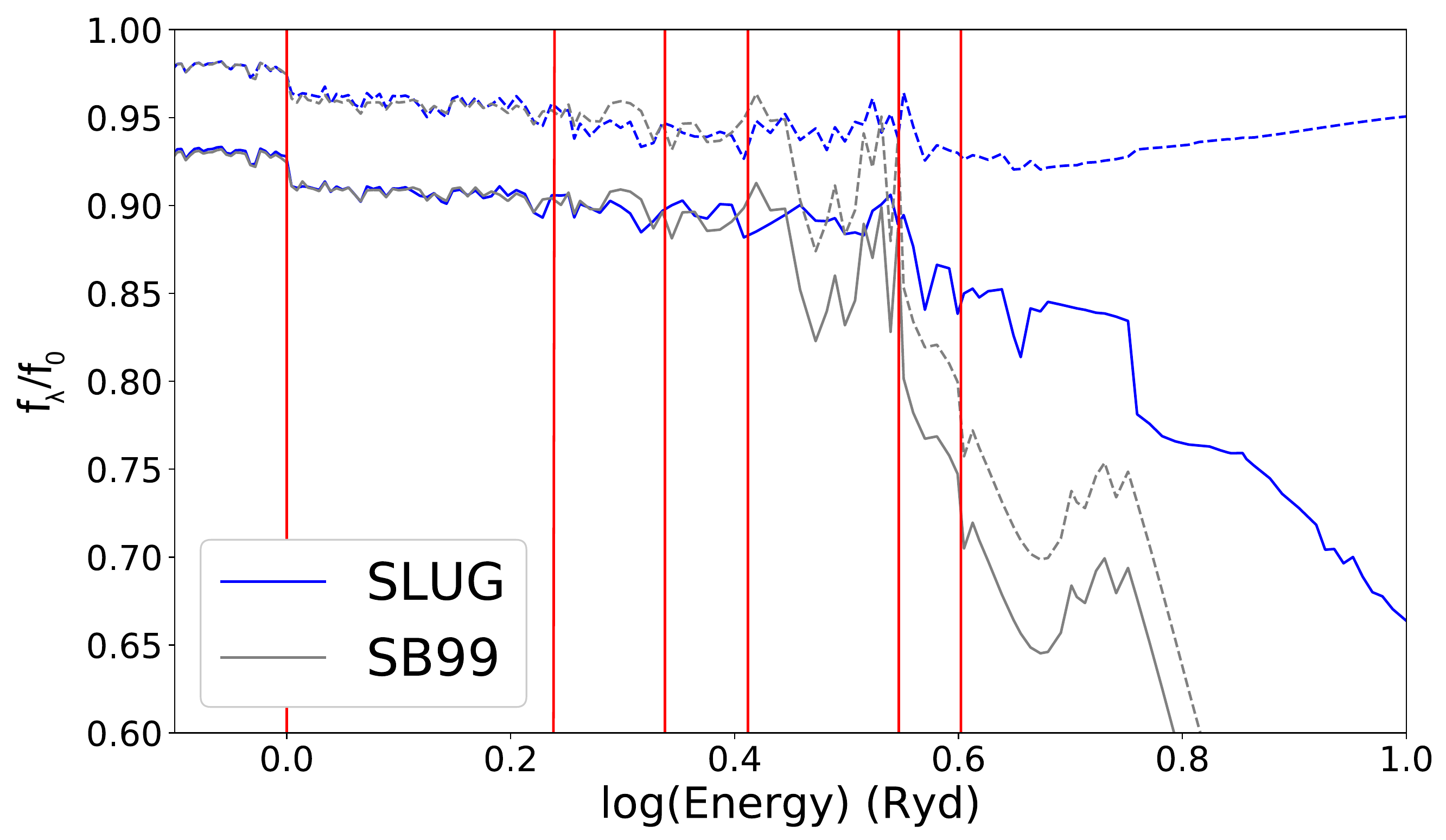}
\caption{Spectra for instantaneous SFH clusters of age 5 Myr at masses of $10^4 M_\odot$ (solid) and $10^6 M_\odot$ (dashed), produced with SB99 and SLUG. All spectra are at a metallicity of $Z = 0.020$. $f_{\lambda}$ represents the individual input specta, and $f_0$ is the blackbody spectrum shown in Figure~\ref{fig:refspec}. The red vertical lines represent the ionisation potentials of important ISM species; from left to right: H$^0$, S$^+$, N$^+$, O$^+$, C$^{2+}$, He$^+$.}
\label{fig:stochasticspec}
\end{figure}

Assuming that the IMF is a pdf like in SLUG, \citet{Cervino2013} and references therein show for cluster masses of $\lesssim 10^4\;M_\odot$ that there is a significant scatter in the mass of the most massive star within the cluster. Hence, at a cluster mass of $10^4\;M_\odot$ and below, variations in the distribution of individual stellar masses in the cluster can lead to very different ionising spectra (and hence a different spectrum for each SPS code execution). From the mass-luminosity relation \citep[first proposed by][]{Eddington1924}, it can be shown that stellar mass and flux are proportional. Hence, large variations in the masses of the stars within the cluster will lead to similarly large variations in the ionising spectra. This is shown in Figure~\ref{fig:stochasticspec} for stellar spectra produced stochastically through SLUG and nonstochastically through SB99 for cluster masses of $10^4 M_\odot$ and $10^6 M_\odot$. Each spectrum produced using SLUG is the result of one Monte Carlo experiment. For SPS codes that populate a stellar cluster simply according to the distribution of the IMF (such as SB99), the stellar spectra of the $10^4$ and $10^6 M_\odot$ clusters have the same shape, with the total luminosity varying by a scale factor. However, if the stellar cluster is populated through the stochastic sampling of the IMF (as in SLUG), the resulting stellar spectra at cluster masses of $10^4$ and $10^6 M_\odot$ will likely be very different.

\paragraph{Binary populations}
\label{sec:binarypop1}

\begin{figure*}[p]
\begin{subfigure}{0.49\textwidth}
\includegraphics[width=1.1\columnwidth]{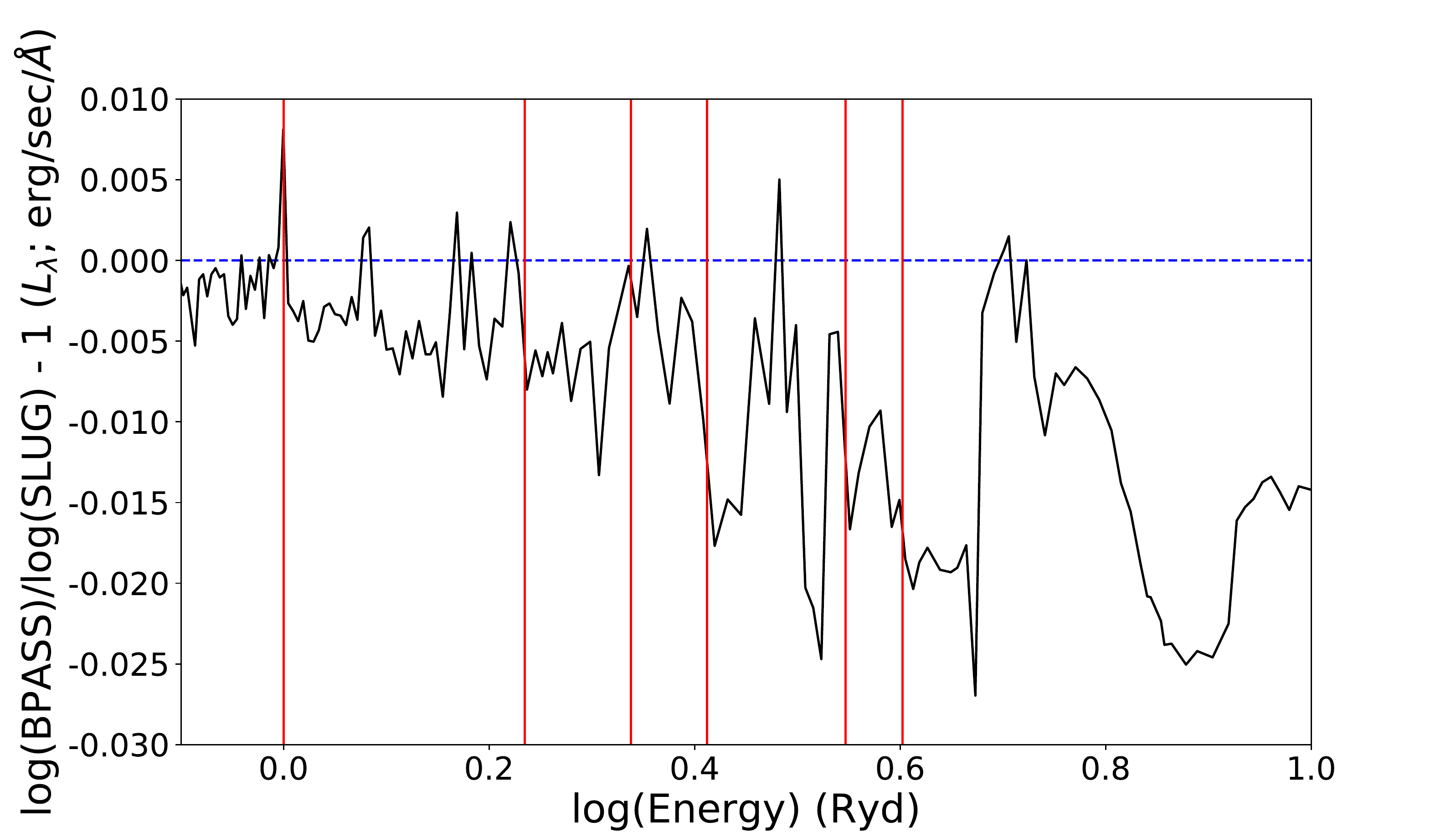}
\caption{1 Myr}
\label{fig:bpassslug_inst1}
\end{subfigure}
\begin{subfigure}{0.49\textwidth}
\includegraphics[width=1.1\columnwidth]{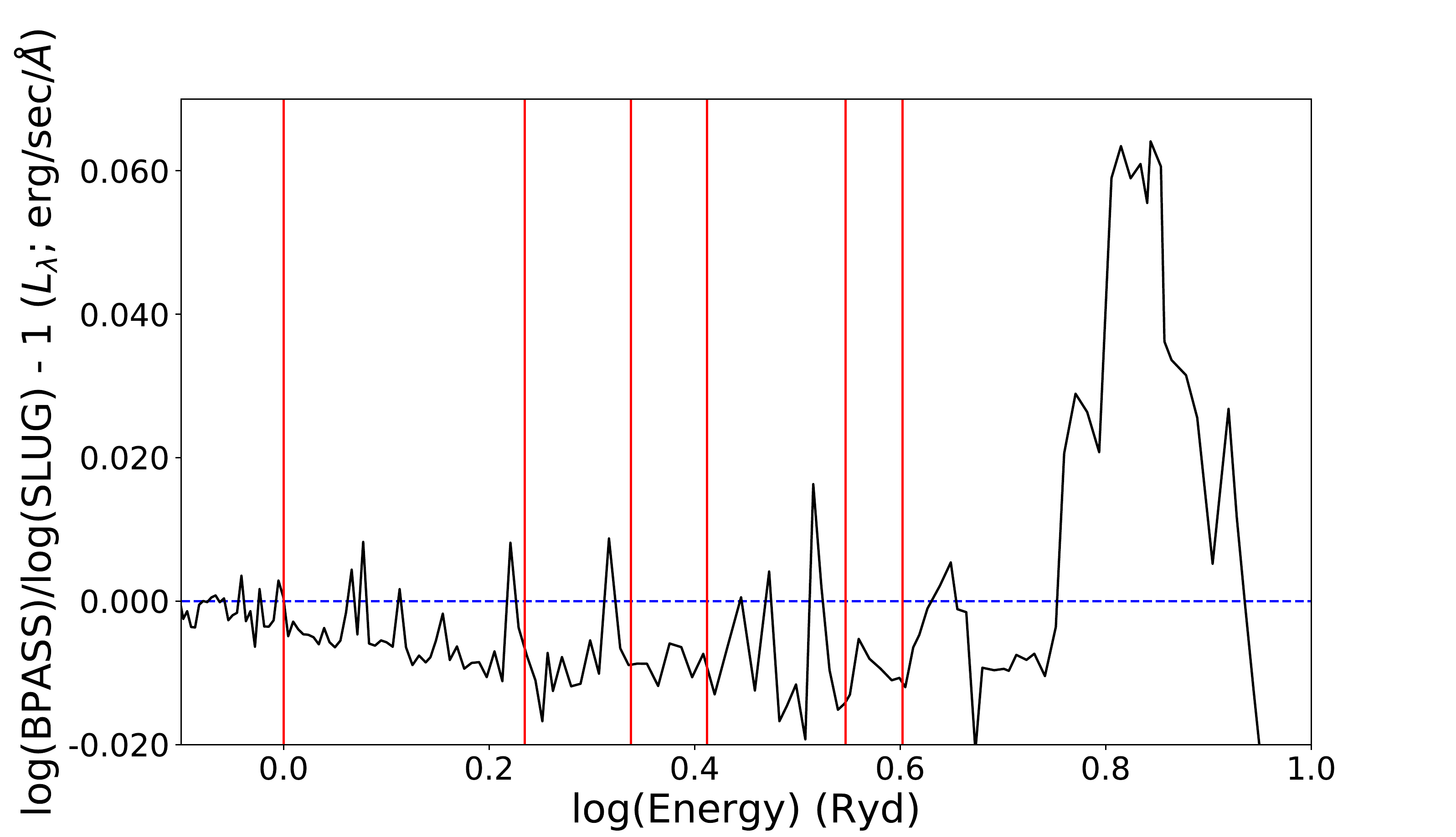}
\caption{2 Myr}
\label{fig:bpassslug_inst2}
\end{subfigure}
\begin{subfigure}{0.49\textwidth}
\includegraphics[width=1.1\columnwidth]{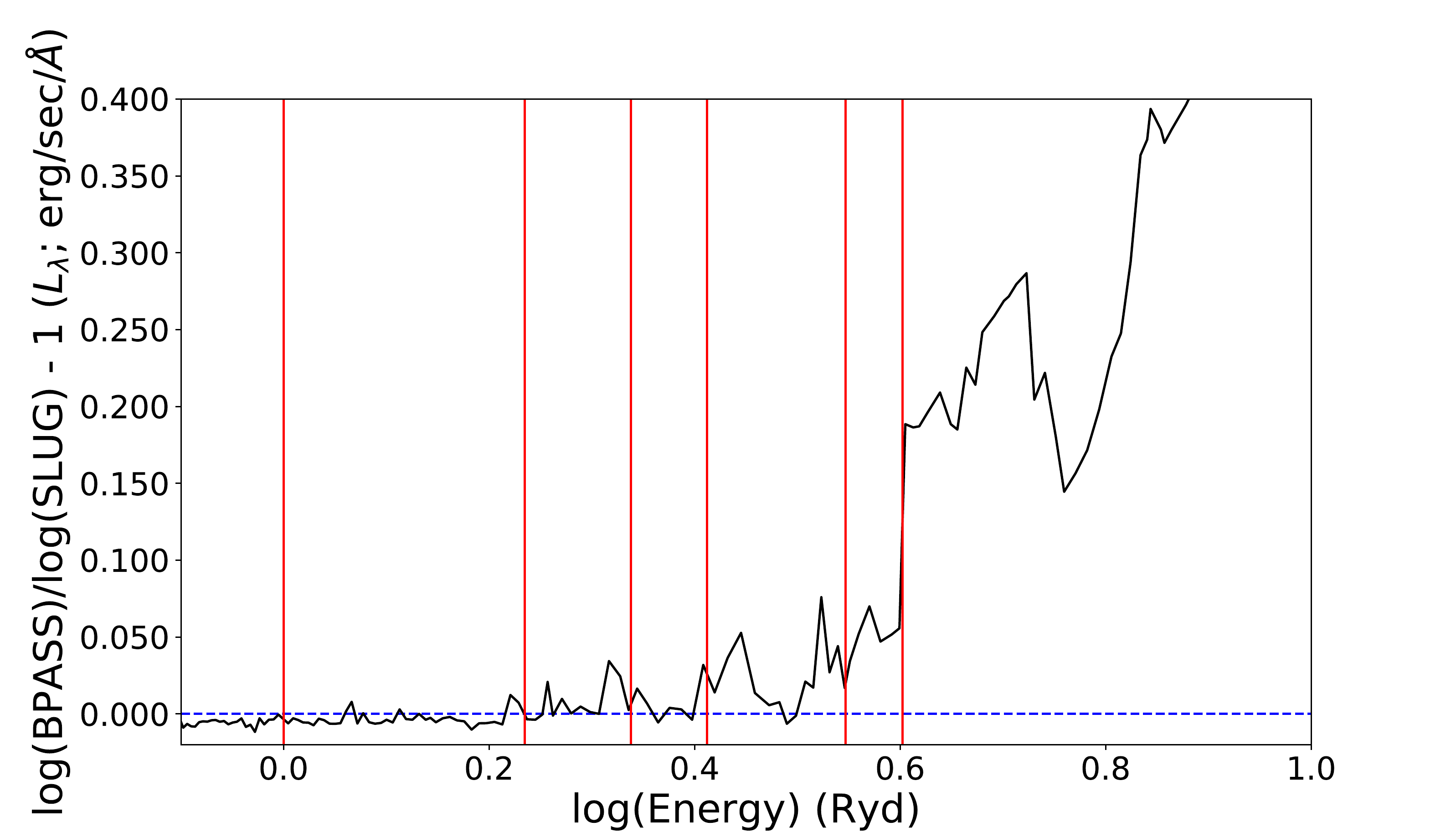}
\caption{3 Myr}
\label{fig:bpassslug_inst3}
\end{subfigure}
\begin{subfigure}{0.49\textwidth}
\includegraphics[width=1.1\columnwidth]{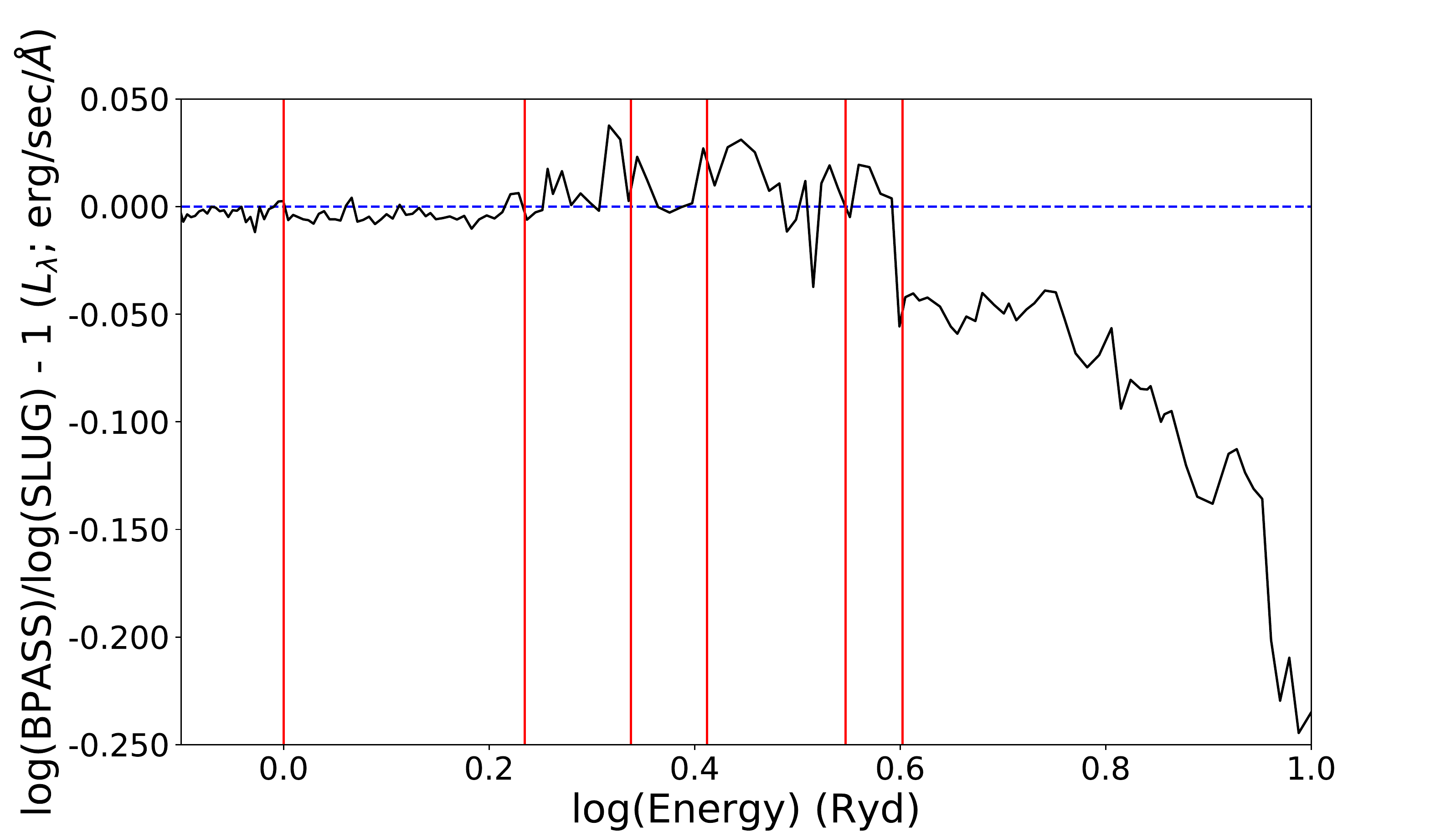}
\caption{4 Myr}
\label{fig:bpassslug_inst4}
\end{subfigure}
\begin{subfigure}{0.49\textwidth}
\includegraphics[width=1.1\columnwidth]{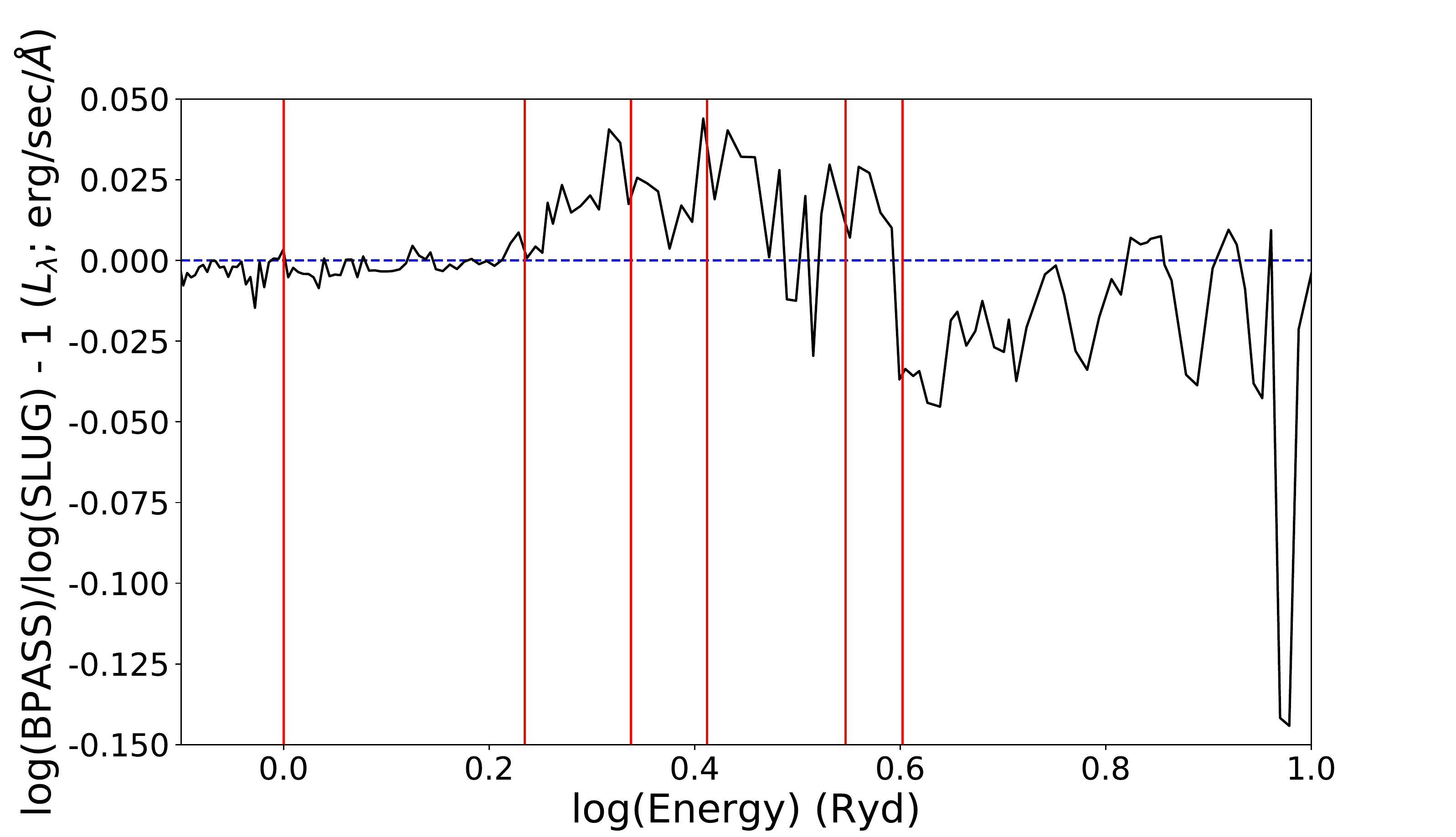}
\caption{5 Myr}
\label{fig:bpassslug_inst5}
\end{subfigure}
\begin{subfigure}{0.49\textwidth}
\includegraphics[width=1.1\columnwidth]{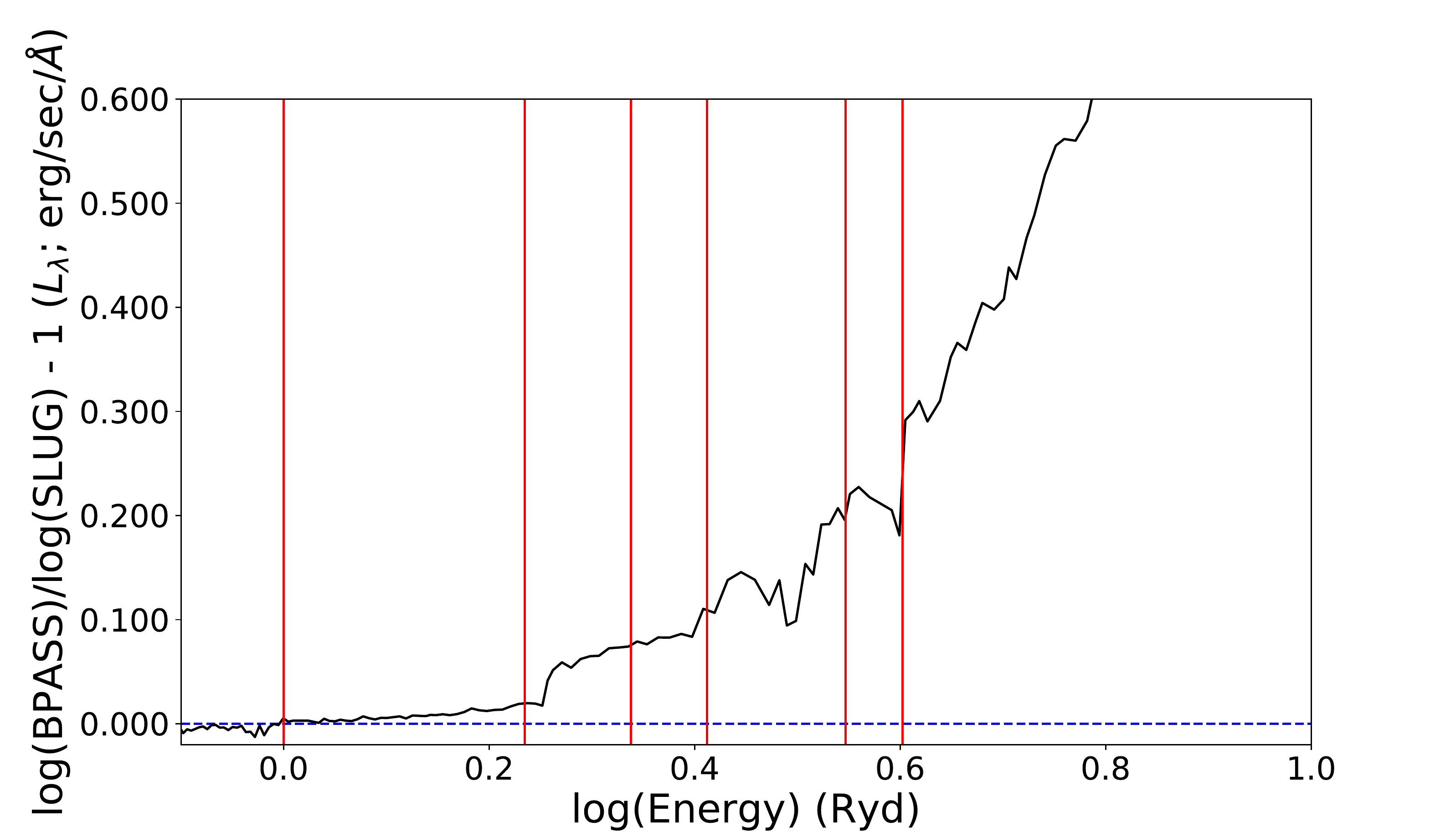}
\caption{6 Myr}
\label{fig:bpassslug_inst6}
\end{subfigure}
\end{figure*}
\begin{figure*}\ContinuedFloat
\begin{subfigure}{0.49\textwidth}
\includegraphics[width=\columnwidth]{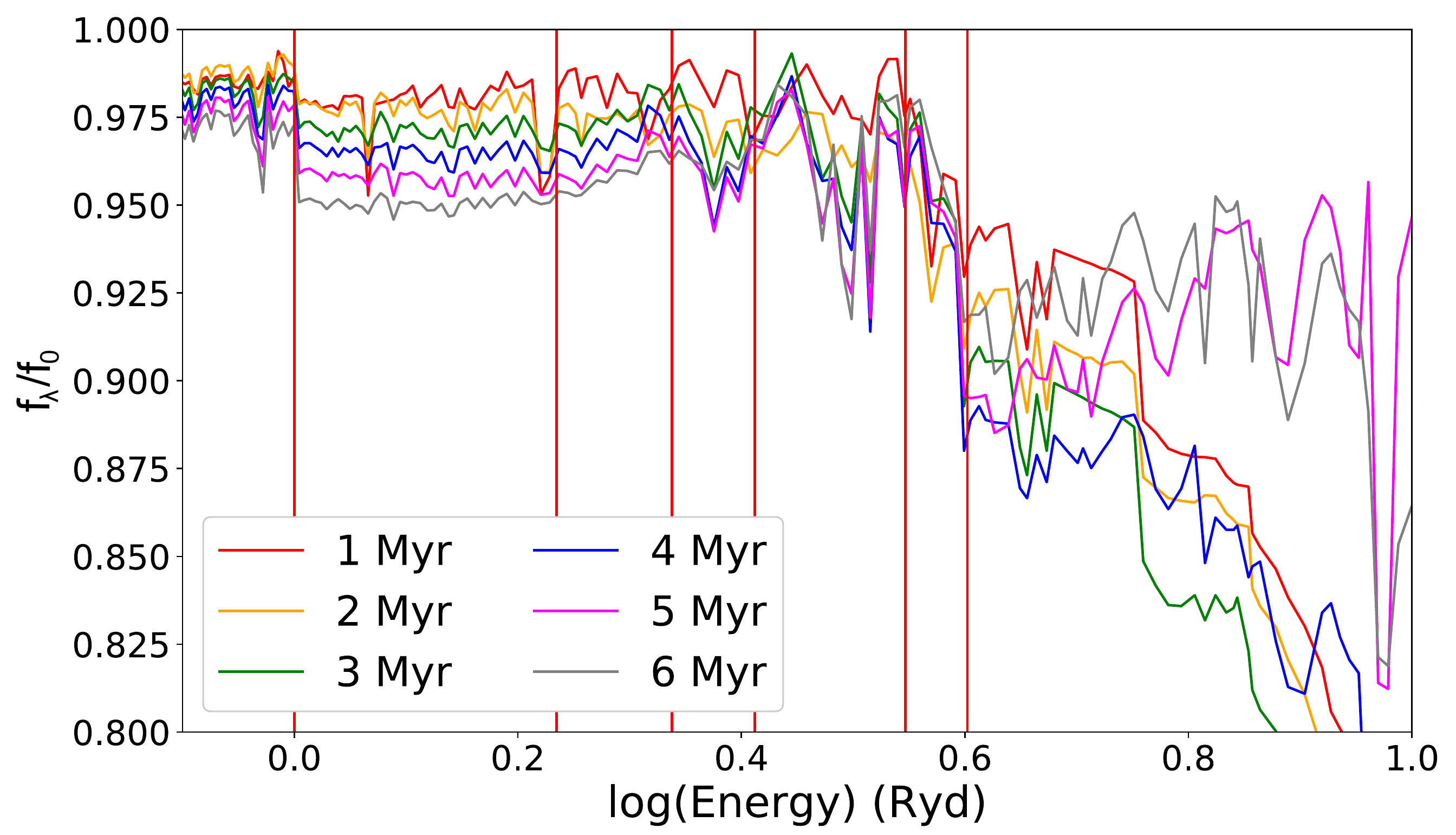}
\caption{Normalised BPASS spectra, 1 Myr to 6 Myr}
\label{fig:bpass_norm_ages}
\end{subfigure}
\caption{Relative difference between spectra of an instantaneous SFH $10^6 M_\odot$ cluster computed with BPASS and SLUG. The relative difference is shown at ages of $1 - 6$ Myr inclusive, in increments of 1 Myr. Instantaneous SFH binary cluster spectra produced with BPASS for each age are shown in panel (g). $f_{\lambda}$ in panel (g) represents the individual input specta, and $f_0$ is the blackbody spectrum shown in Figure~\ref{fig:refspec}. The red vertical lines represent the ionisation potentials of important ISM species; from left to right: H$^0$, S$^+$, N$^+$, O$^+$, C$^{2+}$, He$^+$.}
\label{fig:bpassslug_age}
\end{figure*}

\begin{figure}
\centering
\includegraphics[width=1.1\columnwidth]{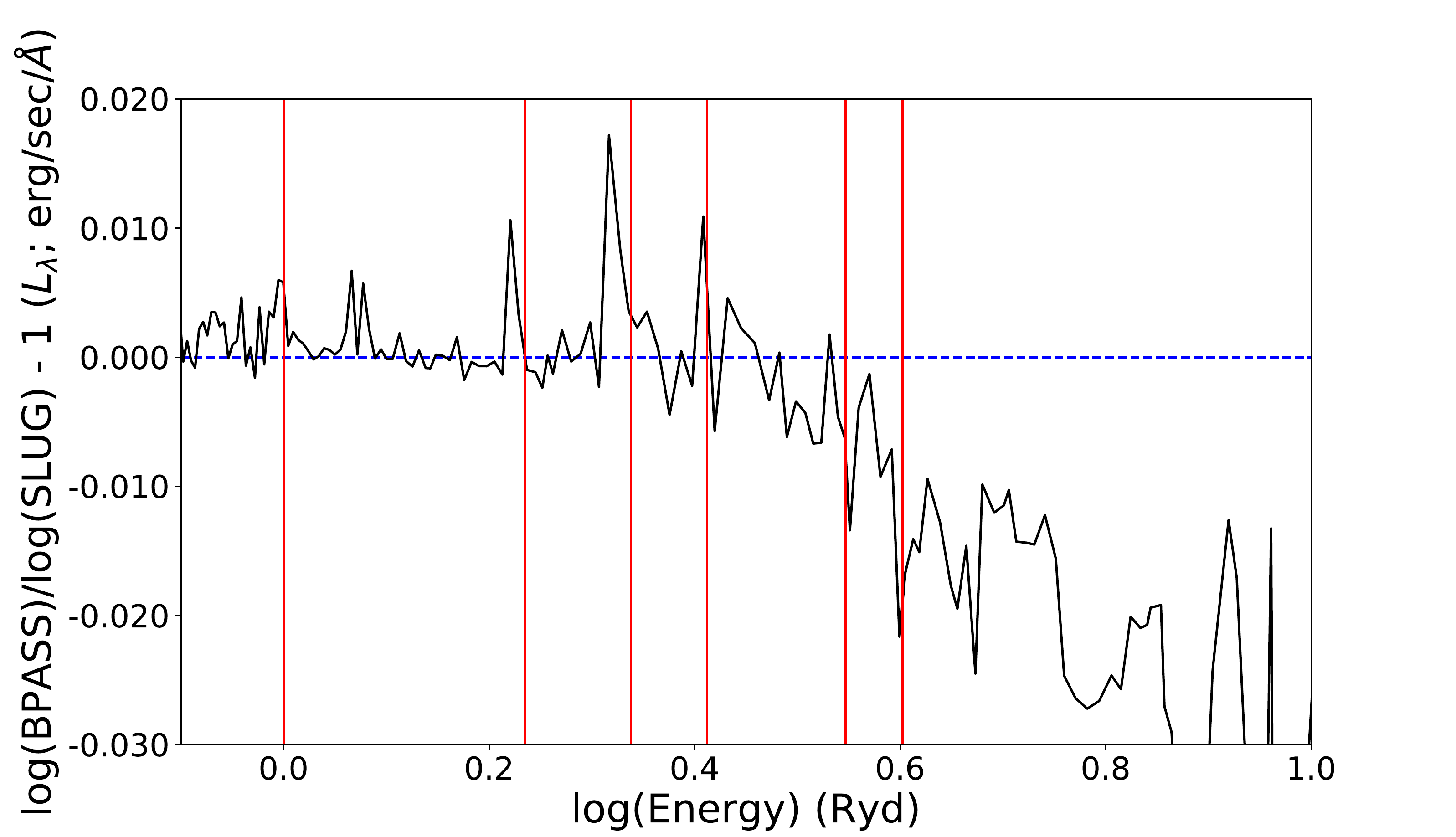}
\caption{Relative difference between the spectra of BPASS and SLUG for a continuous SFH cluster at $Z = 0.020$ and at an age of 5 Myr. The red vertical lines represent the ionisation potentials of important ISM species; from left to right: H$^0$, S$^+$, N$^+$, O$^+$, C$^{2+}$, He$^+$.}
\label{fig:bpassslug_cont}
\end{figure} 

\begin{figure}
\centering
\includegraphics[width=1.1\columnwidth]{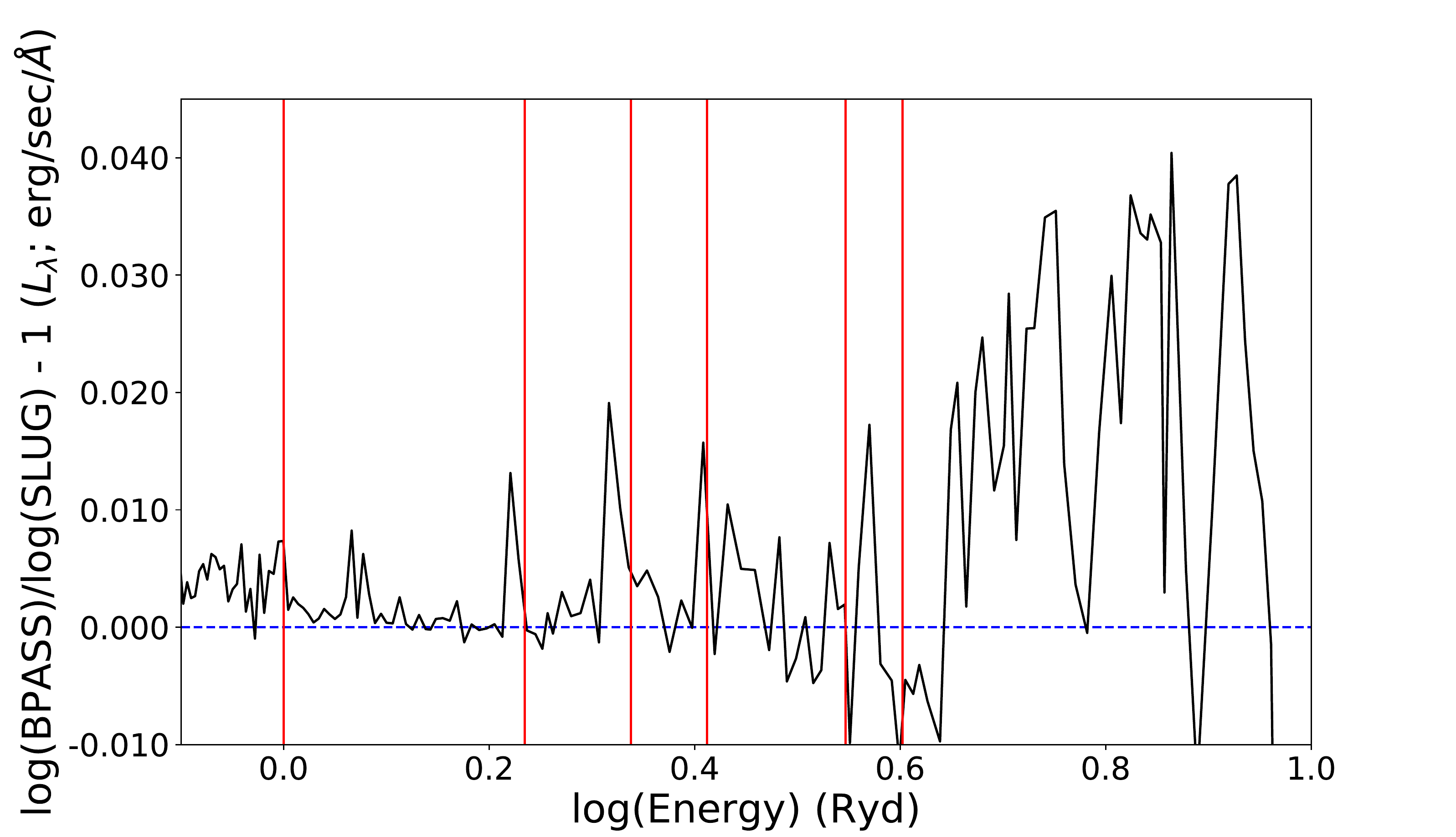}
\caption{Relative difference between the spectra of BPASS and SLUG for a continuous SFH cluster at $Z = 0.020$ and at ages for which the stellar spectrum has reached equilibrium (5 Myr for SLUG, and 10 Myr for BPASS, seen in Figure~\ref{fig:bpassspectra_cont}). The red vertical lines represent the ionisation potentials of important ISM species; from left to right: H$^0$, S$^+$, N$^+$, O$^+$, C$^{2+}$, He$^+$.}
\label{fig:bpassslug_cont_eq}
\end{figure} 

We compare spectra produced using both BPASS and SLUG, showing the relative difference between instantaneous SFH spectrum luminosities with cluster age in Figure~\ref{fig:bpassslug_age}, and for a continuous SFH cluster at 5 Myr in Fgure~\ref{fig:bpassslug_cont}. Seen in Figure~\ref{fig:bpassslug_age}, neither spectrum is systematically harder at all ages. The luminosity of spectra from SLUG increases drastically between cluster ages of 3 and 4 Myr, coinciding with the emergence of W-R stars in the cluster. As a result, SLUG spectra at these ages are harder than those from BPASS. However, as high-mass stars and W-R stars reach the end of their lives at ${\sim} 5-6$ Myr in the single-star population produced with SLUG, W-R stars in the binary population produced with BPASS still continue to emerge, continuing to increase the far-UV (FUV) photon flux. Hence, BPASS spectra are harder than SLUG spectra at ages beyond 5 Myr. The relative difference in the spectra for continuous SFH populations between SLUG and BPASS shown in Figure~\ref{fig:bpassslug_cont} shows BPASS to produce a larger flux for photons of lower energies, with SLUG providing a harder flux for photons with energies  $\gtrsim 0.6$ in log($E$/ryd). Ultimately, SLUG and BPASS use different models for the stellar tracks and atmospheres, which inevitably lead to differences in the output spectra. Treatments for W-R stars differ between the two SPS codes owing to the use of disparate atmospheres to model W-R star atmospheres; hence relative differences in the spectra between the two SPS codes are to be expected, particularly in the FUV part of the spectrum. Differences in synthesis procedure such as the aformentioned stellar track sampling, interpolation method, and stellar atmosphere grid sampling are also very important in the differing spectra produced using the two codes, for all other identical input parameters.

The high-energy region of the spectrum $({\sim} 0.6 - 1$ in log($E$/ryd)) produced by BPASS is still relevant up to an age of 6 Myr, seen in Figures~\ref{fig:bpassslug_inst6} and~\ref{fig:bpass_norm_ages}. This is in stark contrast to the spectra produced by SLUG up to similar cluster ages, where luminosity decreases rapidly in the same region of the spectrum as a result of high-mass star and W-R star deaths. We only consider BPASS cluster ages up to and including 6 Myr in order to directly compare with the clusters produced using SLUG. However, we note that significant ionising radiation produced by clusters simulated with BPASS is present from clusters of ages beyond 6 Myr. \citet{Wofford2016} show that BPASS models at varying metallicity continue to produce ionising radiation at cluster ages beyond 10 Myr, stating three main processes for the sustainment of ionising radiation at later ages. Firstly, stars in binary systems may be ``rejuvenated' towards the end of their lifetimes through mass transfer as a result of Roche lobe overflow \citep[RLOF; e.g.][and references therein]{DT2007}. Hence, the rejuvenated star appears younger and may evolve similarly to a younger star of its new mass \citep[although this is not always the case;][]{DT2007}. Second, and similarly to rejuvenation through mass transfer, binary stars may be rejuvenated through mergers, thus forming a single star that appears younger than the original age of the binary system \citep[e.g.][and references therein]{Schneider2016}. These two rejuvenation processes cause higher-mass stars to be present at later ages of the simulation, far beyond the time frames expected with single stellar models. The third process is envelope removal from RLOF, leading to low-luminosity W-R stars and helium stars at later ages than found in single stellar clusters. Note that envelope removal will typically lead to low-luminosity W-R stars of high-$Z$ and/or high-luminosity (and subsequently high mass) owing to a thin hydrogen layer remaining, which requires removal from sufficiently high stellar winds \citep[and references therein]{Smith2014,Trani2014}. 

\begin{figure*}
\begin{subfigure}{0.49\textwidth}
\includegraphics[width=\columnwidth]{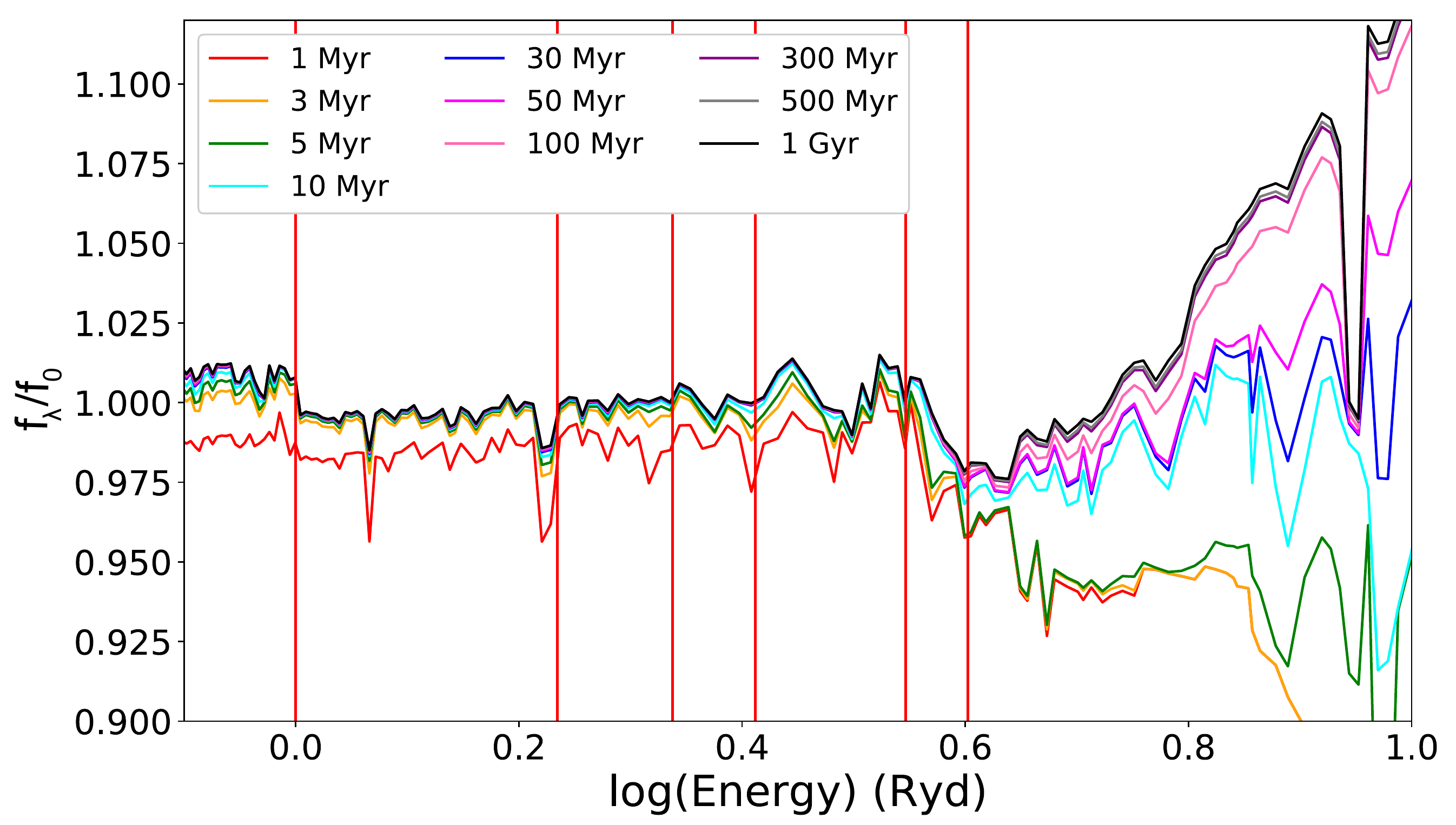}
\caption{Continuous SFH, $1 M_\odot/\mathrm{yr}$}
\label{fig:bpassspectra_cont}
\end{subfigure}
\begin{subfigure}{0.49\textwidth}
\includegraphics[width=\columnwidth]{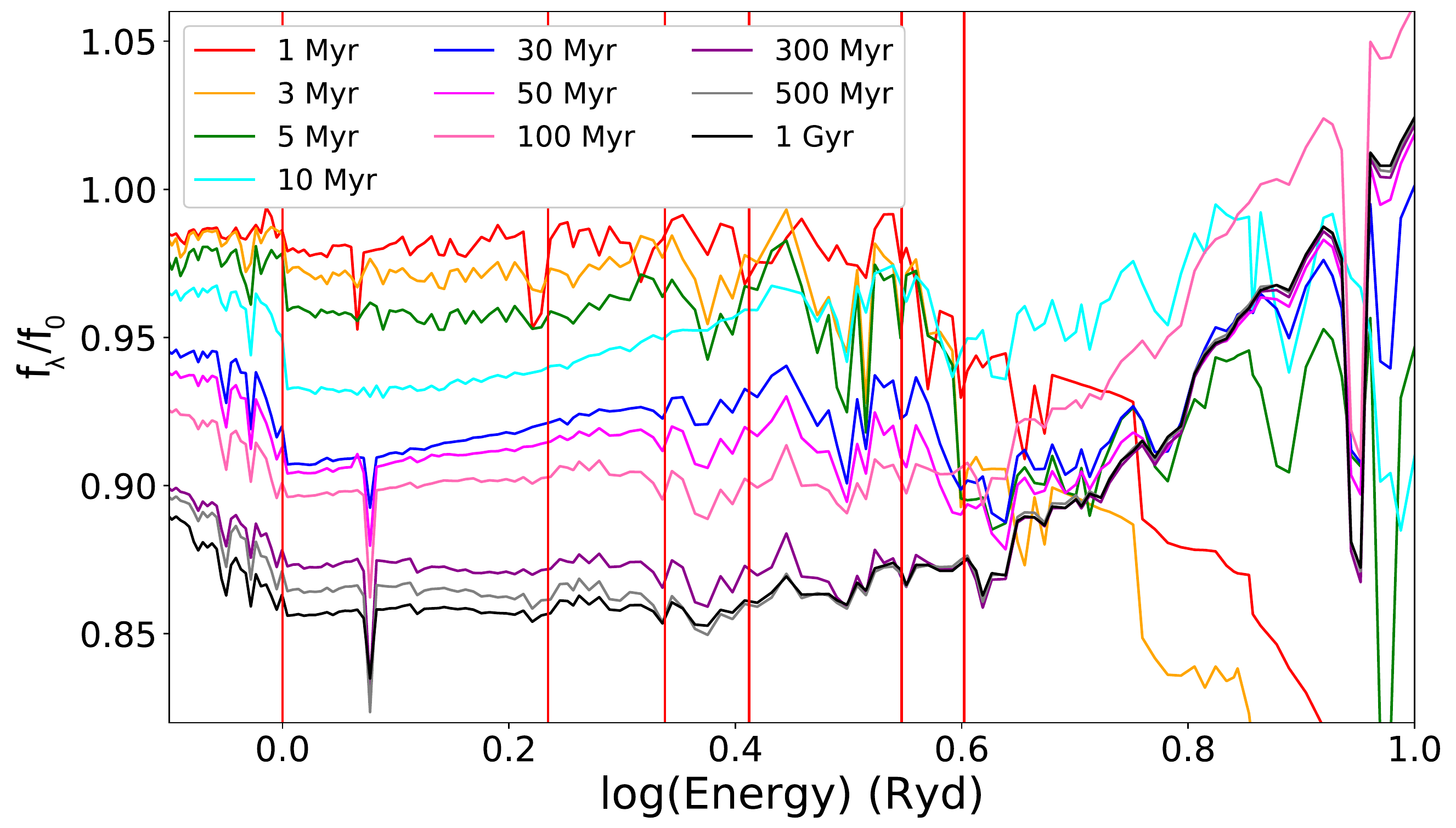}
\caption{Instantaneous SFH, $10^6 M_\odot$ cluster}
\label{fig:bpassspectra_inst}
\end{subfigure}
\caption{Spectra of clusters produced with BPASS assuming a continuous and instantaneous SFH in panels (a) and (b), respectively. $f_{\lambda}$ represents the individual input specta, and $f_0$ is the blackbody spectrum shown in Figure~\ref{fig:refspec}. The red vertical lines represent the ionisation potentials of important ISM species; from left to right: H$^0$, S$^+$, N$^+$, O$^+$, C$^{2+}$, He$^+$.}
\label{fig:bpassspectra}
\end{figure*}

We adopt an age of 5 Myr for a stellar cluster undergoing constant star formation at a rate of $1 M_\odot\;\mathrm{yr}^{-1}$ for our fiducial model because 5 Myr is the age at which the spectrum from a continuous SFH stellar population reaches equilibrium \citep{Kewley2001}. Also, when considering a single stellar cluster, we neglect the spectra from ages beyond 6 Myr owing to the considerable decrease in ionising radiation at these ages \citep[e.g.][and Figure~\ref{fig:instspec}]{Wofford2016}. We note, however, that these ages associated with continuous and instantaneous SFH are based on single stellar clusters. Shown in Figure~\ref{fig:bpassspectra} are spectra produced using BPASS for clusters assuming both continuous and instantaneous SFHs. It is evident from Figure~\ref{fig:bpassspectra} that binary populations reach equilibrium at much later times than single-star populations, for the same three reasons (rejuvenation, mergers, and envelope removal) mentioned earlier. From Figure~\ref{fig:bpassspectra_cont}, it can be seen that different regions of the spectrum produced by a continuous SFH binary population reach equilibrium at different ages. Up until the He \textsc{ii} continuum (below ${\sim} 0.6$ in log($E$/ryd)), the shape of the spectrum stabilises at an age of roughly 10 Myr. At energies within the He \textsc{ii} continuum, however (above ${\sim} 0.6$ in log($E$/ryd)), the spectrum only begins to approach a constant shape and luminosity at ages approaching 1 Gyr. This is a result of the unpredictability of W-R and He star emergence in binary clusters, which drastically increase the flux in the FUV part of the spectrum. 

Figure~\ref{fig:bpassslug_cont_eq} shows the relative difference in the spectra between the two SPS codes for a metallicity of $Z = 0.020$, at ages for which the stellar spectrum has reached equilibrium (5 Myr for SLUG, 10 Myr for BPASS). At these ages, the spectrum produced from BPASS is almost uniformly more luminous for all photon energies than that produced using SLUG, supporting the findings of \citet{Stanway2016}.

W-R and He star emergence and impact are seen more clearly when considering the evolution of a single binary cluster (Figure~\ref{fig:bpassspectra_inst}). For high-metallicity ($Z = 0.014$) BPASS models, \citet{Wofford2016} show a gradual decrease in the number of H$^0$- and He$^0$-ionising photons with increasing age. However, distinct increases in the number of He$^+$-ionising photons emitted by the binary cluster are seen at ${\sim} 10$ Myr and again at ${\sim} 80$ Myr (their Figure 2). This is supported through Figure~\ref{fig:bpassspectra_inst}, showing distinct increases in the spectrum luminosity at ages of 10 Myr and 80 Myr for energies $> 0.6$ in log($E$/ryd), while showing a constant gradual decrease in the spectrum luminosity for lower energies. 

\subsection{BPT Diagrams}
\label{sec:modelcomp2}
\subsubsection{SFH and Age}
\label{sec:age2}

\begin{figure}
\centering
\includegraphics[scale=0.52]{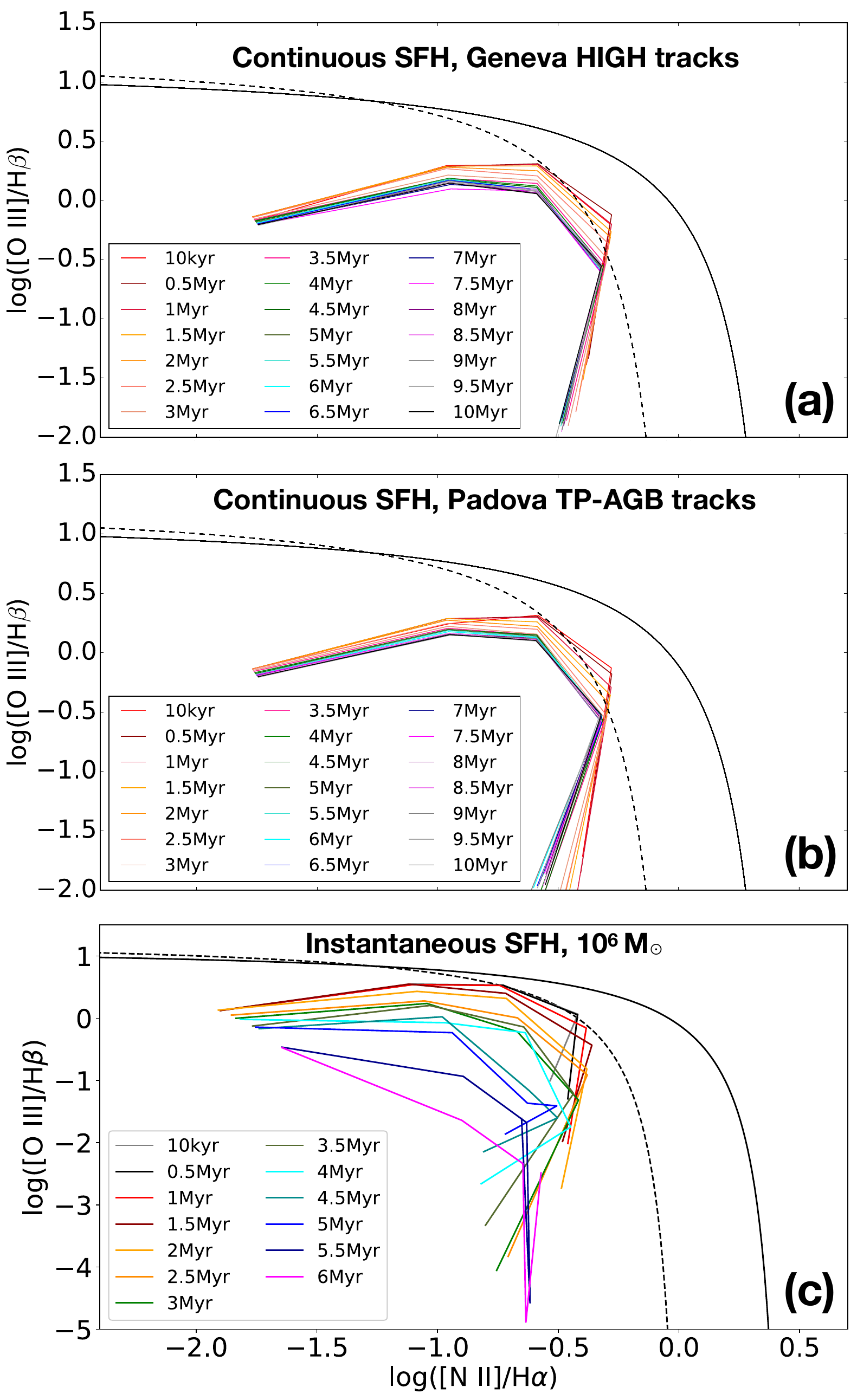}
\caption{The $\mathrm{log}(q(N)/(\mathrm{cm}\;  \mathrm{s}^{-1})) = 7.25$ branch shown at varying ages for continuous SFH clusters in panels (a) and (b), and an instantaneous SFH cluster of mass $10^6 M_\odot$ in panel (c). Panel (a) is a continuous SFH cluster using the Geneva HIGH stellar tracks, while panel (b) uses the Padova TP-AGB tracks. Ages for the continuous SFH cluster extend from 0 Myr (10 kyr) to 10 Myr in 0.5 Myr increments. Ages for the instantaneous SFH cluster range from 0 Myr (10 kyr) to 6 Myr in 0.5 Myr increments.}
\label{fig:sfhgrids}
\end{figure}

Figure.~\ref{fig:sfhgrids} shows the $\mathrm{log}(q(N)/(\mathrm{cm}\; \mathrm{s}^{-1})) = 7.25$ branch of the photoionisation models for both a continuous and instantaneous SFH cluster at varying ages. We show variation with age in a continuous SFH cluster using both the Geneva HIGH tracks and the Padova TP-AGB tracks in Figures~\ref{fig:sfhgrids}(a) and (b), respectively. In general, the [O \textsc{iii}]/H$\beta$ and [N \textsc{ii}]/H$\alpha$ emission-line ratios decrease with age when assuming a continuous SFH. This decrease in the emission-line ratios corresponds to aging stellar populations that produce a softer spectrum, as well as continually emerging young stellar populations that produce a higher flux in hydrogen recombination lines such as H$\alpha$ and H$\beta$. Eventually, however, the shape and position of the photoionisation models stabilises, as the stellar population reaches equilibrium. The age at which this stabilisation occurs differs for the two sets of tracks. The age of stabilisation for models using the Geneva HIGH tracks was found to be 8 Myr by \citet{Kewley2001}. This age agrees with our models using the Geneva HIGH tracks in Figure~\ref{fig:sfhgrids}a. \citet{Kewley2001} also find the age of stabilisation for models using the Padova tracks to be 6 Myr, albeit using an earlier version of the stellar track models described by \citet{Bressan1993}. Figure~\ref{fig:sfhgrids}b appears to show different ages of stabilisation for different model metallicities. At low metallicity, the models only truly stabilise at 10 Myr. At higher metallicity ($Z = 0.020$), the age of stabilisation is roughly 6 Myr. At the maximum metallicity used in the Padova tracks ($Z = 0.050$), the position and evolution of the grids are never seen to stabilise for the ages used. If desiring a stellar age for which models using the Padova TP-AGB tracks are stable, an age of 10 Myr should be chosen. Sources such as \citet{CL2001} and \citet{Feltre2016} show that in general 10 Myr is the age at which 99.9\% of ionising photons have been released for a single stellar generation; thus, no further evolution in the shape of the ionising spectrum is seen past 10 Myr.

With no new stellar populations emerging within the cluster, the instantaneous SFH models never reach a stabilisation point. The emission-line flux ratios for the models shown in Figure~\ref{fig:sfhgrids}c continually decrease with age, excluding the ages for which the hardness of the spectrum increases with the emergence of W-R stars. W-R stars are more abundant in higher-metallicity environments, believed to be the result of mass-loss rate dependence on metallicity \citep{Crowther2007,Mokiem2007,Georgy2015}. Hence, the BPT emission-line flux ratios show a larger increase as the metallicity in the models increases, due to the presence of W-R stars. Considering the ages at which W-R stars do not contribute to the spectra, the emission-line flux ratios decrease with age faster for increasing metallicity. Higher-metallicity stars spend a larger fraction of photons ionising the more abundant metals in their atmospheres. Therefore, there are fewer ionising photons emitted from stars in order to ionise and excite the surrounding ISM for increasing metallicity \citep{Snijders2007}. 

\subsubsection{Stellar Evolutionary Tracks}
\label{sec:tracksbpt}

Model grids incorporating both the Geneva HIGH and Padova TP-AGB stellar tracks for a continuous SFH population at an age of 5 Myr can be seen together in Figure~\ref{fig:threecompgrids}a. Overall, the grids are extremely similar, with the largest difference apparent at the high-metallicity end. Recall that the two sets of stellar tracks differ in their allocation of highest metallicity. The Geneva tracks use a high metallicity of $Z = 0.040$, whereas the Padova tracks use a final metallicity of $Z = 0.050$. Therefore, a difference in the emission-line flux ratios at the high-metallicity end of the model grids is expected owing to the increased ionisation of stellar atmospheric metals and hence inhibition to ionise and excite the surrounding ISM \citep{Snijders2007}. The Geneva and Padova tracks also differ at the low-metallicity end ($Z = 0.001$ and 0.0004 respectively), yet the difference in the emission-line flux ratios between the two tracks is negligible. The reasoning for this is computational, rather than physical. The lowest metallicity available for use in both the Kurucz and Pauldrach atmospheres is $Z = 0.001$. Hence, SLUG proceeds with the calculation of the ionising spectrum using this value, setting the low-metallicity end of the Padova tracks equal to that of the Geneva tracks. At the common metallicities between the two sets of tracks, the emission-line flux ratios from the Padova tracks are higher than those for the Geneva tracks. This modest difference in emission-line flux ratio between the Padova and Geneva tracks is a result of the slighly increased temperature \citep[${\sim} 0.03$ dex;][]{VL2005} present in the Padova tracks. An increase in temperature enhances the collisional rate in the nebula, leading to an increase in the fluxes of collisionally excited lines such as [O \textsc{iii}] and [N \textsc{ii}]. Differences may be larger for instantaneous bursts in the W-R phase.

\subsubsection{Stellar Atmospheres}
\label{sec:atmsbpt}

The photoionisation model grids using each of the atmospheres we consider are shown in Figure~\ref{fig:threecompgrids}b. A slight increase in emission-line ratios is seen in the Kurucz + Hillier grid when compared to the Kurucz grid, following the increased W-R strength added by \citet{HM1998}. The largest difference between these two atmospheres seen in Figure~\ref{fig:threecompgrids}b is ${\sim} 0.1$ dex in both emission-line ratios, seen at higher metallicities. Since W-R stars are likely to be found in higher-metallicity environments \citep[and references therein]{Leitherer1999,Crowther2006,Crowther2007,Georgy2015}, the increase in the emission-line ratios primarily towards the higher-metallicity end of the grid is expected. However, a maximum increase in emission-line flux ratio of 0.1 dex is still small, despite a factor of $2 - 5$ increase in the strength of the flux emitted from W-R stars in the Hillier atmospheres. With ${\sim} 10$\% the amount of W-R stars as O stars in the cluster \citep{Leitherer1999}, the W-R flux increase added by \citet{HM1998} is noticeable, yet small nonetheless. Hence, changes in emission-line strength are not significant. 

The improvements to line blanketing made by \citet{Pauldrach2001} overall lead to a cooler nebula when the Pauldrach atmospheres are added to the Kurucz atmospheres; the ionic temperatures of O$^{2+}$ and N$^+$ decrease by ${\sim} 3$\% at a metallicity of $Z = 0.020$ and ${\sim} 6$\% at a metallicity of $Z = 0.040$. This decrease in temperature is small, yet it nevertheless results in a noticeable difference in the emission-line ratios in Figure~\ref{fig:threecompgrids}b. For almost all metallicities in the models (with the exception of $Z = 0.001$, where the effects of line blanketing are negligible), the difference in the [N \textsc{ii}]/H$\alpha$ line ratios between the Kurucz and Kurucz + Pauldrach atmospheres becomes larger, towards a maximum of ${\sim} 0.2$ dex. The lower temperature of the nebula results in a lower collisional excitation rate, leading to a decreased flux of [N \textsc{ii}] provided by the Kurucz + Pauldrach atmospheres. The [O \textsc{iii}]/H$\beta$ line ratios, however, tend to increase with metallicity when Pauldrach atmospheres are added, up to $Z = 0.040$, where they experience a sharp decrease (${\sim} 0.4$ dex). The spectra in Figure~\ref{fig:specplot} show that adding Pauldrach atmospheres results in a small increase in the luminosity of the spectrum at the ionisation potential of O$^{+}$, leading to an increased abundance of O$^{2+}$ ions in the nebula. Meanwhile, the spectra show no such feature at the ionisation potential of N$^0$, resulting in no significant increase in the abundance of N$^+$ ions in the nebula. Despite the cooler nebular temperature, the increase in the number of O$^{2+}$ ions provides a larger [O \textsc{iii}] flux and hence [O \textsc{iii}]/H$\beta$ ratio. At a metallicity of $Z = 0.040$, the effects of line blanketing become more prominent. When combined with the decrease in temperature already expected from an increase in the abundance of metals, the collisional excitation rate decreases further, resulting in lower [O \textsc{iii}]/H$\beta$ ratios than seen when using the Kurucz atmospheres.

The model produced using the combination of all three atmospheres very closely resembles the model produced using the Kurucz + Pauldrach atmospheres. With a W-R/O ratio of ${\sim} 10$\%, the emission-line strength increase from the Hillier atmospheres is only relevant for ${\sim} 10\%$ of the total stars at higher metallicities. Hence, the W-R atmospheric additions from the Hillier atmospheres are dwarfed by the OB atmospheric additions from the Pauldrach atmospheres, resulting in a small difference in emission-line ratios between the Kurucz + Pauldrach atmospheres and the Kurucz + Hillier + Pauldrach atmospheres.

\subsubsection{SPS Codes}
\label{sec:spscodesbpt}

\paragraph{Single stellar populations}

\begin{figure}
\centering
\includegraphics[width=1.1\columnwidth]{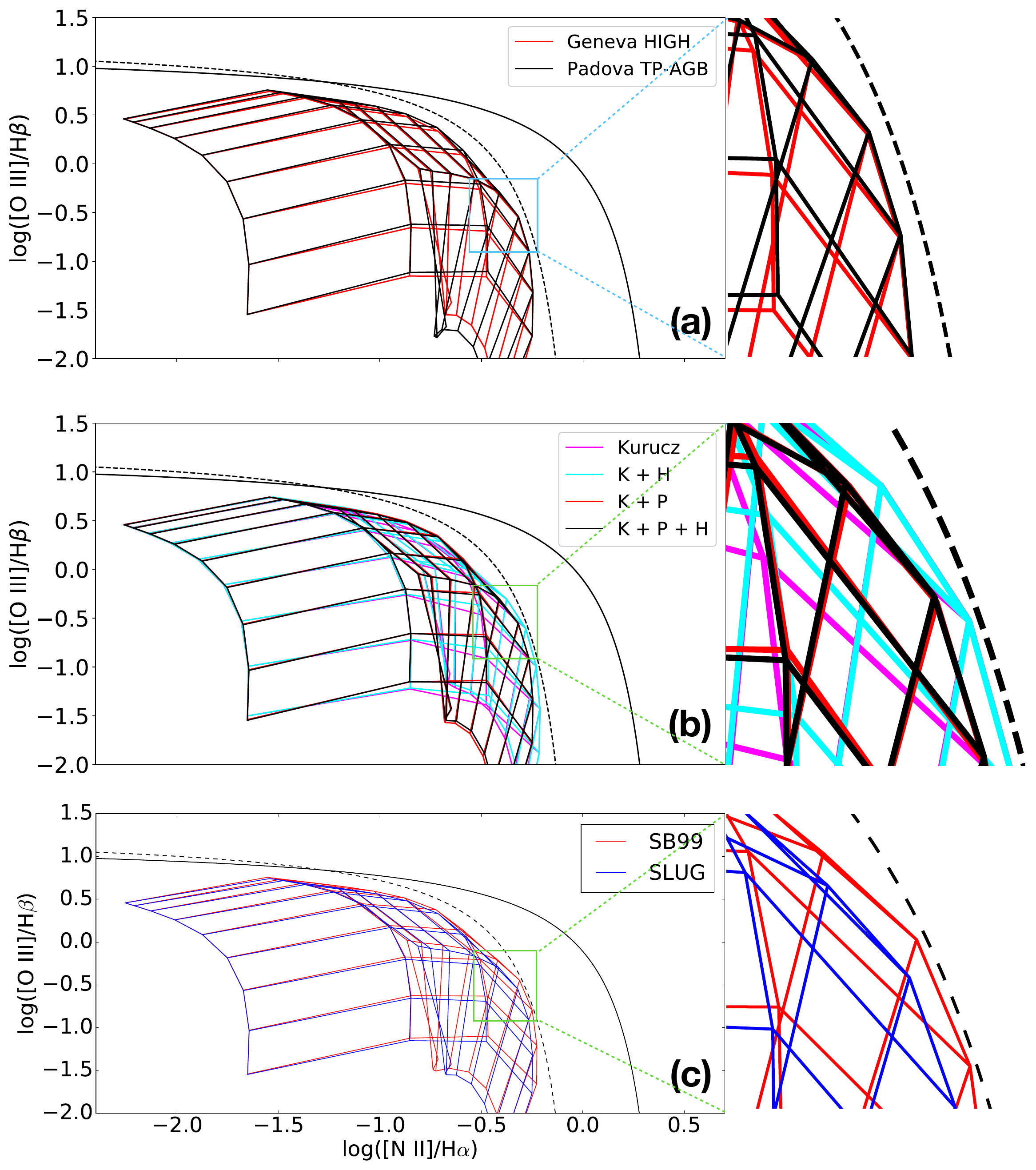}
\caption{(a) Model grids produced using both the Geneva HIGH tracks and the Padova TP-AGB tracks for a continuous SFH population at age 5 Myr. (b) Model grids produced using different stellar atmospheres for a continuous SFH population at 5 Myr. The atmospheres included in the models are Kurucz, Kurucz + Hillier (K + H), Kurucz + Pauldrach (K + P), and Kurucz + Hillier + Pauldrach (K + H + P). (c) Model grids produced using SLUG and SB99. All parameters associated with these grids are fiducial.}
\label{fig:threecompgrids}
\end{figure}

Shown in Figure~\ref{fig:threecompgrids}c are the two grids produced using SLUG and SB99. Overall the grids are similar, with a difference of ${\sim} 0.1$ dex at high metallicity in the [O \textsc{iii}]/H$\alpha$ and [N \textsc{ii}]/H$\beta$ line ratios, and negligible difference at low metallicity. A systematic offset is present in Figure~\ref{fig:threecompgrids}c, as a result of SB99 producing a slightly harder stellar spectrum for a continuous population at 5 Myr (Figure~\ref{fig:reldiff}). Fig.~\ref{fig:threecompgrids}c shows that SLUG and SB99 produce models that agree to within 0.1 dex on the BPT diagram. This offset is within the error range typically assumed through photoionisation modelling \citep[${\sim} 0.1$ dex; e.g.][]{Kewley2001}.

\subparagraph{Stochastic sampling of IMF}

\begin{figure}
\centering
\includegraphics[width=1.1\columnwidth]{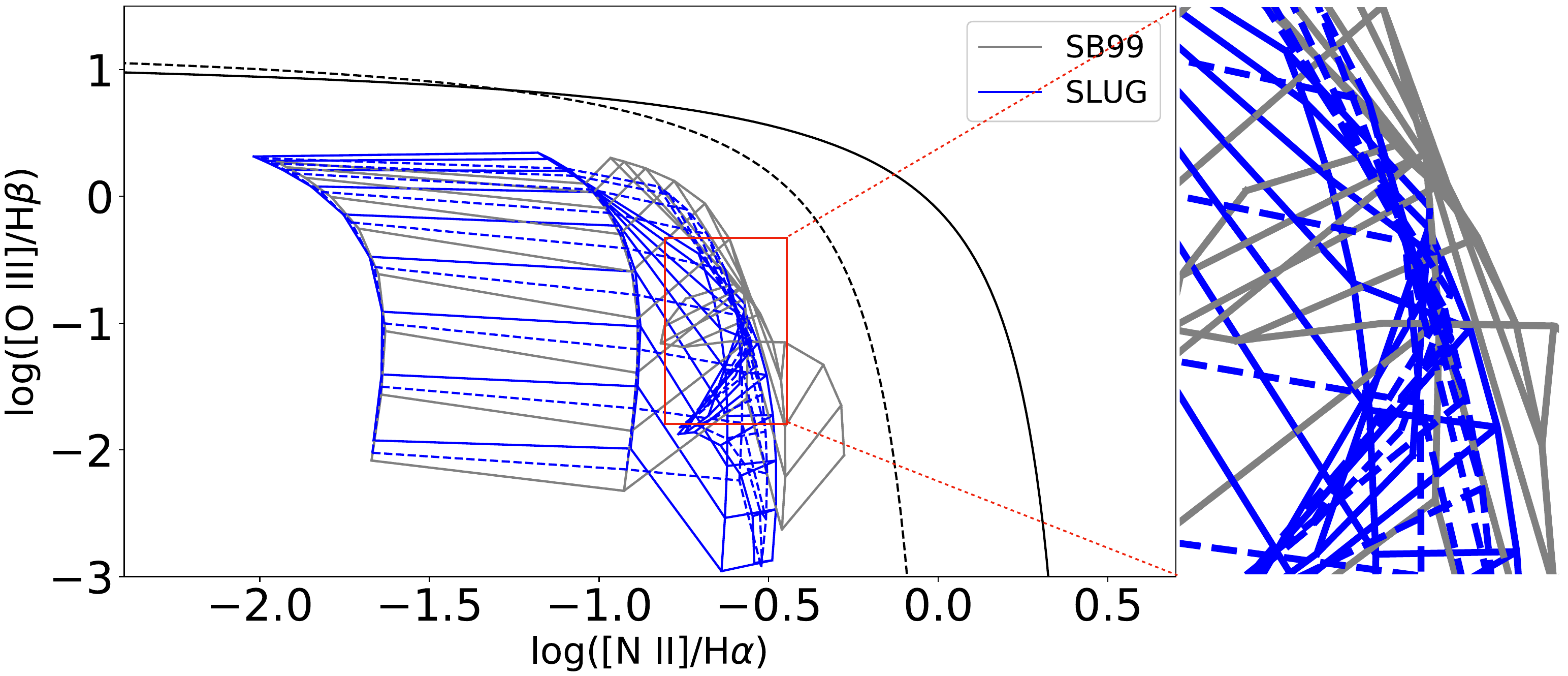}
\caption{Instantaneous SFH grids of ages 5 Myr produced by SLUG and SB99, for cluster masses of $10^4 M_\odot$ (solid) and $10^6 M_\odot$ (dashed). The stochastic sampling of the IMF demonstrated by SLUG produces spectra of different shapes for clusters at $10^4$ and $10^6 M_\odot$, leading to differences in the emission-line ratios. The lack of stochastic IMF sampling shown by SB99 leads to identically shaped ionising spectra for the $10^4 \;M_\odot$ and $10^6 \;M_\odot$ clusters, as shown in Figure~\ref{fig:stochasticspec}.}
\label{fig:stochasticbpt}
\end{figure}

The photoionisation grids for $10^4 M_\odot$ and $10^6 M_\odot$ clusters produced using SLUG and SB99 are shown in Figure~\ref{fig:stochasticbpt}. The varying shape of the stellar spectrum between clusters of masses $10^4 M_\odot$ and $10^6 M_\odot$ when produced stochastically with SLUG leads to differences in emission-line ratios and hence photoionisation grids of varying shape. The differences in the emission-line ratios for the grids of differing cluster mass produced using SLUG range from $< 0.1$ to ${\sim} 0.5$ dex. Conversely, the grids that use spectra produced using SB99 are identical. Figure~\ref{fig:stochasticbpt} shows that clusters of mass $10^4$ and $10^6 M_\odot$ synthesised using SB99 produce spectra that are identical in shape, differing only by a scale factor in luminosity. Luminosity is negligible in the creation of BPT photoionisation grids, because both BPT line ratios contain emission lines from the Balmer sequence. The flux of Balmer lines is directly proportional to the ionsing radiation emitted by the stellar cluster \citep[e.g.][and references therein]{Dopita2002}. Hence, calculating the emission-line ratios with lines from the Balmer series removes luminosity dependence on the BPT diagram. Thus, the $10^4 \;M_\odot$ and $10^6\;M_\odot$ grids have identical values of emission-line ratios for each grid point, leading to equal grids.

\paragraph{Binary populations}

\begin{figure}
\centering
\includegraphics[width=1.05\columnwidth]{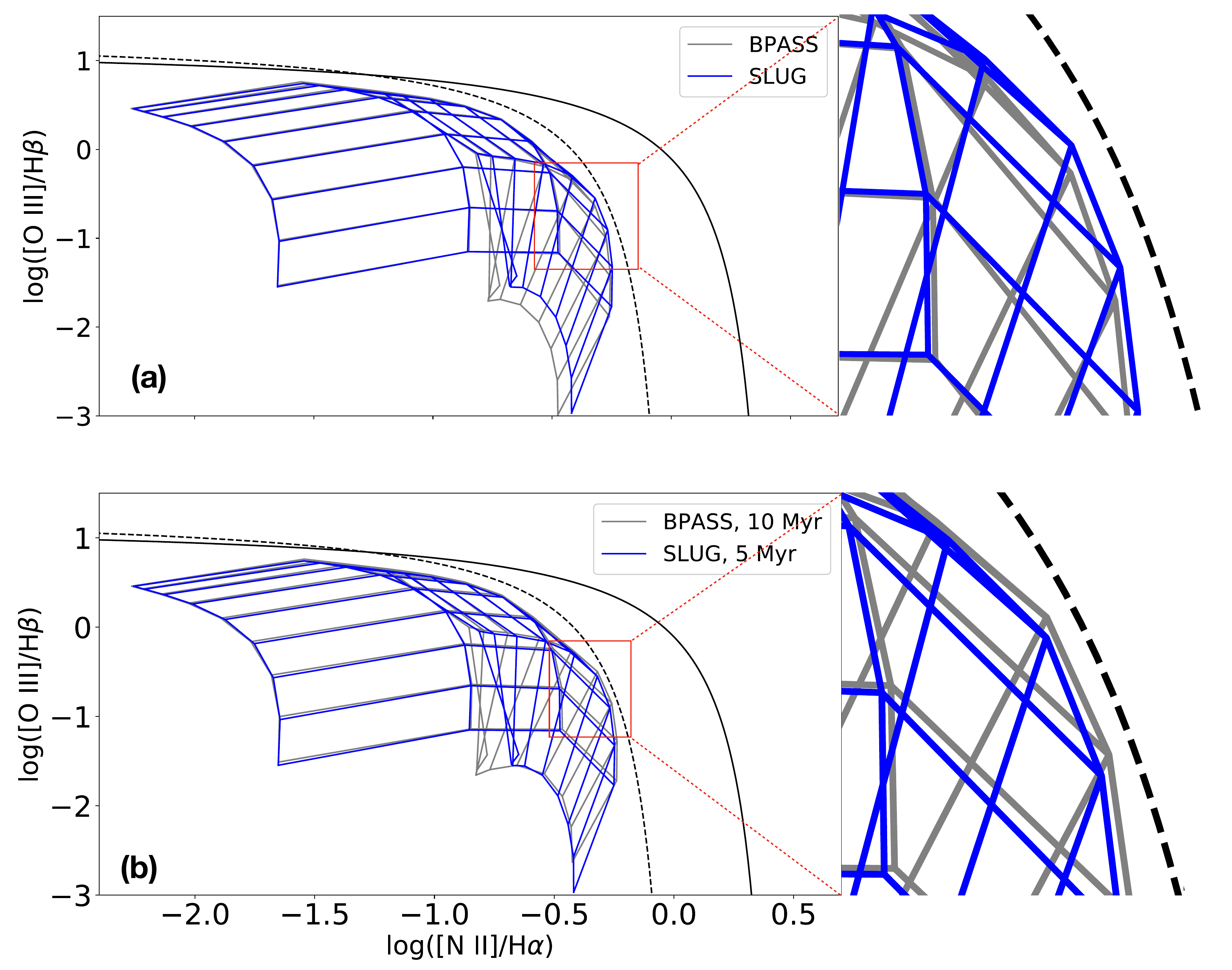}
\caption{(a) Model grids produced using BPASS and SB99, both assuming a continuous SFH at an age of 5 Myr. (b) Model grids produced using BPASS and SB99, using SLUG and BPASS continuous SFH input spectra at ages for which the respective stellar cluster reaches equilibrium. All remaining parameters associated with these grids are fiducial.}
\label{fig:bpassslugcomp}
\end{figure}

Shown in Figure~\ref{fig:bpassslugcomp}a are model grids produced using input spectra from BPASS and SLUG, using the remaining fiducial parameter set. Overall, the differences in emission-line ratios for a 5 Myr old continuous SFH cluster are small between BPASS and SLUG, with the largest differences occurring at high metallicity. The largest difference in the emission-line ratios between the two grids is ${\sim} 0.1$ dex in the [N \textsc{ii}]/H$\alpha$ ratio and ${\sim} 0.2$ dex in the [O \textsc{iii}]/H$\beta$ ratio, with SLUG producing the larger emission-line ratios. Figure~\ref{fig:bpassslugcomp}a shows that, in general, at an age of 5 Myr for both single-star and binary clusters, the spectrum produced with SLUG produces emission-line ratios that are slightly greater than those produced by BPASS. However, at an age of 5 Myr, the binary cluster has yet to reach equilibrium. As shown in Section~\ref{sec:binarypop1}, the age at which the continuous SFH binary cluster simulated using BPASS reaches equilibrium is at ${\sim} 10$ Myr. Figure~\ref{fig:bpassslugcomp}b shows model grids produced using input spectra from SLUG and BPASS, for ages at which both the single-star and binary clusters are at equilibrium (5 Myr for SLUG, 10 Myr for BPASS). At these ages, the emission-line ratios from the BPASS model are consistently higher (albeit by small amounts; less than 0.1 dex in emission-line ratio) than those from the SLUG model across all metallicities covered in the models. This is consistent with the findings of \citet{Stanway2016}, concerning a boosted ionising photon flux in binary populations when compared to single-star populations. A harder ionising radiation field leads to a higher average nebular temperature and hence an increase in the collisional rate of ions, increasing the flux of collisionally excited lines such as [O \textsc{iii}] and [N \textsc{ii}].

\subsubsection{Pressure and Density}

We vary the $P/k$ value by varying the initial electron density of the H \textsc{ii} region, assuming an initial temperature of 8000K. This was partially explored by \citet{Nicholls2014c}, where they show that a higher $P/k$ value increases the strength of certain emission-line ratios. Here we explore this further, sampling a pressure ranging from $P/k$ values of $8 \times 10^3\; \mathrm{cm}^{-3}\;\mathrm{K}$ to $8 \times 10^7\; \mathrm{cm}^{-3}\;\mathrm{K}$ in steps of 1 dex, corresponding to initial densities of $n = 1, 10, 100, 1,000$, and $10,000 \;\mathrm{cm}^{-3}$ (assuming an inner initial temperature of $T_{e}= 8000 \;\mathrm{K}$). It should be noted that a density of $n = 10,000\;\mathrm{cm}^{-3}$ is far higher than the densities observed in local galaxies (see Section~\ref{sec:pandd}), and hence the models produced assuming a density of $n = 10,000\;\mathrm{cm}^{-3}$ do not apply to the SDSS sample.

\begin{figure}
\centering
\includegraphics[width=1.1\columnwidth]{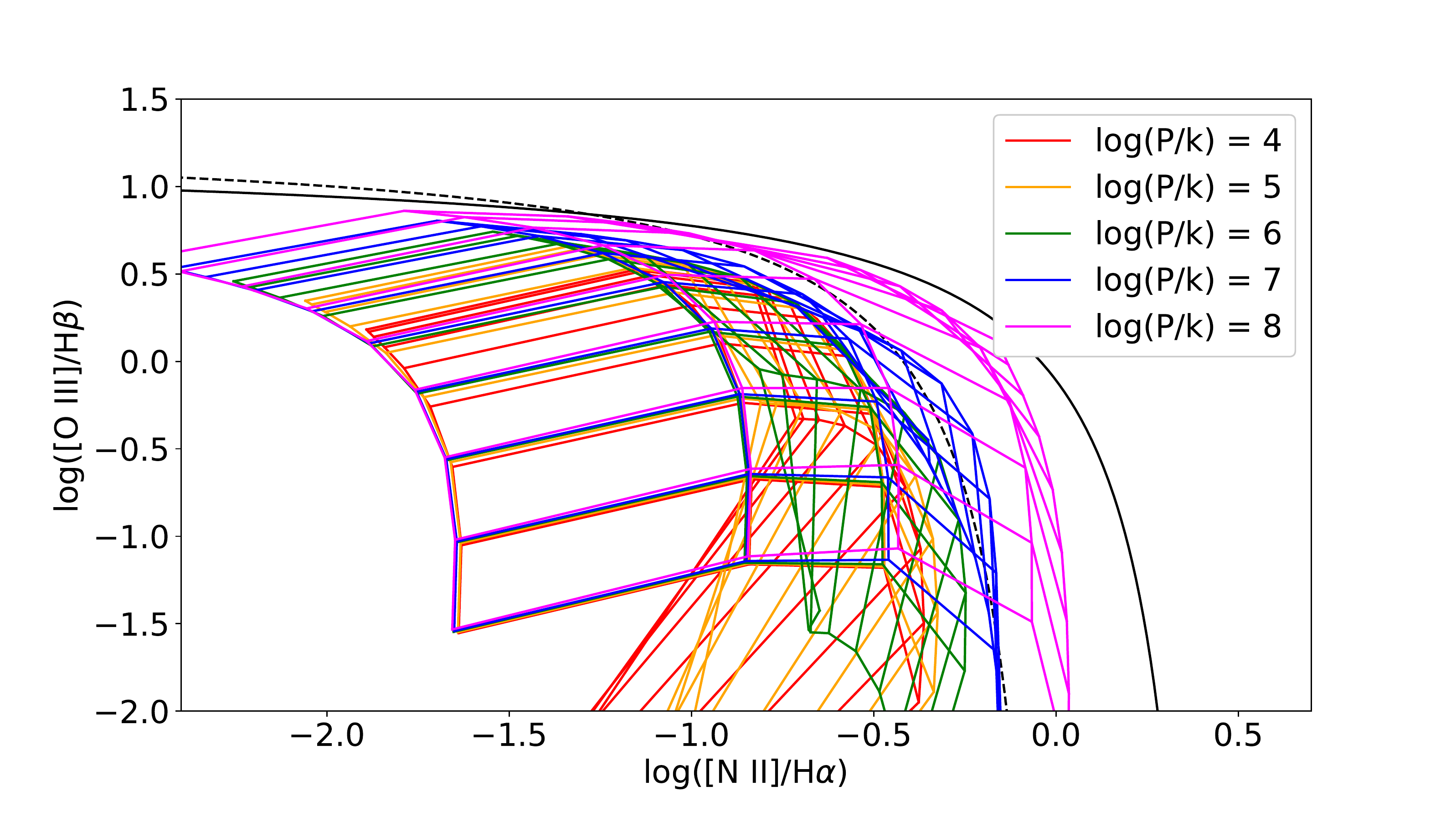}
\caption{Model grids of varying pressure for a continuous SFH population at age 5 Myr, from $P/k = 8 \times 10^{3}$ (log[($P/k$) ${\sim}$ 4) to $P/k = 8 \times 10^{7}\;\mathrm{cm}^{-3}\;\mathrm{K}$ (log($P/k$) ${\sim}$ 8), in increments of 1 dex.}
\label{fig:densall}
\end{figure}

Seen in Figure~\ref{fig:densall}, the BPT line ratios increase with pressure in general. The increase in line ratio with pressure is more noticeable at high metallicity. At high metallicity, the nebula is more susceptible to the effects of cooling, due to an increased abundance of coolants. At very low pressures (log($P/k) {\sim} 5$ and below), the density in the nebula is still below the critical density of several strong cooling lines (namely, [C \textsc{ii}] 157.7$\mu$m, [N \textsc{ii}] 205.3$\mu$m, and [O \textsc{iii}] 88.4$\mu$m; e.g. \citealt{Abdullah2017}), allowing cooling to occur, which decreases the temperature at high metallicity. Hence, the strength of collisionally excited lines such as [N \textsc{ii}] and [O \textsc{iii}] is weakened. An increase in pressure, however, begins to suppress the fine-structure far-infrared cooling lines through collisional de-excitation, which increases the temperature. Our models show the [C \textsc{ii}] 157.7$\mu$m and [N \textsc{ii}] 205.3$\mu$m lines to be the dominant cooling lines affected, with both decreasing in flux by a factor of 100 from a pressure of $8 \times 10^3 \;\mathrm{cm}^{-3}\;\mathrm{K}$ to $8 \times 10^7 \;\mathrm{cm}^{-3}\;\mathrm{K}$. The rise in temperature then increases the strength of collisionally-excited lines, leading to increased line ratios on the BPT diagram. Due to the low abundance of coolants at low metallicity, this effect is still present, yet much weaker.

\begin{figure}
\centering
\includegraphics[width=1.1\columnwidth,height=7cm]{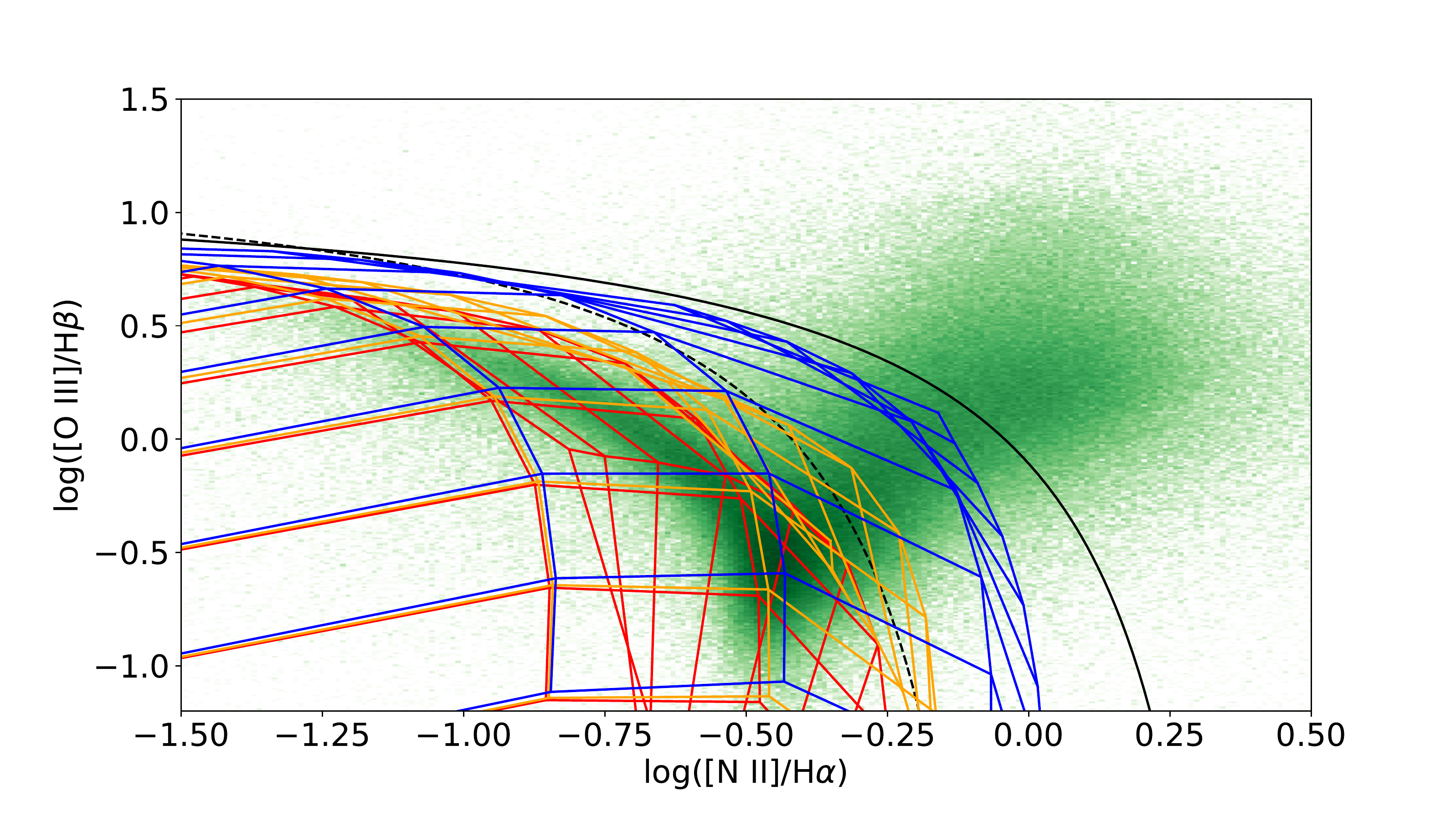}
\caption{Grids of the three highest pressures (red $\Rightarrow P/k = 8 \times 10^5$, orange $\Rightarrow P/k = 8 \times 10^6$, blue $\Rightarrow P/k = 8 \times 10^7$) from Fig.~\ref{fig:densall} plotted with the SDSS DR7 sample. The models show that the SDSS star-forming galaxies are well described by models with pressure ${\sim} 8 \times 10^5\;\mathrm{cm}^{-3}\;\mathrm{K}$. All parameters are fiducial.}
\label{fig:diffpres}
\end{figure}

Shown in Figure~\ref{fig:diffpres} are the three models of the highest pressure taken from Figure~\ref{fig:densall}, superimposed on the SDSS data. Figure~\ref{fig:diffpres} shows that the star-forming datapoints in the SDSS sample are well described by models with log($P/k$) ${\sim} 6$. 

\subsubsection{Boundedness}
\label{sec:boundbpt}

\begin{figure}
\centering
\includegraphics[height=5cm,width=1.1\columnwidth]{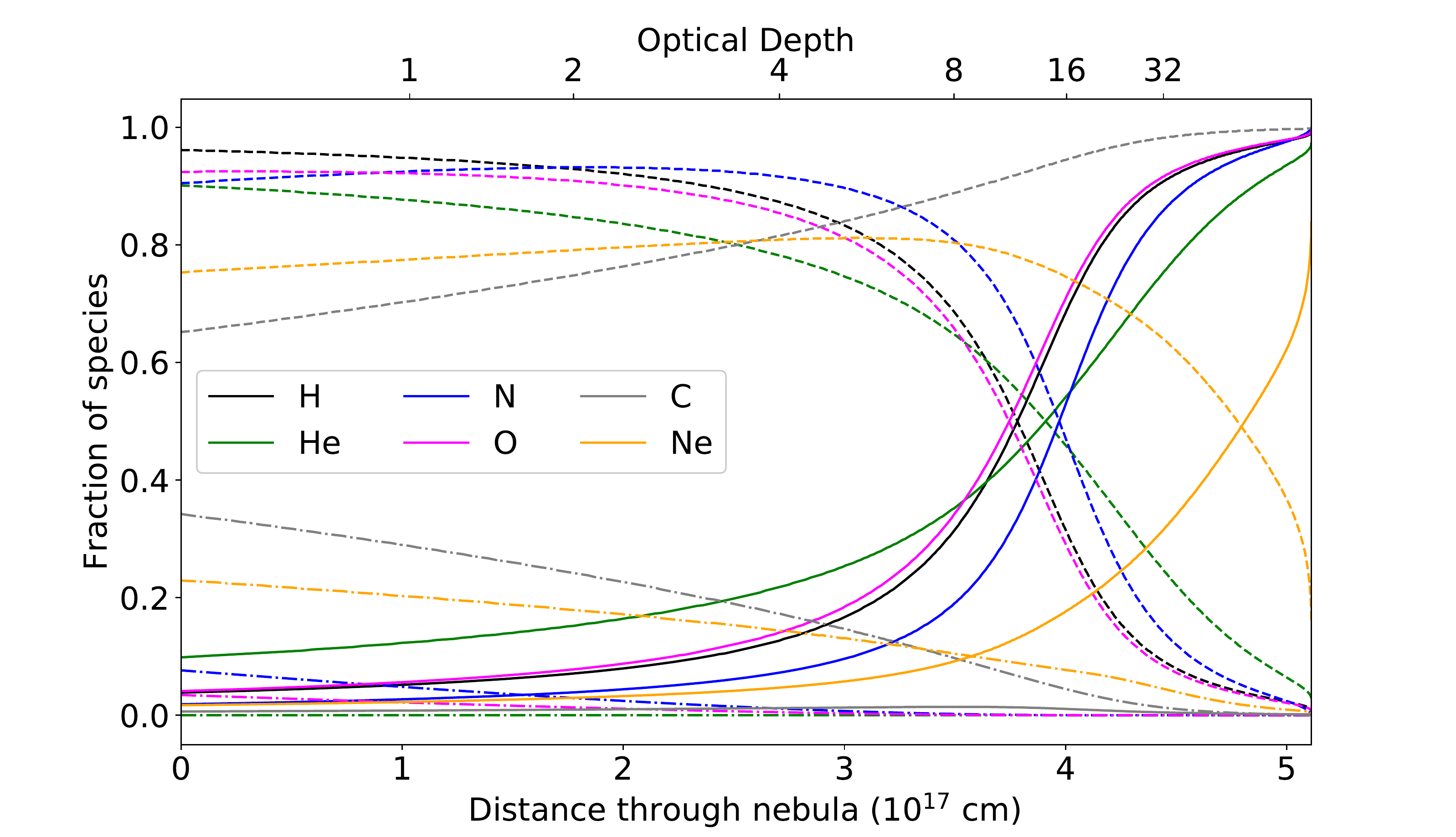}
\caption{Ion fractions of species within the H \textsc{ii} region as a function of distance from the ionising source. The corresponding optical depth is also shown. Solid lines show neutral (\textsc{i}) species, dashed lines show singly ionised (\textsc{ii}) species, and dot-dashed lines show doubly ionised (\textsc{iii}) species. Parameters for this plot are $\mathrm{log}(q) = 6.5$ and $Z = 0.004$. This particular model has been truncated with 1\% of H \textsc{ii} remaining. All other parameters are fiducial.}
\label{fig:allion99_650_Z020}
\end{figure}

\begin{figure}
\centering
\includegraphics[width=1.05\columnwidth]{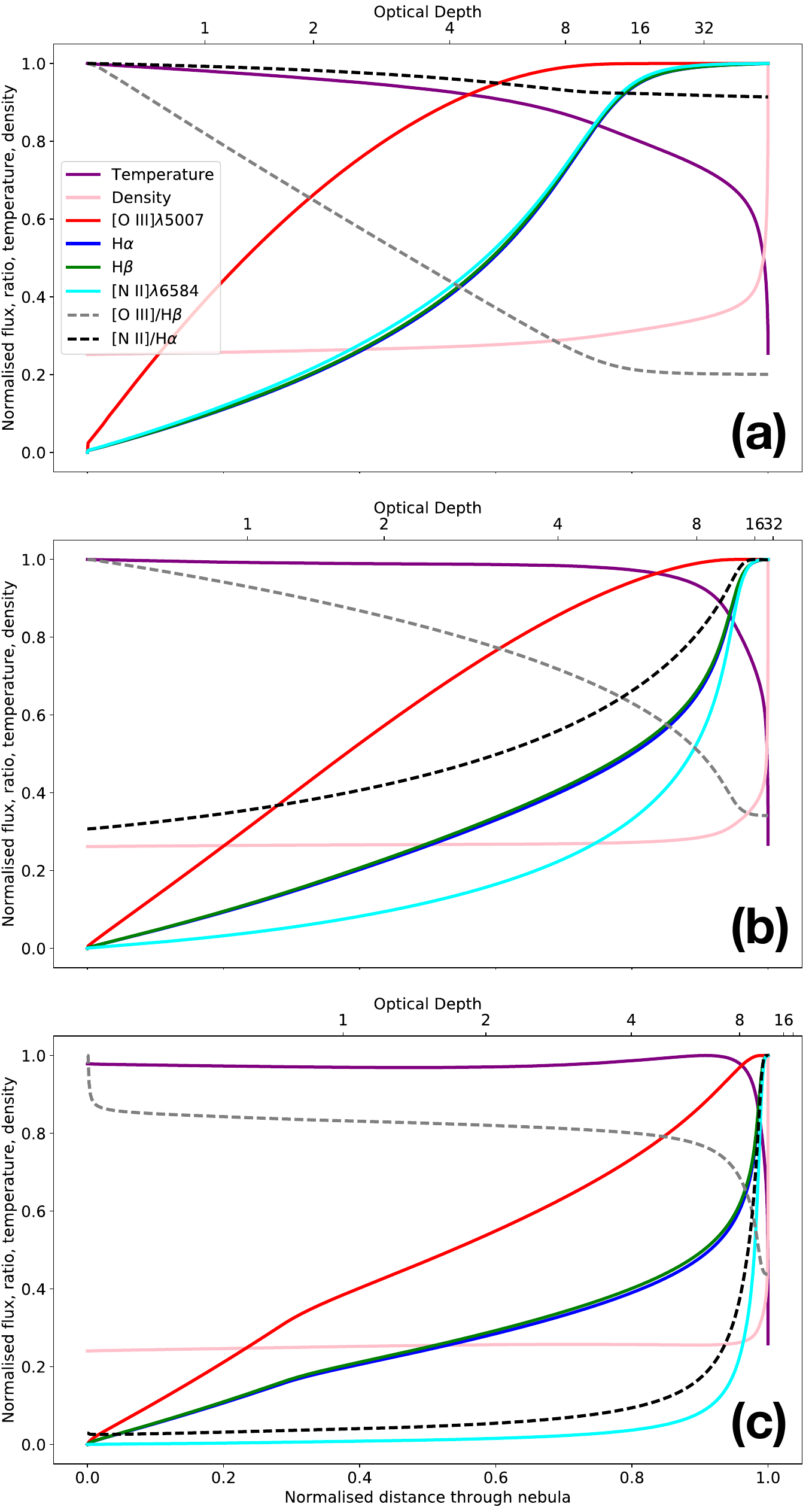}
\caption{Cumulative growth of the BPT emission lines and emission-line ratios as a function of distance through the nebula, normalised to their maximum value, for $\mathrm{log}(q) = 6.5, 7.5$, and $8.5$ in panels (a), (b), and (c), respectively. The normalised temperature and density distribution throughout the nebula are also shown. The metallicity for all plots is set to $Z = 0.004$. All other parameters are identical to Figure~\ref{fig:allion99_650_Z020}.}
\label{fig:flux_optdepth_all3}
\end{figure}

We explore the changes in emission-line ratios as the models progress from radiation-bounded situations to density-bounded ones. To do this, we study the evolution of the models as a function of optical depth, as well as the evolution as a function of hydrogen recombination percentage at the point of truncation. Both sets of models can be explained through the same process. Shown in Fig.~\ref{fig:allion99_650_Z020} are the relative ionic fractions of individual species as a function of distance through the H \textsc{ii} region for $\mathrm{log}(q) = 6.5$ and $Z = 0.004$. Figure~\ref{fig:allion99_650_Z020} shows that truncating the model at earlier values of hydrogen recombination means truncating at a smaller distance into the nebula, which corresponds to a lower optical depth within the H \textsc{ii} region. The models that vary with hydrogen recombination percentage allow us to study the differences along the ionisation front in the H \textsc{ii} region much more finely than we can with the optical depth models.

\begin{figure}
\centering
\includegraphics[width=1.05\columnwidth]{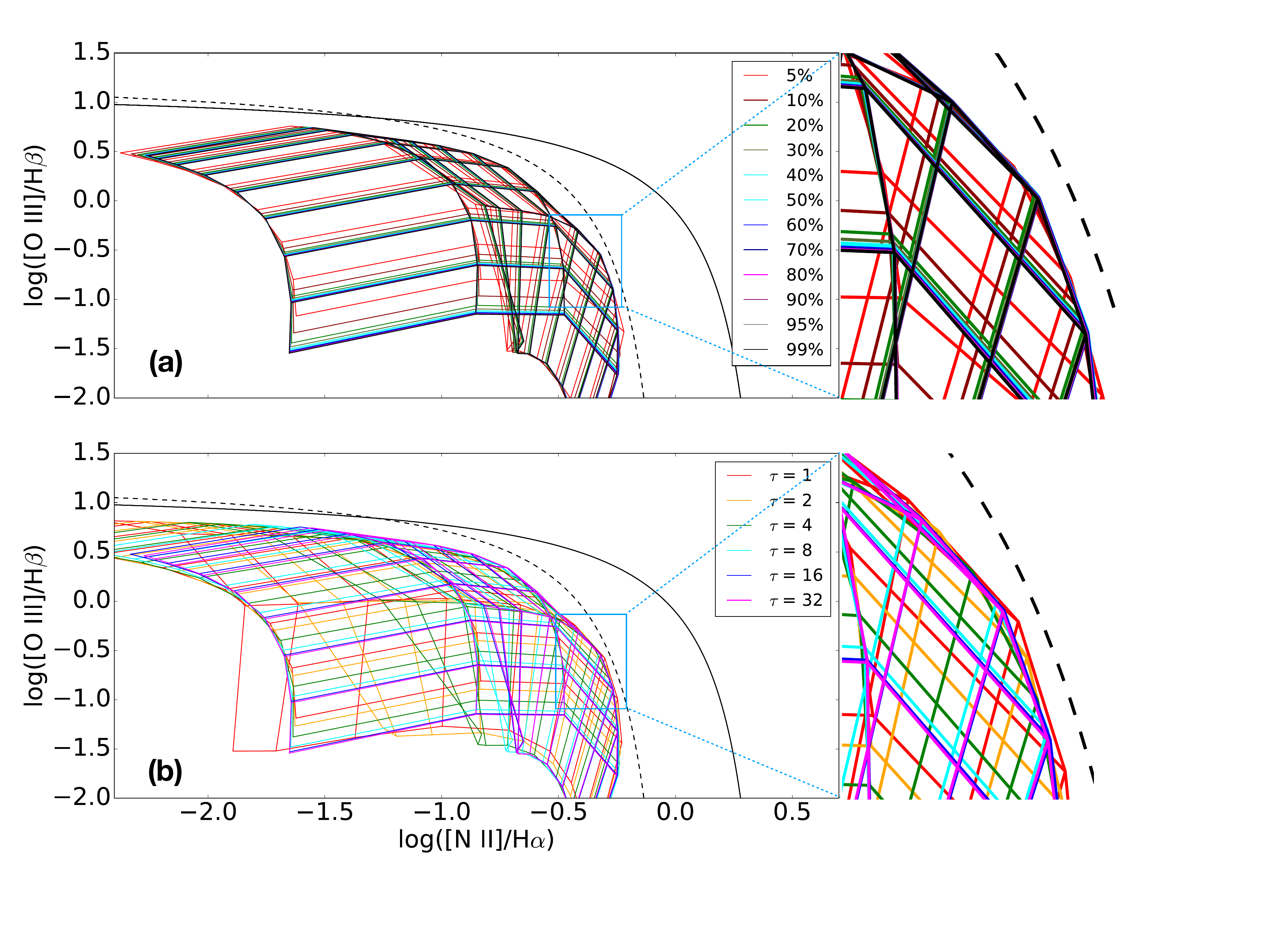}
\caption{(a) Model grids computed with varying H \textsc{ii} ionisation bounds, given as percentages. The percentage given is the fraction of H \textsc{ii} that has undergone recombination to H \textsc{i}. (b) Model grids computed with varying optical depths at the threshold of hydrogen. The optical depth values used are identical to those used by \citet{Nicholls2014c}.}
\label{fig:boundcomp}
\end{figure}

Figure~\ref{fig:flux_optdepth_all3} shows the cumulative growth of the BPT emission lines as a function of distance through the nebula, normalised to their maximum value. It also shows the resulting emission-line ratios (also normalised to their maximum value) if the nebula was to be truncated at a given radius, as in the case of density-bounded H \textsc{ii} regions. The trends seen in the variation of emission-line ratios with optical depth are reflected in the shape and position of the model grids, shown in Figures~\ref{fig:boundcomp}(a) and (b) for hydrogen recombination percentage and optical depth variation, respectively. The optical depth grids in Figure~\ref{fig:boundcomp}(b) use the same values of $\tau$ as \citet{Nicholls2014c}, although differences between this work and that presented in \citet{Nicholls2014c} include the $P/k$ value, input spectrum, and version of \textsc{mappings} used \citep[\citealt{Nicholls2014c} use the \textsc{mappings iv} photoionisation code; see][for details]{Dopita2013,Nicholls2014c}. In general, Figure~\ref{fig:flux_optdepth_all3} shows that the [O \textsc{iii}]/H$\beta$ ratio continues to increase as the optical depth and hydrogen recombination percentage lower. This trend is independent of the value of ionisation parameter, although it is far more noticeable at lower values. The [N \textsc{ii}]/H$\alpha$ ratio, however, is seen to either increase or decrease at lower values of optical depth and hydrogen recombination, with a strong dependence on the ionisation parameter. Lower values of the ionisation parameter ($\mathrm{log}(q/(\mathrm{cm}\;\mathrm{s}^{-1})) \sim 6.5$) lead to an increase in the [N \textsc{ii}]/H$\alpha$ ratio with decreasing optical depth and hydrogen recombination percentage, whilst at higher values of the ionisation parameter ($\mathrm{log}(q/(\mathrm{cm}\;\mathrm{s}^{-1})) \gtrsim 6.75$), the opposite is seen, with the [N \textsc{ii}]/H$\alpha$ ratio decreasing with lowering optical depth.

\begin{figure}
\centering
\includegraphics[width=1.1\columnwidth]{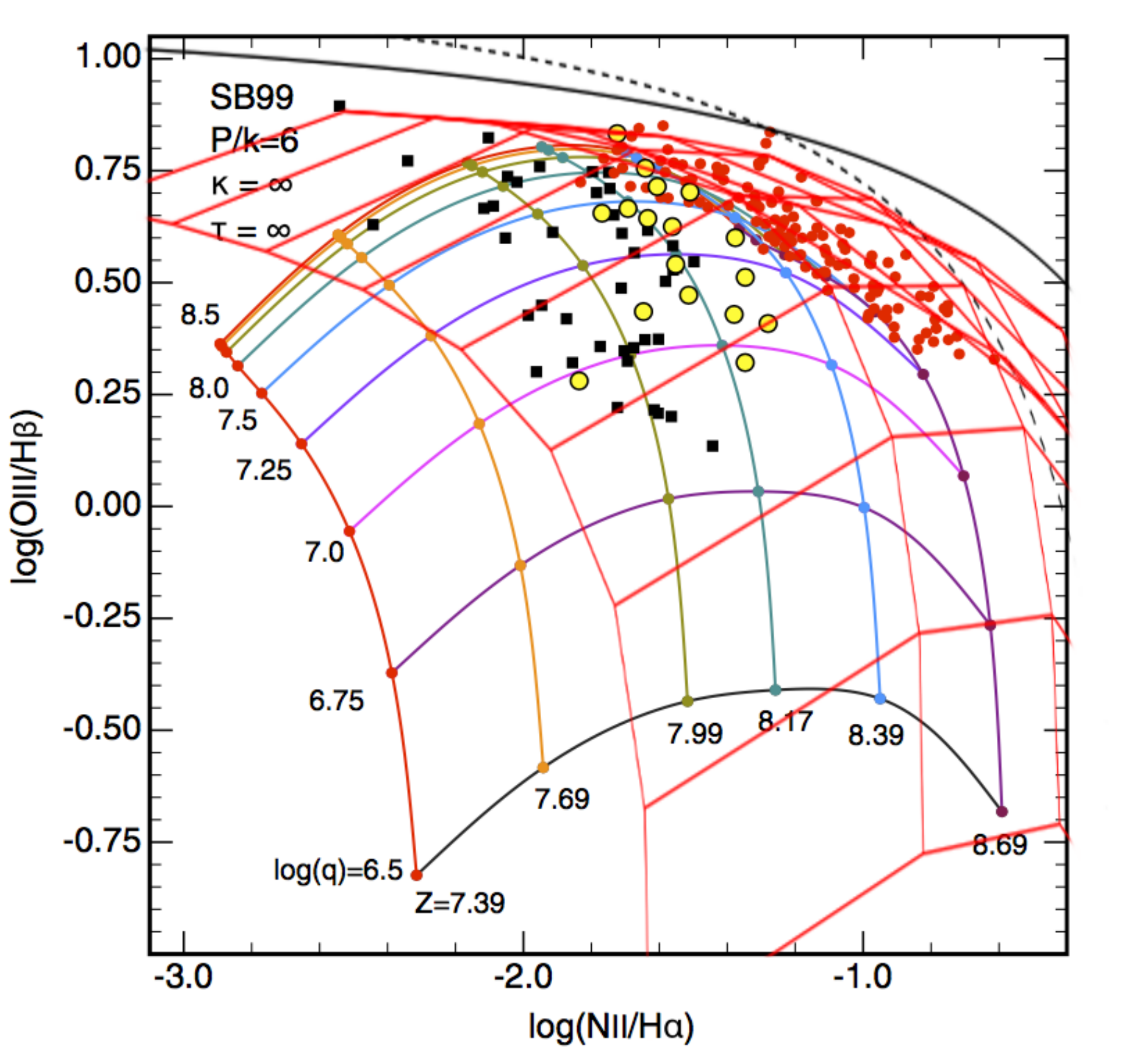}
\caption{Density-bounded ($\tau = 1$) grid (red) with pressure $P/k {\sim} 8 \times 10^7\;\mathrm{cm}^{-3}\;\mathrm{K}$, corresponding to an initial electron density $n = 10,000 \;\mathrm{cm}^{-3}$ under the assumption of an initial electron temperature of 8000K, shown over the top of the SIGRID sample and grid taken from \citet{Nicholls2014c} (multicoloured). The yellow points show the SIGRID sample from \citet{Nicholls2014c}, the black points show the low-metallicity DR7 SDSS subset from \citet{Izotov2012}, and the red points show the DR3 SDSS subset from \citet{Izotov2006}. Metallicity values $Z$ are in units of $12 + \mathrm{log(O/H)}$. Ionisation parameter values are consistent with those used in this paper. All other parameters are fiducial. The density-bounded model better fits the data and helps explain the off-grid data points in \citet{Nicholls2014c}. The very high value of log$(P/k) \sim 8$ is likely to be unphysical and arises as a result of an X-ray deficit in the model ionising spectra.}
\label{fig:nichollsgrid}
\end{figure}

\begin{figure}
\centering
\includegraphics[width=1.05\columnwidth]{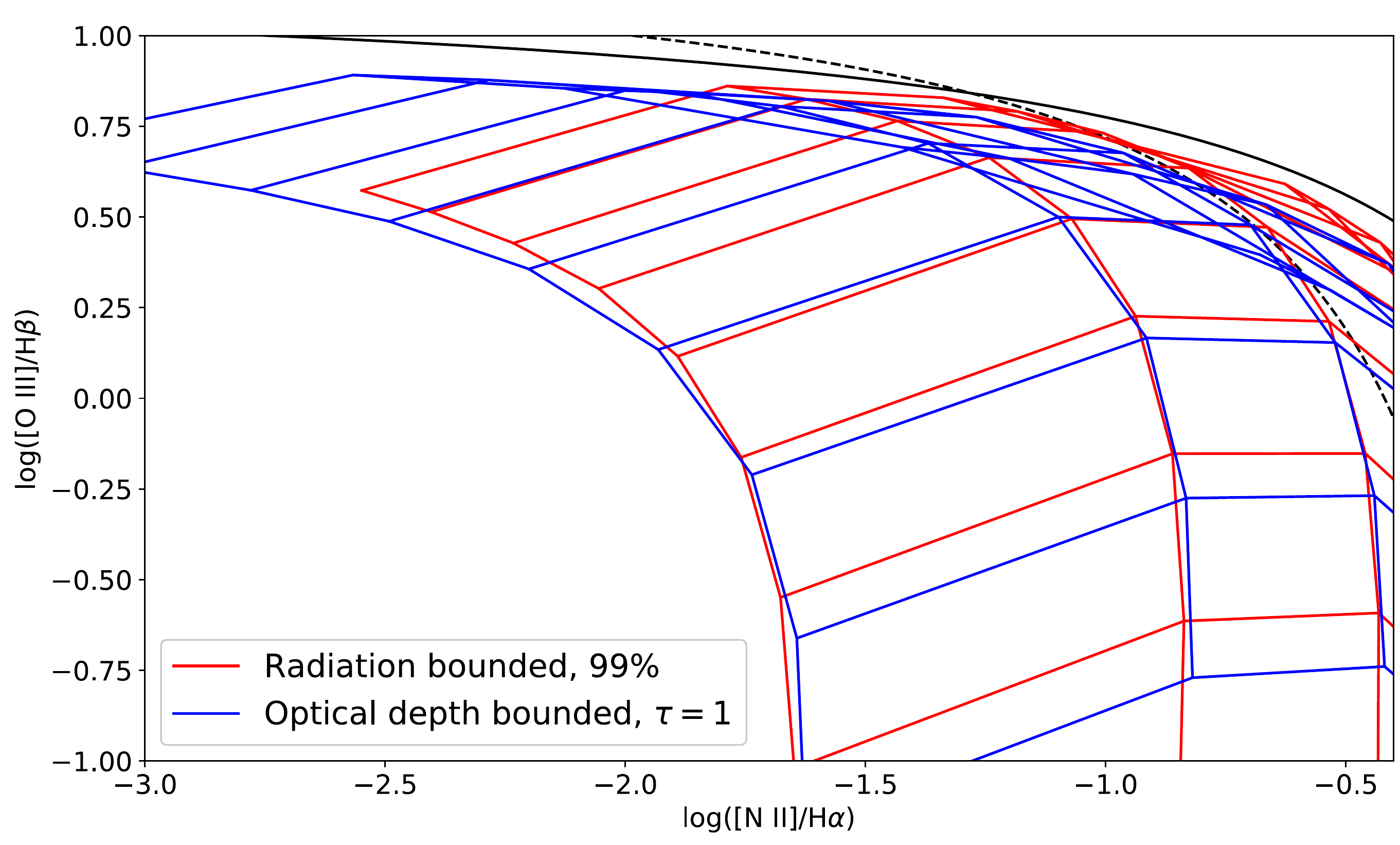}
\caption{Isobaric models with pressure values of $P/k = 8 \times 10^7\;\mathrm{cm}^{-3}\;\mathrm{K}$, assuming a density-bounded regime for the blue grid and radiation-bounded regime for the red grid. The density-bounded model is truncated at an optical depth of $\tau = 1$, and the radiation-bounded model is truncated at 99\% of hydrogen recombination. The very high value of log$(P/k) \sim 8$ is likely to be unphysical and arises as a result of an X-ray deficit in the model ionising spectra.}
\label{fig:radvsoptdepth}
\end{figure}

Upon failure to reproduce the emission-line ratios for all data points in their combined sample, \citet{Nicholls2014c} suggest that the off-grid points are the result of H \textsc{ii} regions that are optically thin. This is supported by the fact that the majority of the off-grid data points are from the SDSS data release 7 (DR7) subsample from \citet{Izotov2012}, which focused on selecting low-metallicity galaxies (12 + log(O/H) between ${\sim} 7.1$ and ${\sim} 7.9$). As mentioned previously, low-metallicity galaxies tend to have a low stellar mass \citep{Tremonti2004}, which corresponds to a higher photon escape fraction from the impact of supernovae \citep{Trebitsch2017}, leading to a lower optical depth. Density-bounded models may provide an explanation for the off-grid data points seen in the SIGRID sample from \citet{Nicholls2014c} and in the low-metallicity sample of SDSS from \citet{Izotov2012}. Seen in Figure~\ref{fig:nichollsgrid} is the SIGRID sample from \citet{Nicholls2014c} (yellow points), the low-metallicity DR7 SDSS subset from \citet{Izotov2012} (black points), and a further DR3 SDSS subset from \citet{Izotov2006} (red points). Also shown in Figure~\ref{fig:nichollsgrid} is a radiation-bounded photoionisation model from \citet{Nicholls2014c} along with our density-bounded photoionisation model, truncated at an optical depth of $\tau = 1$ and with a pressure of $P/k = 8 \times 10^7 \;\mathrm{cm}^{-3}\;\mathrm{K}$. Figure~\ref{fig:nichollsgrid} shows that the density-bounded model describes all of the data points in the \citet{Nicholls2014c} SIGRID sample, \citet{Izotov2006} sample, and \citet{Izotov2012} sample simultaneously, suggesting that the metal-poor galaxies in both samples contain optically thin H \textsc{ii} regions. A density-bounded regime is necessary to encompass all data points from the \citet{Nicholls2014c} combined sample. We show this in Figure~\ref{fig:radvsoptdepth}, by comparing the model truncated at $\tau = 1$ from Figure~\ref{fig:nichollsgrid} with a model truncated at 99\% of hydrogen recombination. The density-bounded model shows both decreases in the [N \textsc{ii}]/H$\alpha$ ratio and increases in the [O \textsc{iii}]/H$\beta$ ratio, sufficient to explain the low-metallicity data points. A pressure of $P/k = 8 \times 10^7 \;\mathrm{cm}^{-3}\;\mathrm{K}$ (log($P/k$) ${\sim} 8$) is necessary to completely encompass all data points seen in Figure~\ref{fig:nichollsgrid}, which is a likely unphysical value for the pressure. D. C. Nicholls et al. (in preparation) note a significant X-ray deficit currently exists in H \textsc{ii} region modelling spectra, which prevents current H \textsc{ii} region models from producing high enough emission-line ratios needed to describe all possible data points -- in particular those from the \citet{Nicholls2014c} combined sample. D. C. Nicholls et al. (in preparation) aim to resolve this issue by including an X-ray excess in the model spectra. 

\subsubsection{Comparing MAPPINGS and CLOUDY}
\label{sec:mvc}

We provide a concentrated comparison between the latest models of the two photoionisation modelling codes, \textsc{mappings} and \textsc{cloudy} \citep{Ferland1996} version 13.03, described in \citet{Ferland2013}. Small-scale comparisons between the two codes have been performed in the past. \citet{Byler2017} compare a constant SFH model grid produced using \textsc{cloudy} with the \textsc{mappings iv} model shown in \citet{Dopita2013}. Both models are matched in ionisation parameter and metallicity (except for the highest metallicity in both grids), as well as gas-phase abundance. The two models show agreement in overall coverage of the star-forming region of the BPT diagram, yet there is significant disagreement between points of equal metallicity and ionisation parameter between the models. \citet{Byler2017} suggest that this is due to the difference in metallicity between the grids at the high-metallicity end, which has an effect on the gas-phase abundance. The \textsc{cloudy} model shown in \citet{Byler2017} ties the gas-phase abundance with the metallicity of the stellar population. Such a large disparity between the two models at the high-metallicity end will ultimately lead to differences in the synthesised stellar population. Differences in the synthesised stellar population between the two models may also arise as a result of the different SPS codes used. The input spectrum for the \textsc{cloudy} model grid is synthesised using the SPS code Flexible Stellar Population Synthesis \citep[FSPS;][]{Conroy2009}, whereas \citet{Dopita2013} use Starburst99 to synthesise their stellar population. As discussed and shown in Sections~\ref{sec:spscodes} and later Section~\ref{sec:spscodesbpt} when comparing the single-star SPS codes SLUG and Starburst99, different SPS codes may lead to variations in the input stellar spectrum, ultimately leading to differences in the emission-line ratios.

\begin{figure}
\centering
\includegraphics[width=1.1\columnwidth]{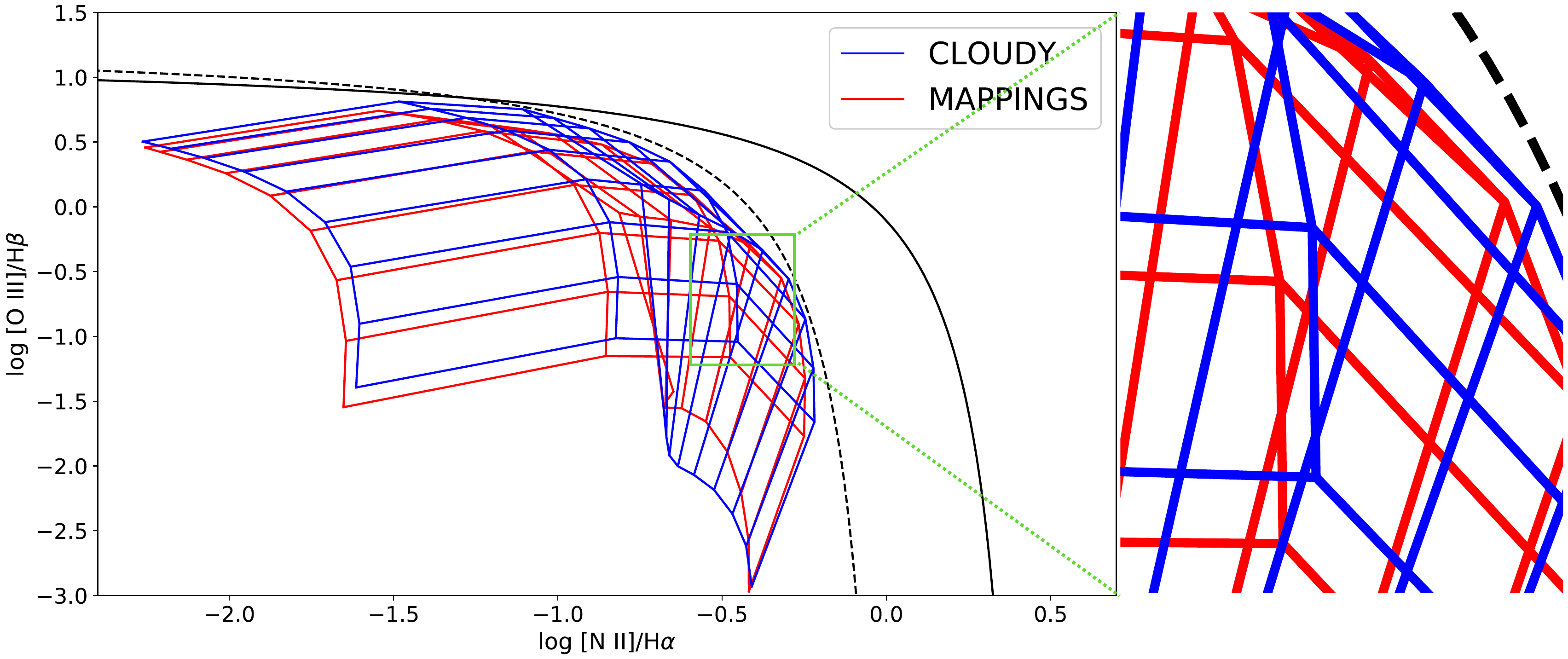}
\caption{Model grids computed using the photoionisation modelling codes \textsc{mappings} and \textsc{cloudy}. All parameters associated with these grids are fiducial.}
\label{fig:mvsc}
\end{figure}

A BPT diagram showing photoionisation grids produced using \textsc{mappings} and \textsc{cloudy} is shown in Figure~\ref{fig:mvsc}. Both the \textsc{mappings} and \textsc{cloudy} models are run with the exact same fiducial inputs, which includes the input ionising spectrum, ionisation parameters, abundances, depletion factors, and pressure. Hence, the differences seen in Figure~\ref{fig:mvsc} are the result of intrinsic differences between the two codes, such as atomic datasets, input physics, and model assumptions. Overall, the difference in emission -ine ratios between the two models is ${\sim} 0.1$ dex on average across all metallicities explored.

\section{Discussion}
\label{sec:disc}
\subsection{Spread of SDSS galaxies}

\subsubsection{Ionising Radiation Field Parameters}

The systematic differences in each of the ionising radiation field parameter variations are small and hence individually do not better explain or cover the spread in the star-forming SDSS galaxies. The average systematic difference in the emission-line ratios between varying model parameters is ${\sim} 0.1$ dex, ranging from $< 0.1$ dex to a maximum of ${\sim} 0.5$ dex more notably in the [O \textsc{iii}]/H$\beta$ ratio at high metallicities in the model grids. As demonstrated by the relative variation in the ionising spectra (Figures~\ref{fig:instspec} - \ref{fig:bpassspectra}), the impacts of stellar atmospheres, stellar tracks, and SPS codes on the ionising radiation spectra are much more visible at higher energies. This indicates that these variations will have a more significant impact on higher-ionisation species and temperature-sensitive emission lines than explored here. Even so, the impact of the uncertainties in modelling ionising spectra of clusters is visible in the BPT. Therefore, we recommend that a ${\sim} 0.1$ dex uncertainty be included when comparing model grids to data and that the most sophisticated model spectra be used when possible. We believe that the current best models include: 

\begin{itemize}
\item The stellar atmosphere combination of Kurucz + Hillier + Pauldrach, which takes the Kurucz atmospheres compiled by \citet{Lejeune1997} and includes updates on the W-R and OB stars' atmospheres from \citet{HM1998} and \citet{Pauldrach2001} respectively.

\item The stellar evolutionary tracks to be used are dependent on the age of the cluster considered. For ages at which the TP-AGB phase is reached (${\sim} 0.1 - 2$ Gyr), the Padova TP-AGB tracks described in \citet{Girardi2000} and \citet{VW1993} should be used. At lower stellar ages such as those considered in this paper, we favour the Geneva tracks for their sensible correction to the definition of effective temperature at the W-R phase. The lack of an effective temperature correction for W-R stars in the Padova tracks leads to the same definition of $T_\mathrm{eff}$ for all stars, resulting in an order of magnitude of ${\sim} 8$ difference in the FUV spectra between the two sets of tracks as calculated by SLUG.

\item SPS modelling still needs a large improvement in the high stellar mass regime. The sparse grid points at high stellar masses that aid in the interpolation of the stellar isochrone result in large uncertainties in the spectrum. Overall, both SPS codes Starburst99 and SLUG are suitable, as the differences produced between the two are negligible. However, the use of the Akima spline in SLUG results in less systematic uncertainties from outliers in the SLUG output. Similarly, the differences in emission-line ratios produced by BPASS are small when compared to single stellar SPS codes.
\end{itemize}

\subsubsection{Intrinsic H \textsc{ii} Region Parameters}

The spread of the SDSS starburst galaxies as modelled by the photoionisation grids seen in Figures~\ref{fig:classicgrid} and~\ref{fig:diffpres} is largely affected by variations in parameters associated with the H \textsc{ii} region. The starburst spread in SDSS, defined as the galaxies below the \citet{Kauffmann2003} line, incorporates a large range of metallicities (${\sim} 0.028 \pm 0.002$ in $Z$) and ionisation parameter (${\sim} 2 \pm 0.13$ dex in log$(q(N)/(\mathrm{cm}\;  \mathrm{s}^{-1}))$). 

Seen in Figure~\ref{fig:diffpres}, models with a pressure of log($P/k$) $\approx 6$, which approximately corresponds to a density of $n = 100$ cm$^{-3}$ cover the majority of the starburst spread. The extent of the $n = 100$ cm$^{-3}$ model grid does not reach the \citet{Kauffmann2003} line, however, and hence there are areas of the star-forming region of the BPT that are yet to be explained. Increasing the density and hence pressure of the H \textsc{ii} region does increase the strength of the emission-line ratios, offering a larger covering range on the BPT diagram. Overall, all galaxies in the SDSS star-forming sequence can be explained with a pressure between $6 \lesssim \mathrm{log}(P/k) \lesssim 7$. However, increasing the pressure of the H \textsc{ii} region to pressures of log($P/k$) ${\sim}$ 7 and above forces the models to include galaxies that may have a possible nonstellar (i.e. AGN) contribution to their emission-line spectrum. A discussion on the various other sources powering the emission from galaxies along the mixing sequence can be found in Section~\ref{sec:discother}.

Different types of H \textsc{ii} region boundedness do not better explain the spread of SDSS galaxies, of the boundedness types we have explored. At the metallicities of the star-forming SDSS galaxies ($0.002 \lesssim Z \lesssim 0.030$), the model grids are all largely similar for different bound situations at the same pressure, seen in Figures~\ref{fig:boundcomp}. Shown in Figure~\ref{fig:diffpres} are three model grids of $\mathrm{log}(P/k) \approx 6 \Rightarrow n = 100$ cm$^{-3}$, $\mathrm{log}(P/k) \approx 7 \Rightarrow n = 1,000$ cm$^{-3}$, and $\mathrm{log}(P/k) \approx 8 \Rightarrow n = 10,000$ cm$^{-3}$, computed in a radiation-bounded situation truncated at 99\% of H \textsc{ii} recombination. 

As explained and shown in Section~\ref{sec:boundbpt} with the use of the combined SIGRID and \citet{Izotov2006,Izotov2012} low-metallicity sample, density-bounded situations only appear applicable at extremely low metallicities. At the metallicities of the star-forming SDSS galaxies, a radiation-bounded situation is sufficient to explain and encompass the spread of galaxies.

Variations in the metallicity and ionisation parameter values used in the models have not been explored in this paper; however, the star-forming sequence in SDSS is adequately explained by the values of metallicity ($Z = 0.001 - 0.040$ inclusive) and ionisation parameter (log($q(N)/(\mathrm{cm}\;  \mathrm{s}^{-1})) = 6.5 - 8.5$ inclusive) used in our models.

\subsubsection{Massive-star Lifetimes}
\label{sec:disclife}

Whilst the stochasticity associated with star formation incorporated in SLUG may be seen as an improvement on the SB99 method, there are still fundamental issues surrounding the creation of an ionising spectrum through SPS. The massive stars that produce the majority of the ionising radiation in stellar populations are rare; hence, IMF sampling is an issue (discussed in Section \ref{sec:spscodes}). These massive stars also evolve rapidly, with short lifetimes. This rapid evolution, along with their rarity and obscuration by their still clearing birth clouds, provides difficulty in obtaining constraining observations, and hence the models are relatively unconstrained. Also, the very strong stellar winds from massive stars require a very finely detailed modelling of their atmospheres. These issues surrounding the modelling of massive stars lead to systematic uncertainties with the resulting emission lines from star forming regions.

\subsection{Further Ionising Processes}
\label{sec:discother}

It is possible that objects that lie below the \citet{Kewley2001} line may have contributions in their emission from non-star-formation-related processes. AGNs, shocks, and DIG all provide a means of altering the strong-line-to-hydrogen-line ratios. Shocks and radiation from AGNs provide an increased collisional rate that increases the strength of the strong, forbidden/collisionally excited lines. 

The theory of DIG is still largely a mystery, with a common explanation not yet having been agreed upon. There is much evidence for the idea of `leaky H \textsc{ii} regions' from density-bounded H \textsc{ii} regions \citep{Ferguson1996, OK1997, Zurita2002}; however, simulations have also shown the formation of DIG through other processes, such as magnetic recombination \citep{Hoffmann2012} and radiative cooling \citep{Bordoloi2016}. 

\section{Conclusions}
\label{sec:conc}

Using \textsc{mappings v}, we have explored the common parameters used in the production of photoionisation model grids. We explore the effects of the variations on the resulting emission-line flux ratios in the optical emission-line diagnostic diagrams. Further, we search the parameter space for models that explain the spread of galaxies in SDSS and in a combined metal-poor sample consisting of data from the SIGRID survey \citep{Nicholls2011} and data from \citet{Izotov2006,Izotov2012}. We find the following:

(i) Variations in the parameters associated with the ionising radiation field (SPS code, stellar atmospheres, stellar evolutionary tracks, IMF) are small, with an average systematic difference of ${\sim} 0.1$ dex in optical emission-line ratio between varying models. In this situation where spectra resulting from varying models are similar, we make the following recommendations:

\begin{itemize}
\item Kurucz + Hillier + Pauldrach atmospheres for the updated W-R and OB stellar input physics.
\item Geneva HIGH mass-loss evolutionary tracks for younger stellar populations due to the W-R effective temperature definition correction. Padova TP-AGB evolutionary tracks for older stellar populations, where the thermally pulsing AGB phase is reached.
\item Either SLUG or SB99 are suitable for modelling single stellar populations. 
\end{itemize}

(ii) Similarly, both photoionisation modelling codes \textsc{mappings} and \textsc{cloudy} are suitable for modelling H \textsc{ii} regions. The difference in emission-line ratios produced between the two codes is small, on the order of ${\sim} 0.1$ dex

(iii) Including the ionising input spectrum parameters listed above, the main star-forming spread of SDSS can be explained by using a photoionisation model with metallicity ranging from $Z = 0.001$ to 0.040 inclusive, ionisation parameter ranging from log($q(N)/(\mathrm{cm}\;  \mathrm{s}^{-1})$) = 6.5 to 8.5 inclusive, and log($P/k) \approx 6$, corresponding to an initial electron density of approximately $n = 100 \;\mathrm{cm}^{-3}$ under the assumption of an initial temperature of 8000 K, and computed assuming a radiation-bounded situation where truncation occurs at 99\% H \textsc{ii} recombination. This is supported by work done by \citet{Kewley2001} that shows the electron density within local galaxies to be $n = 350 \;\mathrm{cm}^{-3}$. 

(iv) The position of high-redshift galaxies on the BPT \citep[e.g.][]{Kewley2013b} is well explained by models at higher pressure. We find star-forming high-redshift galaxies to be well explained by photoionisation models containing a pressure value of $\mathrm{log}(P/k) \approx 7$. For comparison, the star-forming sequence of SDSS is well explained by models with $\mathrm{log}(P/k) \approx 6$. This supports work showing high-redshift galaxies to be observed with a higher electron density \citep[e.g.][]{Kashino2017,Onodera2016}.

(v) Low-metallicity galaxies can be explained using a density-bounded H \textsc{ii} region model rather than a radiation-bounded one. From the mass-metallicity relation, these low-metallicity galaxies have lower stellar masses and hence contain a lower gravitational binding energy, leading to an increase in the number of escaped photons. Our density-bounded models show better agreement with the low-mass combined sample from the SIGRID survey \citep{Nicholls2011} and  \citet{Izotov2006,Izotov2012}, when compared to the radiation-bounded model used by \citet{Nicholls2014c}.

Our future work includes using the findings of this paper to model H \textsc{ii} regions, in order to separate star-forming emission from other sources, such as AGNs and shocks. The theoretical nature of the H \textsc{ii} region models allows us to discern physical properties about the star-forming regions of a galaxy, in addition to quantifying the amount of star-formation in IFU spaxels.

\section*{Acknowledgements}

Parts of this research were conducted by the Australian Research Council Centre of Excellence for All Sky Astrophysics in 3 Dimensions (ASTRO 3D), through project No. CE170100013. B.G. gratefully acknowledges the support of the Australian Research Council as the recipient of a Future Fellowship (FT140101202).




\bibliographystyle{aasjournal}
\bibliography{2016} 





\end{document}